\documentclass[reprint,amsmath,amssymb, aps]{revtex4-2}
\usepackage{graphicx,epstopdf,siunitx,braket,xspace,amssymb,dsfont,bm,natbib,xfrac,nicefrac,miller,gensymb}
\usepackage{xr}
\usepackage{todonotes}
\usepackage[colorlinks=true, allcolors=blue]{hyperref}
\DeclareMathSymbol{\mhyphen}{\mathord}{AMSa}{"39}
\newcommand{\TRion}[0]{\ensuremath{T_{2\mathrm{n+}}^{*}}\xspace}
\newcommand{\Sb}[0]{$^{123}$Sb\xspace}
\newcommand{\Ph}[0]{$^{31}$P\xspace}
\newcommand{\gamman}[0]{\ensuremath{\gamma_{\mathrm{n}}}\xspace}
\newcommand{\gammae}[0]{\ensuremath{\gamma_{\mathrm{e}}}\xspace}

\newcommand{\VDCDG}[0]{\ensuremath{V_{\mathrm{DC}}^{\mathrm{DG1}}}\xspace}

\RequirePackage[T1]{fontenc}
\RequirePackage[utf8]{inputenc}
\RequirePackage{lmodern}
\DeclareSIUnit \dbm {dBm}
\graphicspath{{./images/}}

\begin{document}

\title{Navigating the 16-dimensional Hilbert space of a high-spin donor qudit with electric and magnetic fields}

\author{Irene Fern\'andez de Fuentes \textsuperscript{1}}
\author{Tim Botzem \textsuperscript{1}}
\author{Mark A. I. Johnson\textsuperscript{1}}
\author{Arjen Vaartjes \textsuperscript{1}}
\author{Serwan Asaad \textsuperscript{1}}
\author{Vincent Mourik \textsuperscript{1}}
\author{Fay E. Hudson \textsuperscript{1}}
\author{Kohei M. Itoh\textsuperscript{2}}
\author{Brett C. Johnson\textsuperscript{3}}
\author{Alexander M. Jakob\textsuperscript{3}}
\author{Jeffrey C. McCallum\textsuperscript{3}}
\author{David N. Jamieson\textsuperscript{3}}
\author{Andrew S. Dzurak\textsuperscript{1}}
\author{Andrea Morello\textsuperscript{1}}
\affiliation{\textsuperscript{1} ARC Centre for Quantum Computation and Communication Technology (CQC2T), School of Electrical Engineering and Telecommunication, University of New South Wales, Sydney, NSW, Australia}
\affiliation{\textsuperscript{2}School of Fundamental Science and Technology, Keio University, Minato City, Yokohama, Japan}
\affiliation{\textsuperscript{3}School of Physics, University of Melbourne, Melbourne, Victoria, Australia}

\date{\today}


\begin{abstract}
Efficient scaling and flexible control are key aspects of useful quantum computing hardware. Spins in semiconductors combine quantum information processing with electrons, holes or nuclei, control with electric or magnetic fields, and scalable coupling via exchange or dipole interaction. However, accessing large Hilbert space dimensions has remained challenging, due to the short-distance nature of the interactions. Here, we present an atom-based semiconductor platform where a 16-dimensional Hilbert space is built by the combined electron-nuclear states of a single antimony donor in silicon. We demonstrate the ability to navigate this large Hilbert space using both electric and magnetic fields, with gate fidelity exceeding 99.8\% on the nuclear spin, and unveil fine details of the system Hamiltonian and its susceptibility to control and noise fields. These results establish high-spin donors as a rich platform for practical quantum information and to explore quantum foundations.
\end{abstract}

\maketitle


For computing purposes, one of the key property of quantum systems is that the dimension $D$ of the computational space -- in this case, the Hilbert space -- can grow exponentially with the number $N$ of physical qubits, i.e. as $D=2^N$. Unlike in a classical computer, where each additional bit simply adds one dimension to the data array, in a quantum computer each qubit multiplies the Hilbert space dimension by two. In practice, whether this is actually the case depends upon creating maximally entangled states with high fidelity, which in turn is a delicate function of the physical layout of the qubits and the details of the interaction between them.

An alternative quantum computing paradigm starts with physical components whose intrinsic Hilbert space dimension is $d>2$, thus called qu\textit{d}its \cite{wang2020qudits}. Using qudits, a $D$-dimensional Hilbert space can be constructed with a factor $\log_2 d$ smaller number of physical units compared to the qubit case. Circuit complexity can be reduced even further; using two-qudit gates, an $N$-dimensional unitary operator $U$ can be simulated using a factor $(\log_2 d)^2$ less gates as compared to its qubit-based counterpart\,\cite{Muthukrishnan2000}. General schemes exist to perform fault-tolerant operations in a way that takes advantage of a larger $d$ \cite{campbell2014}, and to compile various quantum algorithms in a resource-efficient way \cite{bullock2005,nikolaeva2021efficient}. Experimental qudit platforms can be found in optics \cite{lu2022bayesian,chi2022programmable}, superconductors \cite{neeley2009emulation,yurtalan2020,goss2022high}, trapped ions \cite{ringbauer2022universal}, atomic ensembles \cite{anderson2015} and molecular magnets \cite{godfrin2018generalized}. 

Here we present a physical platform for high-dimensional qudit encoding in a silicon nanoelectronic device. Silicon quantum devices \cite{zwanenburg2013silicon} host spin qubits that combine exceptionally long coherence times \cite{veldhorst2014addressable}, exceeding 30 seconds in nuclear spins \cite{muhonen2014storing}, one- and two-qubit gate fidelities above 99\% \cite{mkadzik2022precision,noiri2022fast,xue2022quantum,mills2022two}, and compatibility with the manufacturing processes that underpin the established semiconductor industry \cite{zwerver2022qubits}. Electron spin qubits can be controlled using both magnetic \cite{pla2012single,veldhorst2014addressable} (Electron Spin Resonance, ESR) and electric \cite{noiri2022fast,xue2022quantum,mills2022two} (Electric Dipole Spin Resonance, EDSR) fields; nuclear qubits are normally driven by Nuclear Magnetic Resonance \cite{pla2013high} (NMR), but quadrupolar nuclei can exhibit Electric \cite{asaad2020coherent} (NER) or even Acoustic \cite{o2021engineering} (NAR) resonances. Magnetic drive lends itself to global control methods, where a spatially extended oscillating magnetic field drives multiple qubits \cite{laucht2015electrically,vahapoglu2022coherent}, whereas electric drive is easier to localise at the nanometre scale.

Our chosen qudit platform is the antimony donor in silicon, $^{123}$Sb:Si. Our initial interest for this system was in the context of fundamental studies on quantum chaos \cite{mourik2018exploring}. The serendipitous discovery of nuclear electric resonance \cite{asaad2020coherent} and the steady development of ideas to use high-spin nuclei in quantum information processing \cite{chiesa2020molecular,gross2021designing,gross2021hardware} highlighted the unique opportunity to use $^{123}$Sb as a qudit which exploits all the benefits and flexibility of silicon quantum electronic devices. 

In this work we show magnetic and electric control over the 16-dimensional Hilbert space of the combined electron and nuclear spin of the $^{123}$Sb donor, benchmark quantum gate fidelities, and provide detailed understanding of the microscopic physics that governs the behaviour of this novel qudit system.

\subsection*{THE ANTIMONY DONOR}

Like phosphorus \cite{pla2012single,pla2013high}, arsenic \cite{franke2015interaction} and bismuth \cite{morley2010initialization}, antimony is a group-V donor in silicon. It behaves as a hydrogenic impurity where the Coulomb potential of the nuclear charge loosely binds an electron in a $1s$-like orbital \cite{zwanenburg2013silicon}. The $^{123}$Sb isotope of antimony possesses a nuclear spin $I=7/2$, with gyromagnetic ratio $\gamma_{\rm n}=5.55$~MHz/T. The non-spherical charge distribution in the nucleus creates an electric quadrupole moment $q_{\rm n}=[-0.49, -0.69]\times 10^{-28}$~m$^2$ \cite{mourik2018exploring}. The $S=1/2$ spin of the donor-bound electron has a gyromagnetic ratio $\gamma_{\mathrm{e}}\approx 27.97$~GHz/T, and is magnetically coupled to the nuclear spin via the Fermi contact hyperfine interaction $A\boldsymbol{\hat{S}}\cdot\boldsymbol{\hat{I}}$, with $A = 101.52$~MHz in bulk silicon. 

The charge state of the donor can be easily modified by placing it in a nanoelectronic device, where metallic electrodes lift the donor electrochemical potential $\mu_{\rm D}$ above the Fermi level of a nearby charge reservoir, thus energetically favouring the weakly bound electron to leave the donor. The resulting ionised (positively charged) $D^+$ donor, placed in a magnetic field $B_0$ oriented along the Cartesian $z$-axis, has the following static Hamiltonian:
\begin{equation}
\label{eq:H_ion}
\hat{\mathcal{H}}_{D^{+}} = -B_{0}\gamman\hat{I}_z+\sum_{\alpha,\beta\in\{x,y,z\}} Q_\mathrm{\alpha\beta} \hat{I}_\mathrm{\alpha}\hat{I}_\mathrm{\beta},
\end{equation}
where $\alpha,\beta=\{x,y,z\}$ are Cartesian axes, $\hat{I}_{\alpha}$ are the corresponding 8-dimensional nuclear spin projection operators,
and $Q_\mathrm{\alpha\beta} = \frac{\mathrm{e}q_\mathrm{n}\mathcal{V}_{\alpha\beta}}{2 I(2I-1)h}$ is the nuclear quadrupole interaction energy, governed by the electric field gradient (EFG) tensor $\mathcal{V}_{\alpha\beta}=\partial^2 V(x,y,z)/\partial \alpha \partial \beta$. The quadrupole interaction introduces an additional orientation-dependent energy shift to the nuclear Zeeman levels (Fig.~1a), allowing for the individual addressability of nuclear states even in the ionised case \cite{ono2013coherent,franke2015interaction}.

In the charge-neutral state $D^{0}$, the system Hamiltonian $\mathcal{H}_{D^{0}}$ becomes a 16-dimensional matrix expressed in terms of the tensor products of the electron and nuclear spin operators:
\begin{equation}
\label{eq:H_neut}
\hat{\mathcal{H}}_{D^{0}} = B_{0}\left(-\gamman\hat{I}_{z}+\gammae \hat{S}_{z}\right)+A\boldsymbol{\hat{S}}\cdot\boldsymbol{\hat{I}}+\sum_{\alpha\beta\in\{x,y,z\}} Q_\mathrm{\alpha\beta} \hat{I}_\mathrm{\alpha}\hat{I}_\mathrm{\beta}.
\end{equation}

\begin{figure*}[htbp]
\includegraphics[width=\textwidth]{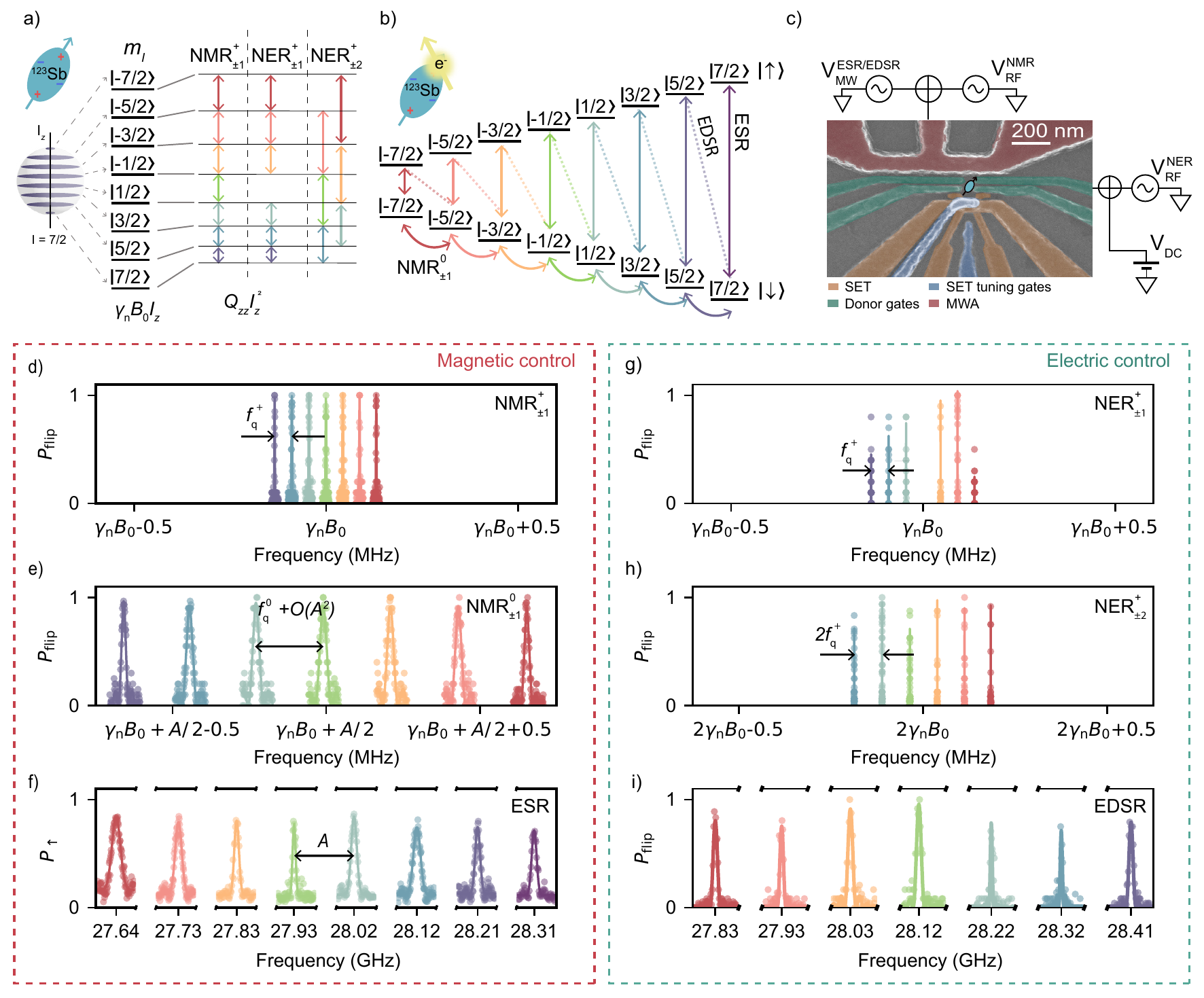}
\caption{\textbf{Spectrum of the \Sb atom.} \textbf{a)} Energy diagram of the ionised \Sb atom. The arrows indicate the allowed spin transitions for the different driving mechanisms, including NMR$^{+}_{\pm 1}$, NER$^{+}_{\pm 1}$ for $\Delta m_I = \pm 1$ and NER$^{+}_{\pm 2}$ for $\Delta m_I = \pm 2$, where $+$ denotes the charge state of the nucleus. The colors of all lines and symbols reflect the initial $\ket{m_I-1}$ state of each spin transition, and are used consistently across this manuscript. The Zeeman energy $\gamman B_0 \hat{I}_z$ ($\approx5.5$~MHz in this work) yields equispaced nuclear levels, but the quadrupole coupling, written for simplicity as $Q_{zz}\hat{I}_z^2$, shifts the resonance frequencies according to $m_I$ and allows their individual addressing. \textbf{b)} Energy diagram of the \Sb atom in the neutral charge state.The NMR$^{0}_{\pm 1}$ transitions are represented by curved arrows, while the ESR is depicted by vertical solid arrows, and the EDSR is indicated by dashed arrows. \textbf{c)} False-coloured scanning electron micrograph of a device identical to the one used for the experiments. The ESR, EDSR and NMR driving signals are applied to the microwave antenna (MWA), whereas the NER drives are applied to one of the open-circuited gates. The green ellipse depicts the approximate location of the implanted donor with respect to the surface gates.  \textbf{d)} Experimental NMR$^+$ spectrum for the ionised donor, showing 7 resonant peaks. The distance between adjacent peaks is given to first order by the quadrupolar splitting $f_{\rm q}^{\rm +}=-44.1(2)$~kHz. \textbf{e)} NMR spectrum for the neutral atom, split by the quadrupolar interaction $f_{\rm q}^{\rm n0}=-52.5(5)$~kHz and second order contributions of the hyperfine interaction $\propto A^{2}/ \gamma_{\mathrm{n}}B_0$. We use the same frequency range in the $x-$axis for panels \textbf{d)} and \textbf{e)} to highlight the effect of the hyperfine interaction on the separation of the resonances in the neutral case. \textbf{f)} ESR spectrum, showing 8 resonance peaks depending on the nuclear projection $m_I$, split to first order by the hyperfine interaction $A$. \textbf{g)} NER$^{+}_{\pm 1}$ spectrum for the ionised donor. The transition $m_{-1/2}\leftrightarrow m_{1/2}$ is forbidden by NER. \textbf{h)} NER$^{+}_{\pm 2}$ spectrum, with frequencies $f_{m_I-2\leftrightarrow m_{I}}^{\rm{NER}^{+}} = f_{m_I-2\rightarrow m_I-1}^{\rm{NER}^{+}}+f_{m_{I}-1\rightarrow m_{I}}^{\rm{NER}^{+}}$. \textbf{i)} EDSR spectrum, showing 7 electron-nuclear resonances that conserve $m_I + m_S$. In all panels from d) to f), the resonance lines are power-broadened.} 
\label{fig:spectrum}
\end{figure*}

We operate the device in a magnetic field $B_0 \approx 1$~T, which ensures that the eigenstates of $\hat{\mathcal{H}}_{D^{+}}$ (Fig.~\ref{fig:spectrum}a) are well approximated by the eigenstates $\ket{m_I}$ of $\hat{I}_z$ ($m_I = -7/2, -5/2 ..., +7/2$) because $\gamman B_0 \gg Q_{\alpha\beta}$, and the eigenstates of $\hat{\mathcal{H}}_{D^{0}}$ (Fig.~\ref{fig:spectrum}b) are approximately the tensor products of $\ket{m_I}$ with the eigenstates $\{\ket{\downarrow},\ket{\uparrow}\}$ of $\hat{S}_z$  because $\gammae B_0 \gg A \gg Q_{\alpha\beta}$. The latter condition implies $\hat{\mathcal{H}}_{D^{0}} \approx B_{0}(-\gamman\hat{I}_{z}+\gammae \hat{S}_{z}) + A\hat{S}_{z}\hat{I}_{z}$ ensuring that the nuclear spin operator approximately commutes with the electron-nuclear interaction. This condition allows for nearly quantum nondemolition (QND) readout of the nuclear spin via the electron spin ancilla \cite{pla2013high} (see Supplementary Information, Section 1 for deviations from QND condition).


A key feature of this work is that coherent transitions between the $^{123}$Sb spin eigenstates can be induced by both magnetic and electric fields, on both the electron and the nuclear spin. Electron spin resonance  (ESR) \cite{pla2012single} is achieved by adding the driving term $\hat{\mathcal{H}}^{\mathrm{ESR}} = B_{1}\gammae \hat{S}_{x} \cos (2\pi f^{\rm ESR}_{m_I} t)$ to $\hat{\mathcal{H}}_{D^{0}}$, where $B_1$ is the amplitude of an oscillating magnetic field at one of the eight resonance frequencies $f^{\rm ESR}_{m_I}$ determined by the nuclear spin projection $m_I$. Similarly, nuclear magnetic resonance (NMR) \cite{pla2013high} requires a magnetic drive term $\hat{\mathcal{H}}^{\mathrm{NMR}} = B_{1}\gamman \hat{I}_{x} \cos (2\pi f^{\rm NMR}_{m_I - 1 \leftrightarrow m_I} t)$, applicable to both the neutral (NMR$^0_{\pm 1}$) and the ionised (NMR$^+_{\pm 1}$) case. The $\pm 1$ subscript indicates that such transitions change the nuclear spin projection by one quantum of angular momentum, i.e. $\Delta m_I = \pm 1$. 

Electrically driven spin transitions can be obtained in two ways. One, involving the combined state of electron and nucleus, is the high-spin generalization of the `flip-flop' transition demonstrated recently in the $I=1/2$ $^{31}$P system \cite{savytskyy2023electrically}. An oscillating electric field $E_1 \cos (2 \pi f_{m_I-1\leftrightarrow m_I}^{\rm EDSR}t)$ induces electric dipole spin resonance transitions (EDSR) in the neutral donor by time-dependently modulating the hyperfine interaction $A(E_1)\hat{S}_{\pm}\hat{I}_{\mp}$ via the Stark effect \cite{pica2014hyperfine}, where the $\pm$ subindices indicate the rising and lowering operators, respectively. This mechanism preserves the total angular momentum of the combined electron-nuclear states. Therefore, the EDSR transitions appear as diagonal (dashed) lines in Fig.~\ref{fig:spectrum}b. The second electrical transition, called nuclear electric resonance (NER) \cite{asaad2020coherent} acts on the nucleus alone. It exploits the modulation of electric quadrupole coupling terms involving the operators $\hat{I}_z \hat{I}_{\pm}$ for transitions with $\Delta m_I = \pm 1$ (NER$_{\pm 1}$), and $\hat{I}_{\pm}^2$ for transitions with $\Delta m_I = \pm 2$ (NER$_{\pm 2}$). The microscopic mechanism by which the electric field  $E_1 \cos (2 \pi f_{m_I-1\leftrightarrow m_I}^{\rm NER}t)$  creates a time-dependent electric field gradient at the nucleus was understood to arise from the distortion of the atomic bond orbitals, in a lattice site lacking point inversion symmetry \cite{asaad2020coherent}. The energy level structure of the neutral and ionised $^{123}$Sb results in a total of 54 resonant transitions, the frequencies of which are listed in Table\,\ref{tab:1}. 


To manipulate and read out the 16-dimensional Hilbert space of the single \Sb, we use a silicon nanoelectronic device as shown in Fig.~\ref{fig:spectrum}c (fabrication details in Supplementary Section 2). The device features a single electron transistor (SET) to read out the spin of the donor-bound electron \cite{morello2010single}, a set of gates to control the electrostatic potential of the donor or drive NER\,\cite{asaad2020coherent}, and a broadband short-circuited microwave antenna used to deliver the $B_1$ field for ESR and NMR. To drive the donor spins electrically at microwave frequencies via EDSR, we exploit the stray electric fields from the microwave antenna.

\subsection*{RESONANCE SPECTRA AND ENERGY LEVEL ADDRESSABILITY}
The spin resonance spectrum of the ionised nucleus is reported in Fig.~\ref{fig:spectrum}d (NMR$^+_{\pm 1}$) and Fig.~\ref{fig:spectrum}g (NER$^+_{\pm 1}$). The spectra are of course identical, except for the absence of the $m_I = -1/2 \leftrightarrow +1/2$ transition in the NER$^+_{\pm 1}$ case, due to the selection rules imposed by modulation of the quadrupole interaction \cite{asaad2020coherent}. The static quadrupole splitting $f_{\rm q}^{+}=-44.1(2)$\,kHz is obtained directly from the distance between adjacent peaks. The presence of a nonzero quadrupole splitting ensures that all pairs of nuclear levels are individually addressable, as required for complete SU(8) control of the qudit \cite{wang2020qudits}. We know the sign of $f_{\rm q}^{+}$ because we are able to deterministically initialise a specific nuclear state $\ket{m_I}$ through a combination of ESR and EDSR transitions (see Supplementary Information, Section 3) and thus identify the $\ket{7/2} \leftrightarrow \ket{5/2}$ transition as the one at the lowest frequency. The numerical value of $f_{\rm q}^{+}$ is close to that observed in a similar device \cite{asaad2020coherent} and is well understood as arising from the EFG produced by static strain in the device as a consequence of the differential thermal expansion of the aluminium gates placed over the silicon \cite{thorbeck2015formation,asaad2020coherent}. 

The NMR frequency for $m_I = -1/2 \leftrightarrow +1/2$ is equal to the Zeeman splitting $\gamman B_0$, without contributions from the quadrupole interaction. This allows us to accurately calibrate the static magnetic field value, $B_0 = 999.5(5)$~mT, which is provided by an array of permanent magnets \cite{adambukulam2021ultra} and thus not precisely known a priori.

When the donor is in the charge neutral state, the NMR$^0_{\pm 1}$ frequencies are shifted equally to first order by the hyperfine interaction, and further split by second-order hyperfine terms $O(A^2) \propto A^2/\gammae B_0$, depending on the nuclear spin projection (see Supplementary Information, Section 4). This can be appreciated in Fig.~\ref{fig:spectrum}e where the frequency axis has been offset by the linear contribution of the hyperfine coupling $A/2$, which is equal for all the transitions. Plotting the NMR$^+_{\pm 1}$ (Fig.~\ref{fig:spectrum}d) and the NMR$^0_{\pm 1}$ (Fig.~\ref{fig:spectrum}e) spectra across the same frequency spread $\approx \pm 1$~MHz highlights that, in the neutral case, the splitting caused by the $O(A^2)$ terms is much larger than $f_{\rm q}^{+}$, proving that all NMR$^0_{\pm 1}$ transitions would be individually addressable even in the absence of quadrupole effects. From the NMR$^0_{\pm 1}$ spectrum we extract $A = 96.584(2)$~MHz and $f_{\rm q}^{0} = -52.5(5)$~kHz (see Supplementary Information, Section 4 for calculation details). The quadrupole splitting thus differs by $\approx 8$~kHz between the neutral and the ionised donor case. This could be due to a small additional EFG contribution from the electron wavefunction, which is itself distorted from its $1s$ symmetry by the local strain \cite{franke2016quadrupolar}


The eight ESR resonances (Fig.~\ref{fig:spectrum}f), each conditional on one of the $m_I$ nuclear spin projections, are split by the hyperfine interaction $A\boldsymbol{\hat{S}}\cdot\boldsymbol{\hat{I}}$. A detailed calculation (see Supplementary Information, Section 5) shows that both first- and second-order terms in $A$ contribute to the ESR frequency splitting, whereas only the resonances conditional on $m_I = \pm 1/2$ are separated by exactly $A$. We also observe the seven expected EDSR flip-flop transitions (Fig.~\ref{fig:spectrum}g), where both the electron and nucleus undergo simultaneous spin flips with $\Delta(m_I + m_S)=0$, driven by the electrical modulation of the hyperfine interaction. 

\subsection*{COHERENT NUCLEAR SPIN CONTROL}
\begin{table*}[htbp!]
     \centering
    \includegraphics[width=\textwidth]{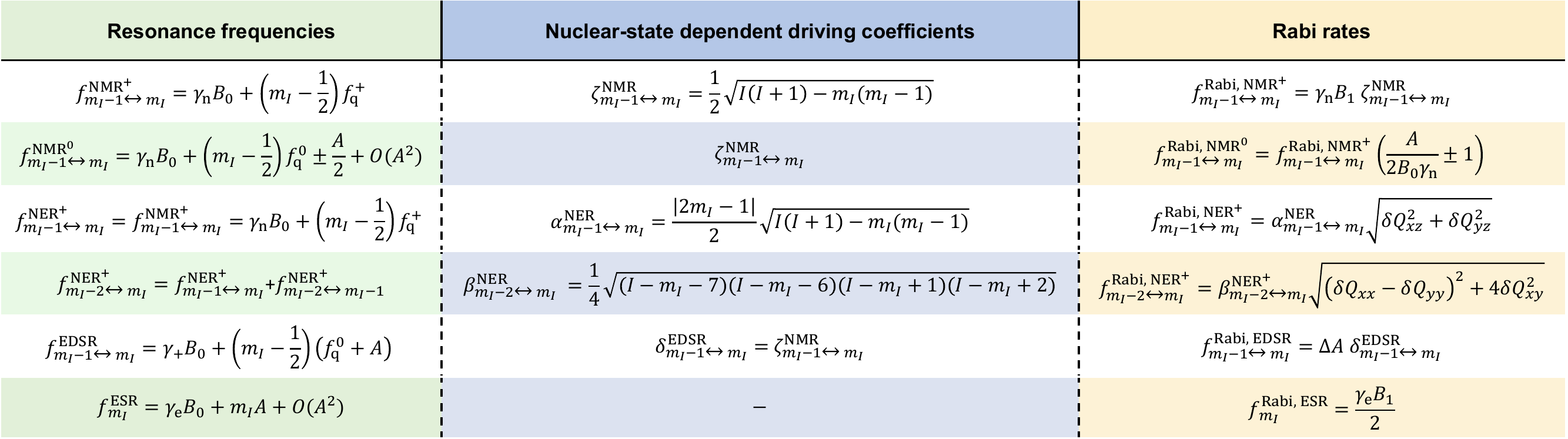}
    \caption{Resonance frequencies, nuclear-state dependent scaling coefficients and Rabi rates  for the different spin driving mechanisms of the \Sb donor, including electric (NER$_{\pm1,\pm2}$, EDSR) and magnetic (NMR, ESR) control. Here $I = 7/2$, $m_I = \{-I,-I+1...,I\}$ and $\gamma_{+} = \gamma_{\rm n}+ \gamma_{\rm e}$, where $\gamma_{\rm n} = 5.55$\,MHz and $\gamma_{\rm e}= 27.97$\,GHz. } 
\label{tab:1}
\end{table*}

\begin{figure*}[htbp!]
     \centering
    \includegraphics[width=\textwidth]{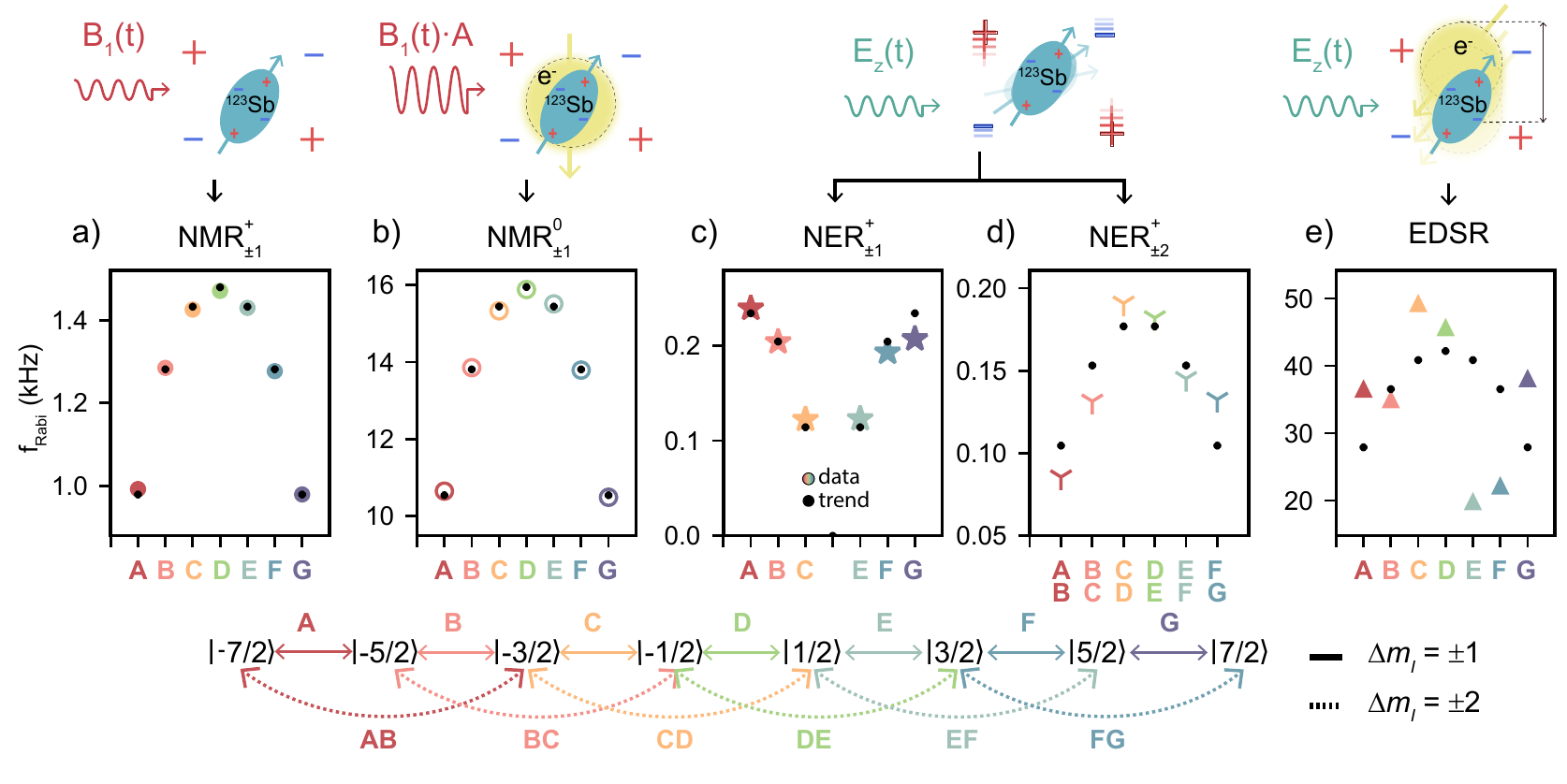}
    \caption{\textbf{Coherent magnetic and electric drive of the \Sb nuclear spin.} \textbf{a)} Rabi trends obtained when driving the ionised nucleus with an oscillating magnetic field, through NMR. \textbf{b)} Trends for NMR on the charge-neutral atom. The hyperfine-enhanced nuclear gyromagnetic ratio yields faster Rabi oscillations for same $B_1$ amplitude. The experiments in panels \textbf{a)} and \textbf{b)} were carried out by applying a voltage of $V^{\rm pp}_{\rm RF} = 50$\,mV to the input of the on-chip antenna. Here we solely account a 10\,dB attenuation occurring at the 4K stage of the dilution refrigerator. \textbf{c), d)} Rabi frequencies obtained by driving the nucleus via NER, through the electrical modulation of the quadrupolar interaction for \textbf{c)} $\Delta m = 1$, and \textbf{d)} $\Delta m=2$. In both cases, this is achieved by applying an oscillating voltage with an amplitude of $V^{\rm pp}_{\rm RF} = 60$~mV to a donor gate. \textbf{e)}  Stray electric fields from the microwave antenna (-6~dBm at source) are used to drive electron-nuclear spin transitions coherently (through EDSR). The physical mechanisms that drive the nuclear spins is illustrated above each panel. We label and color code the nuclear spin transitions using the diagram below the panels.} 
\label{fig:nuclear_drive}
\end{figure*}

Having identified all the resonance frequencies of the $^{123}$Sb system, we demonstrate five different methods of driving coherent rotations on the nuclear spin qudit, including NMR for the ionised (NMR$^{+}_{\pm 1}$) and neutral (NMR$^{0}_{\pm 1}$) atom, ionised NER$^{+}_{\pm 1, \pm 2}$, and EDSR (Fig.\,\ref{fig:nuclear_drive}). A notable feature of magnetic and electric drive in high-spin systems is the dependence of the Rabi frequencies on the nuclear spin number $m_I$, which arises from the distinct transition matrix elements in the driving operators\,\cite{asaad2020coherent}. Table~\ref{tab:1} summarizes the nuclear-spin dependent scaling coefficients and driving amplitudes for the different driving mechanisms.

With magnetic (NMR) drive, the oscillating magnetic field $B_1 \cos(2\pi f^{\rm NMR}_{m_I - 1 \leftrightarrow m_I}t)$ couples to the nuclear spin via the off-diagonal matrix elements of the $\hat{I}_x$ spin operator. Therefore, the Rabi rates are expected to increase for smaller $|m_I|$, in both the ionised and neutral case, as observed in the data in Fig.~\ref{fig:nuclear_drive}~a,b. We find the neutral donor Rabi rates to be enhanced with respect to the ionised case by a factor $f_{\mathrm{Rabi}}^{\mathrm{NMR0}}/f_{\mathrm{Rabi}}^{\mathrm{NMR+}}=10.776(8)$, which is consistent with a hyperfine-enhanced nuclear gyromagnetic ratio~\cite{sangtawesin2016hyperfine}. This is a consequence of electron-nuclear state mixing through the transverse term of the hyperfine interaction, $A \hat{S}_x\hat{I}_x$, which effectively creates an additional driving field of magnitude $\frac{A B_1}{2\gamman B_0}$ along the $x$ axis, adding to the external $B_1$. Using the measured values of $A$ and $B_0$, this mechanism predicts an increase in Rabi rates $f_{\mathrm{Rabi}}^{\mathrm{NMR^0}} = f_{\mathrm{Rabi}}^{\mathrm{NMR^+}}(1+\frac{A}{2\gamman B_0}) \approx 9.6$. The slight discrepancy with the measured enhancement is likely due to a different frequency response of the driving circuitry at $f^{\rm NMR^+} \approx 5.5$~MHz and $f^{\rm NMR^0} \approx 54$~MHz.

For electrical drive with $\Delta m_I = \pm 1$ (NER$^+_{\pm 1}$), the relevant transition matrix elements come from  the quadrupolar interaction involving the operators: $\hat{I}_x\hat{I}_z$, $\hat{I}_z\hat{I}_x$, $\hat{I}_y\hat{I}_z$, $\hat{I}_z\hat{I}_y$. The fastest Rabi rates in this case are found at larger $|m_I|$, whereas the $\ket{-1/2}\leftrightarrow \ket{1/2}$ transition is completely forbidden. This behaviour is reflected in Fig.~\ref{fig:nuclear_drive}d, showing the expected decreasing Rabi rates for lower $|m_I|$, and the missing value for the middle transition. The `double transition' NER$^+_{\pm 2}$ is obtained by modulating quadratic terms of the form $\hat{I}_{\alpha}\hat{I}_{\alpha}$ with $\alpha,\beta \in \{x,y\}$ whose matrix elements are larger for lower $|m_I|$, thus similar to NMR. This is confirmed by the data in Fig.~\ref{fig:nuclear_drive}e.

The nuclear spin can be driven electrically at microwave frequencies via EDSR, through the modulation of the hyperfine interaction \cite{tosi2017silicon,savytskyy2023electrically}. In this case, the trends are expected to match those obtained for NMR. We use the stray electric fields from the microwave antenna to drive EDSR, and extract the Rabi frequencies for all the flip-flop transitions (Fig.~\ref{fig:nuclear_drive}j). Here, the measured Rabi frequencies show no clear trend because of the strongly frequency-dependent response of the microwave antenna in the range of $f^{\rm EDSR}_{m_I-1\leftrightarrow m_I} \approx 28$~GHz. This is also evident in the different Rabi frequencies obtained for ESR (see Supplementary Information, Section 6), where no dependence in nuclear spin number is expected (Tab.~\ref{tab:1}). The observed $f^{\rm Rabi, EDSR}_{5/2\leftrightarrow 7/2}/\delta_{5/2\leftrightarrow 7/2}^{\rm EDSR} \approx 28$~kHz is obtained using $V_{\rm MW}^{\rm pp} = 300$~mV of driving amplitude at the source (or $V_{\rm MW}^{\rm pp} \approx 30$~mV at the input of the antenna, accounting for the $\approx 20$\,dB attenuation along the line~\cite{savytskyy2023electrically}). In a device with $^{31}$P donors and a dedicated open-circuited antenna to deliver microwave electric fields, a similar value of $f^{\rm EDSR}_{\rm Rabi}$ required $V_{\rm MW}^{\rm pp} = 3$~V \cite{savytskyy2023electrically} at the source. As we discuss below, this is an indication that the hyperfine Stark shift in $^{123}$Sb is much larger than in $^{31}$P.

\subsection*{ELECTRICAL TUNABILITY OF THE RESONANCE FREQUENCIES}

The $^{123}$Sb Hamiltonians, Eqs.~(\ref{eq:H_ion},\ref{eq:H_neut}), contain terms that depend on the electric field applied to the donor, which itself depends on the DC voltages applied to the gates, $V_{\rm DC}$. For the ionised donor, the only electrically-tunable term is the nuclear quadrupole interaction, which depends on the applied voltage through the linear quadrupole Stark effect (LQSE)\cite{armstrong1961linear,asaad2020coherent}. The shift of the NMR$^+_{\pm 1}$ resonance as a function of the DC voltage on donor gate 1, \VDCDG, obeys the relation
\begin{equation}
\Delta f^{\rm NMR^{+}}_{m_I-1\leftrightarrow m_I} =  \left(m_I-\frac{1}{2}\right)\Delta{f_{\rm q}^{+}},
\end{equation}
where $\Delta f_{\rm q}^{+} = (\partial f_{\rm q}^{+}/\partial V) \cdot \Delta V_{\rm DC}^{\rm DG1}$. In this device, we measure $\partial f_{\rm q}^{+}/\partial V=-2.07(2)\,\mathrm{kHz/V}$ (Supplementary Fig.~S14).

\begin{figure}[!ht]
\includegraphics[width=0.5\textwidth]{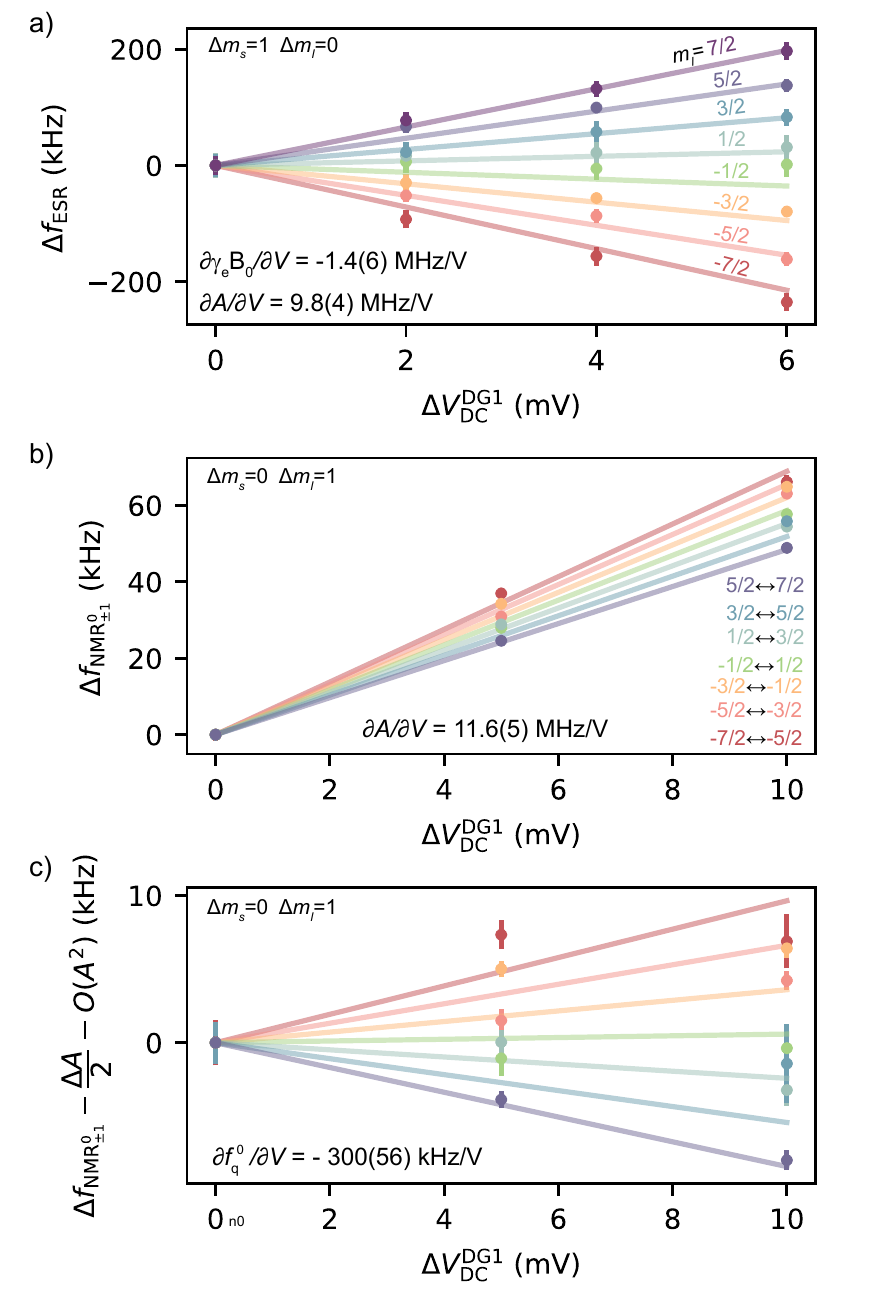}
\caption{\textbf{Stark effect.} \textbf{a)} Stark shift on the ESR resonance frequencies as a function of gate voltage variation, denoted by $\Delta$\VDCDG, for all nuclear spin projections $m_{I}$. \textbf{b)} Stark shift on the neutral NMR resonance frequencies as a function of $\Delta$\VDCDG, for all nuclear spin transitions $\ket{m_I-1}\leftrightarrow \ket{m_{I}}$. \textbf{(c)} The NMR Stark shift when substracting the linear and second order hyperfine contributions is shown to highlight the nuclear dependent trends arising from the LQSE. The solid lines in all panels are obtained numerically by solving Eq.\,\ref{eq:H_neut} as a function of \VDCDG using the experimentally obtained Stark effect parameters. } 
\label{fig:stark_effect}
\end{figure}

In the neutral donor, electric fields additionally affect the electron gyromagnetic ratio $\gamma_{\rm e}$ and the hyperfine coupling $A$ through the Stark effect \cite{pica2014hyperfine,laucht2015electrically}. The ESR frequency shifts as a function of gate voltage as:
\begin{equation}
\Delta f_{\mathrm{ESR}} = \Delta \gamma_{\mathrm{e}}B_{0}+2m_{I}\Delta A,
\end{equation}
where $\Delta \gammae$ and $\Delta A$ describe a change in the coupling parameters as a function of \VDCDG. The factor $m_I$ indicates that the eight ESR frequencies shift at different rates for a change in $A$, whereas a change in $\gamma_{\rm e}$ causes all frequencies to move by the same amount. The clear fan-out of the ESR frequencies in Fig.~\ref{fig:stark_effect}a shows that the hyperfine Stark shift is the dominant effect here. A fit to the data yields $\partial\gamma_{\mathrm{e}}B_{0}/\partial V = -1.4(6)$ ~MHz/V and  $\partial A/\partial V = 9.8(4)$~MHz/V (See Supplementary Information, Section 7). The hyperfine Stark shift is a factor $\approx 10$ larger than was observed in a \Ph donor device \cite{laucht2015electrically}. A similar enhancement, albeit for the quadratic Stark effect, was found with multi-valley effective mass models and experiments conducted on bulk donors in silicon \cite{pica2014hyperfine}. This qualitatively explains why we were able to coherently drive the flip-flop transitions with the stray electric field generated at the ESR antenna, more efficiently than in a \Ph device with a dedicated electrical antenna  \cite{savytskyy2023electrically}. Furthermore, we verify that the donor under study operates in a regime where the hyperfine Stark shift is linear in voltage (Supplementary Information, Section 8).

The neutral NMR$^0_{\pm 1}$ frequencies are voltage-dependent through the hyperfine Stark shift $\Delta A$ and the LQSE $\Delta f_{\rm q}^0$:
\begin{equation}
\begin{split}
\Delta f^{\rm NMR^{0}}_{m_I-1\leftrightarrow m_I} =  \left(m_I-\frac{1}{2}\right)\Delta{f_{\rm q}^{0}}\pm \frac{1}{2}\Delta A + \\
+ g_{m_I-1\leftrightarrow m_I}\frac{2A}{\gammae B_0}\Delta A,
\end{split}
\end{equation}
where the last term corresponds to second-order corrections to the hyperfine interaction, which are comparable in magnitude to the LQSE. The factor $(m_I - 1/2)$ preceding $\Delta f_{\rm q}^{0}$ and the coefficient $ g_{m_I-1\leftrightarrow m_I}$ are now responsible for making $\Delta f^{\rm NMR^{0}}_{m_I-1\leftrightarrow m_I}$ depend on the nuclear spin transition. From the data in Fig.~\ref{fig:stark_effect}c we extract $\partial f_{\mathrm{q}}^{0}/\partial V=-300(56)$~kHz/V and $\partial A/\partial V=11.57(45)$~MHz/V (Supplementary Information, Section 7). The slight difference between the estimated $\partial A/\partial V$ extracted from the data in Fig.\ref{fig:stark_effect}\,a and Fig.\ref{fig:stark_effect}\,b may be attributed to variations in the DC voltage settings between measurements, potentially impacting the electron's wavefunction sensitivity to electric fields \,\cite{tosi2017silicon}.

Because the shift in resonance frequencies is dominated by $\Delta A/2$, in Fig.~\ref{fig:stark_effect}\,c we plot $\Delta f^{\rm NMR^{0}}_{m_I-1\leftrightarrow m_I}-\Delta A/2-O(A^2)$ to highlight the contribution of the LQSE to the nuclear spin dependent fan-out (Fig.~\ref{fig:stark_effect}\,c).
Notably, the value obtained for LQSE in the ionised nucleus is two orders of magnitude smaller than the one obtained for the neutral atom. This observation could be used in the future to refine and validate \emph{ab initio} models of the nuclear quadrupole interaction.

\subsection*{DECOHERENCE: MAGNETIC AND ELECTRIC NOISE}

The key property of $^{31}$P donor qubits is their exceptionally long coherence times \cite{muhonen2014storing}, largely due to their weak sensitivity to electric fields. The ionised nucleus is strictly unaffected by electric fields due to its spin $I=1/2$. Moving to a heavier donor like $^{123}$Sb, with larger hyperfine Stark shifts and a nuclear electric quadrupole moment, raises the question of whether this will deteriorate spin coherence.

\begin{figure}[ht]
\includegraphics[width=0.5\textwidth]{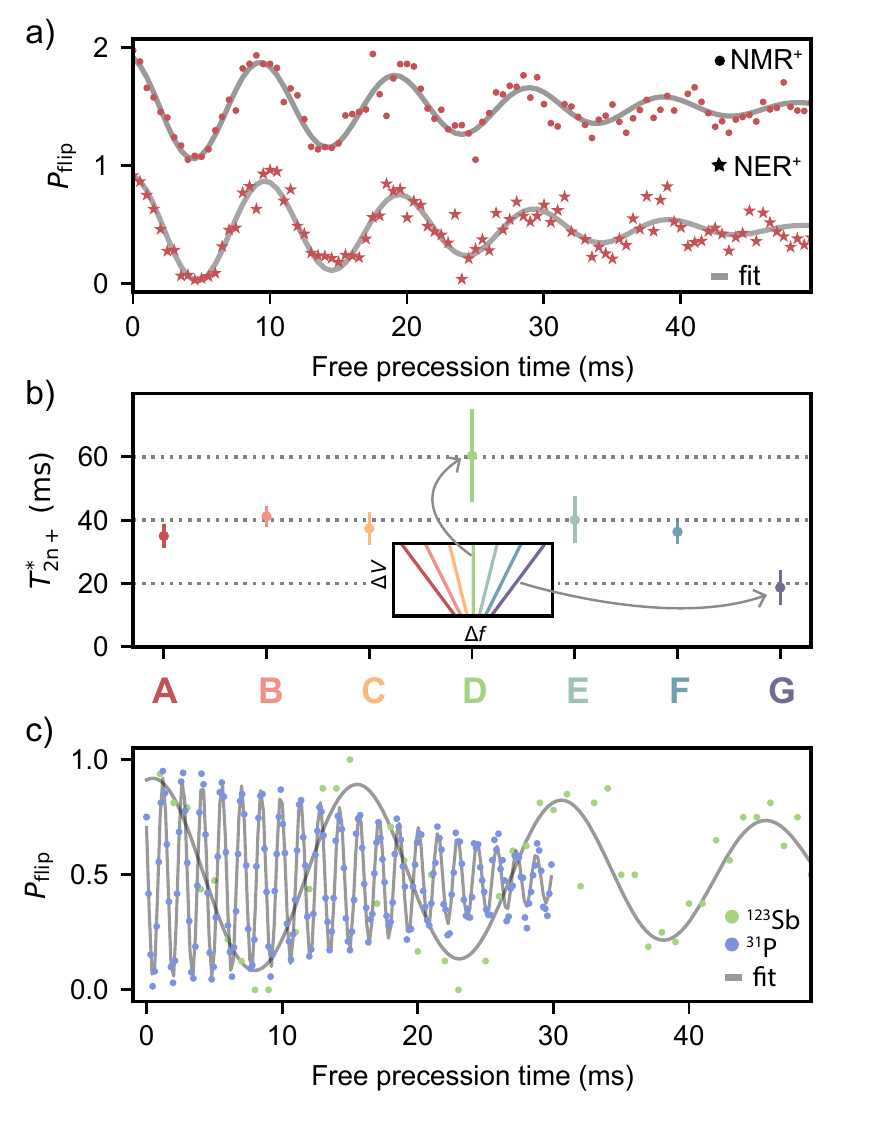}
    \caption{\textbf{Electric and magnetic noise on the ionised nucleus}. \textbf{a)} Ramsey decay for the transition $\ket{-7/2}\leftrightarrow \ket{-5/2}$ using NMR and NER. $\TRion\approx 29$\,ms in both cases, indicating no effect of the driving mechanism on the dephasing rates.  \textbf{b)} Dephasing times \TRion measured with a Ramsey sequence for all $m_I$, showing an increased \TRion for the $\ket{-1/2}\leftrightarrow\ket{1/2}$ transition. The duration of the Ramsey experiments lasted for a period of 3 hours. The inset depicts the linear quadrupole Stark effect on the resonance frequencies, to illustrate that the inner transition is unaffected by electric fields. The large errorbar for D is attributed to a
    lower fidelity in state preparation (Supplementary Information, section 1). \textbf{c)} Superimposed Ramsey decays for ionized \Ph and \Sb nuclei, both measured on the electric-field insensitive $\ket{-1/2}\leftrightarrow\ket{1/2}$ transitions, showing a shorter $T_2^*$ for the \Ph nucleus, in proportion to its larger gyromagnetic ratio.} 
    \label{fig:decoherence}
\end{figure}

Focussing on the ionised nucleus, we first verify that the driving mechanism does not affect the dephasing time $T_{\rm 2n+}^*$. Fig.~\ref{fig:decoherence}a compares two Ramsey experiments on the $\ket{-7/2} \leftrightarrow \ket{-5/2}$ transition where the $\pi/2$ pulses were delivered using either NMR or NER. We found near-identical values $T_{\rm 2n+}^* = 29.4(3)$~ms with NMR and $T_{\rm 2n+}^* = 29.8(3)$~ms with NER. This is intuitively expected because the Ramsey experiment probes the free evolution of the spin, in the absence of drives. However, this result indicates that the application of strong AC electric fields needed to drive NER does not destabilise the electrical environment of the nucleus in a noticeable way \cite{franke2017multiple}.

The ionised $^{123}$Sb nucleus offers a unique opportunity to rigorously distinguish magnetic from electric contributions to the noise that affects the spin coherence. Since $f_{m_I - 1 \leftrightarrow m_I}^{\rm NMR^+} = \gamma_n B_0 + (m_I - 1/2)f_{\rm q}^+$, quadrupole shifts caused by electric fields do not affect the coherence of the $\ket{-1/2} \leftrightarrow \ket{1/2}$ transition \cite{franke2015interaction}, i.e. the spin-1/2 nuclear subspace behaves exactly like a \Ph donor nucleus ($I=1/2$) would. Fig.~\ref{fig:decoherence}\,b shows the dephasing times $T_{\rm 2n+}^{*}$ as a function of $m_I$ (Fig.~\ref{fig:decoherence}\,b) for all transitions, measured using NMR. The $\ket{-1/2} \leftrightarrow \ket{1/2}$ transition has a $\approx 1.5 \times$ longer coherence than the outer transitions. The ionised \Sb nucleus thus couples measurably to electric field noise, but the coherence degradation is only by a factor of order unity in this type of devices, despite the fact that decoherence channels of magnetic origin are already minimised by the use of an isotopically purified $^{28}$Si substrate. By comparison, a factor $\sim 10$ degradation in $T_2^{\rm H}$ between the inner and the outer transitions was observed in experiments on ensembles of near-surface As$^+$ donors in natural Si \cite{franke2015interaction}, indicating that the electrical and charge noise level in our devices is remarkably benign.

In this particular device we co-implanted a small dose of \Ph donors, and we were able to address one of them. This allowed us to measure the dephasing time of two different donor species in the same device (Fig.~\ref{fig:decoherence}c). The ionised \Ph donor nucleus has only one NMR transition, $\ket{-1/2}\leftrightarrow \ket{1/2}$, for which we found $T_{\rm 2n+P}^* = 24.5(5)$~ms. Taking the ratio of $T_{\rm 2n+}^*$ for the same transition in \Sb yields $T_{2n+\rm{Sb}}^{*}/T_{2n+\rm{P}}^{*} = 2.5(6) $, in agreement with the ratio of the nuclear gyromagnetic ratios $\gamma_{\rm{n,P}}/\gamma_{\rm{n,Sb}} = 3.1$, where $\gamma_{\rm{n,P}} = 17.23$~MHz/T. A small discrepancy could be caused by a different distribution of residual $^{29}$Si spins around each donor.

\subsection*{GATE FIDELITIES}
 In preparation for future work on qudits \cite{wang2020qudits} and logical qubits \cite{gross2021designing} encoding on the \Sb system, we used gate set tomography (GST) \cite{mkadzik2022precision,nielsen2021gate} to benchmark the performance of one-qubit gates. We chose the qubit basis as the $\ket{0} = \ket{-5/2}$ and $\ket{1} = \ket{-7/2}$ states of the ionised donor nucleus, and assessed the performance of the $X_{\pi/2}$,$Y_{\pi/2}$ and $\mathds{I}$ gates, for both magnetic (NMR$_{\pm 1}^+$) and electric (NER$_{\pm 1}^+$) drive. The $X_{\pi/2}$ and $Y_{\pi/2}$ gates represent half rotations of the spin around the Bloch sphere, achieved through simple rectangular-envelope pulses modulating an oscillating driving field in resonance with the qubit Larmor frequency. The idle gate $\mathds{I}$ employs a far off-resonance stimulus that does not drive the qubit, but delivers the same power to the device as the other gates. This helps reducing context-dependent errors, where the frequency of the qubit or the readout contrast in the charge sensor are affected by the presence of absence of a driving field \cite{savytskyy2023electrically,undseth2023hotter}. The results are shown in Table \ref{tab:2}. All driven gates have average fidelity higher than 99.4\%, with errors dominated by coherent effects, i.e. inaccurate rotation angles (see Supplementary Materials, section 10).
\begin{table}[htpb]
\begin{tabular}{|c|c|c|c|c|}
\hline
 Gate & \multicolumn{2}{|c|}{Pulse duration} & \multicolumn{2}{|c|}{Average gate fidelity} \\
\hline
 & NMR& NER & NMR & NER\\
 \hline
$\mathds{I}$ & 255 $\rm{\mu s}$& 1.2087 ms & 99.42(30)\% & 98.35(44)\% \\
\hline
$X_{\pi/2}$ & 255 $\rm{\mu s}$ & 1.2087 ms& 99.82(24)\%&
99.76(26)\%\\
\hline
$Y_{\pi/2}$ & 255 $\rm{\mu s}$ & 1.2087 ms&
99.88(25)\% &
99.96(27)\% \\
\hline

\end{tabular}

    \caption{\textbf{Gate set tomography results.}} 
\label{tab:2}
\end{table}
\subsection*{CONCLUSIONS}

We have presented the experimental demonstration of coherent control of the electron and nuclear states of a single $^{123}$Sb donor atom, ion-implanted in a silicon nanoelectronic device. The combined Hilbert space of the atom spans 16 dimensions, and can be accessed using both electric and magnetic control fields. The exquisite spectral resolution afforded by the weak spin decoherence allowed us to extract detailed information on the value and the tunability of the Hamiltonian terms that determine the atom's quantum behaviour. The nuclear spin already shows gate fidelities exceeding 99\% regardless of the drive mechanism.

Future work will focus on exploiting the large Hilbert space for the creation of Schr\"{o}dinger cat states \cite{gupta2023robust}, with applications in quantum sensing \cite{chalopin2018quantum} and quantum foundations \cite{zaw2022detecting}. The relation between lattice strain and nuclear quadrupole interaction will be exploited to demonstrate nuclear acoustic resonance \cite{o2021engineering}, and to use the $^{123}$Sb atom as a local probe for strain in semiconductor nanoscale devices \cite{corley2023nanoscale}. For quantum information processing, an exciting prospect is the encoding an error-correctable logical qubit in the $I=7/2$ nuclear spin \cite{gross2021designing}. Multiple nuclei could be further entangled using the same electron-mediated two-qubit gates already demonstrated in $^{31}$P \cite{mkadzik2022precision}. The high tunability of the hyperfine coupling observed in our experiment bodes well for the prospect of using electric-dipole coupling in a flip-flop qubit architecture \cite{tosi2017silicon}.

\subsection*{ACKNOWLEDGMENTS}
We acknowledge discussions with R. Blume-Kohout, K. Rudinger, T. Proctor  and C. I. Ostrove. The research was funded by an Australian Research Council Discovery Project (grant no. DP210103769), the US Army Research Office (contract no. W911NF-17-1-0200), and the Australian Department of Industry, Innovation and Science (grant no. AUSMURI000002). We acknowledge support from the Australian National Fabrication Facility (ANFF). I.F.d.F. and A.V. acknowledge support from the Sydney Quantum Academy. All statements of fact, opinion, or conclusions contained herein are those of the authors and should not be construed as representing the official views or policies of the U.S. Army Research Office or the U.S. government. The U.S. government is authorized to reproduce and distribute reprints for government purposes notwithstanding any copyright notation herein.

\subsection*{AUTHOR CONTRIBUTIONS}

I.F.d.F. led the experiments and analyzed the data, T.B. assisted with the experiments and data interpretation under A.M. supervision. M.A.I.J. and S.A. supported with the measurement software experiment design. A.V. performed part of the NER measurements. I.F.d.F., F.E.H. and V.M. fabricated the device with supervision from A.M. and A.S.D., on an isotopically enriched $^{28}$Si wafer supplied by K.M.I.. B.C.J., A.M.J, J.C.McC. and D.N.J. designed and performed the ion implantation. I.F.d.F. and A.M. wrote the manuscript with input from all authors. 

\subsection*{COMPETING INTERESTS}
A.M., S.A. and V.M. are inventors on a patent related to this work, describing the use of high-spin donor nuclei as quantum information processing elements (application no. AU2019227083A1, US16/975,669, WO2019165494A1). A.S.D. is a founder, equity holder, director, and CEO of Diraq Pty Ltd. The other authors declare that they have no competing interests. 

\subsection*{ADDITIONAL INFORMATION}
Correspondence and requests for materials should be addressed to Andrea Morello.\\
Online supplementary information accompanies this paper.\\
All data needed to evaluate the conclusions in the paper are present in the paper and/or the Supplementary Materials. All the data and analysis scripts supporting the contents of the manuscript, can be downloaded from the following repository: 
\url{https://datadryad.org/stash/share/yypCpKkL1wniO7Il3EFKbSDYfDka9qIcQznHX5ssyvs}.
 
\bibliography{main_AM}

\begin{thebibliography}{50}%
\makeatletter
\providecommand \@ifxundefined [1]{%
 \@ifx{#1\undefined}
}%
\providecommand \@ifnum [1]{%
 \ifnum #1\expandafter \@firstoftwo
 \else \expandafter \@secondoftwo
 \fi
}%
\providecommand \@ifx [1]{%
 \ifx #1\expandafter \@firstoftwo
 \else \expandafter \@secondoftwo
 \fi
}%
\providecommand \natexlab [1]{#1}%
\providecommand \enquote  [1]{``#1''}%
\providecommand \bibnamefont  [1]{#1}%
\providecommand \bibfnamefont [1]{#1}%
\providecommand \citenamefont [1]{#1}%
\providecommand \href@noop [0]{\@secondoftwo}%
\providecommand \href [0]{\begingroup \@sanitize@url \@href}%
\providecommand \@href[1]{\@@startlink{#1}\@@href}%
\providecommand \@@href[1]{\endgroup#1\@@endlink}%
\providecommand \@sanitize@url [0]{\catcode `\\12\catcode `\$12\catcode
  `\&12\catcode `\#12\catcode `\^12\catcode `\_12\catcode `\%12\relax}%
\providecommand \@@startlink[1]{}%
\providecommand \@@endlink[0]{}%
\providecommand \url  [0]{\begingroup\@sanitize@url \@url }%
\providecommand \@url [1]{\endgroup\@href {#1}{\urlprefix }}%
\providecommand \urlprefix  [0]{URL }%
\providecommand \Eprint [0]{\href }%
\providecommand \doibase [0]{https://doi.org/}%
\providecommand \selectlanguage [0]{\@gobble}%
\providecommand \bibinfo  [0]{\@secondoftwo}%
\providecommand \bibfield  [0]{\@secondoftwo}%
\providecommand \translation [1]{[#1]}%
\providecommand \BibitemOpen [0]{}%
\providecommand \bibitemStop [0]{}%
\providecommand \bibitemNoStop [0]{.\EOS\space}%
\providecommand \EOS [0]{\spacefactor3000\relax}%
\providecommand \BibitemShut  [1]{\csname bibitem#1\endcsname}%
\let\auto@bib@innerbib\@empty
\bibitem [{\citenamefont {Wang}\ \emph {et~al.}(2020)\citenamefont {Wang},
  \citenamefont {Hu}, \citenamefont {Sanders},\ and\ \citenamefont
  {Kais}}]{wang2020qudits}%
  \BibitemOpen
  \bibfield  {author} {\bibinfo {author} {\bibfnamefont {Y.}~\bibnamefont
  {Wang}}, \bibinfo {author} {\bibfnamefont {Z.}~\bibnamefont {Hu}}, \bibinfo
  {author} {\bibfnamefont {B.~C.}\ \bibnamefont {Sanders}},\ and\ \bibinfo
  {author} {\bibfnamefont {S.}~\bibnamefont {Kais}},\ }\bibfield  {title}
  {\bibinfo {title} {Qudits and high-dimensional quantum computing},\ }\href
  {https://doi.org/10.3389/fphy.2020.589504} {\bibfield  {journal} {\bibinfo
  {journal} {Frontiers in Physics}\ }\textbf {\bibinfo {volume} {8}},\ \bibinfo
  {pages} {589504} (\bibinfo {year} {2020})}\BibitemShut {NoStop}%
\bibitem [{\citenamefont {Muthukrishnan}\ and\ \citenamefont
  {Stroud}(2000)}]{Muthukrishnan2000}%
  \BibitemOpen
  \bibfield  {author} {\bibinfo {author} {\bibfnamefont {A.}~\bibnamefont
  {Muthukrishnan}}\ and\ \bibinfo {author} {\bibfnamefont {C.~R.}\ \bibnamefont
  {Stroud}},\ }\bibfield  {title} {\bibinfo {title} {Multivalued logic gates
  for quantum computation},\ }\href
  {https://doi.org/10.1103/PhysRevA.62.052309} {\bibfield  {journal} {\bibinfo
  {journal} {Physical Review A}\ }\textbf {\bibinfo {volume} {62}},\ \bibinfo
  {pages} {052309} (\bibinfo {year} {2000})}\BibitemShut {NoStop}%
\bibitem [{\citenamefont {Campbell}(2014)}]{campbell2014}%
  \BibitemOpen
  \bibfield  {author} {\bibinfo {author} {\bibfnamefont {E.~T.}\ \bibnamefont
  {Campbell}},\ }\bibfield  {title} {\bibinfo {title} {Enhanced
  {F}ault-{T}olerant {Q}uantum {C}omputing in $d$-{L}evel {S}ystems},\ }\href
  {https://doi.org/10.1103/PhysRevLett.113.230501} {\bibfield  {journal}
  {\bibinfo  {journal} {Physical Review Letters}\ }\textbf {\bibinfo {volume}
  {113}},\ \bibinfo {pages} {230501} (\bibinfo {year} {2014})}\BibitemShut
  {NoStop}%
\bibitem [{\citenamefont {Bullock}\ \emph {et~al.}(2005)\citenamefont
  {Bullock}, \citenamefont {O'Leary},\ and\ \citenamefont
  {Brennen}}]{bullock2005}%
  \BibitemOpen
  \bibfield  {author} {\bibinfo {author} {\bibfnamefont {S.~S.}\ \bibnamefont
  {Bullock}}, \bibinfo {author} {\bibfnamefont {D.~P.}\ \bibnamefont
  {O'Leary}},\ and\ \bibinfo {author} {\bibfnamefont {G.~K.}\ \bibnamefont
  {Brennen}},\ }\bibfield  {title} {\bibinfo {title} {Asymptotically {O}ptimal
  {Q}uantum {C}ircuits for $d$-{L}evel {S}ystems},\ }\href
  {https://doi.org/10.1103/PhysRevLett.94.230502} {\bibfield  {journal}
  {\bibinfo  {journal} {Physical Review Letters}\ }\textbf {\bibinfo {volume}
  {94}},\ \bibinfo {pages} {230502} (\bibinfo {year} {2005})}\BibitemShut
  {NoStop}%
\bibitem [{\citenamefont {Nikolaeva}\ \emph {et~al.}(2021)\citenamefont
  {Nikolaeva}, \citenamefont {Kiktenko},\ and\ \citenamefont
  {Fedorov}}]{nikolaeva2021efficient}%
  \BibitemOpen
  \bibfield  {author} {\bibinfo {author} {\bibfnamefont {A.~S.}\ \bibnamefont
  {Nikolaeva}}, \bibinfo {author} {\bibfnamefont {E.~O.}\ \bibnamefont
  {Kiktenko}},\ and\ \bibinfo {author} {\bibfnamefont {A.~K.}\ \bibnamefont
  {Fedorov}},\ }\bibfield  {title} {\bibinfo {title} {Efficient realization of
  quantum algorithms with qudits},\ }\href {https://arxiv.org/abs/2111.04384}
  {\bibfield  {journal} {\bibinfo  {journal} {arXiv preprint arXiv:2111.04384}\
  } (\bibinfo {year} {2021})}\BibitemShut {NoStop}%
\bibitem [{\citenamefont {Lu}\ \emph {et~al.}(2022)\citenamefont {Lu},
  \citenamefont {Myilswamy}, \citenamefont {Bennink}, \citenamefont {Seshadri},
  \citenamefont {Alshaykh}, \citenamefont {Liu}, \citenamefont {Kippenberg},
  \citenamefont {Leaird}, \citenamefont {Weiner},\ and\ \citenamefont
  {Lukens}}]{lu2022bayesian}%
  \BibitemOpen
  \bibfield  {author} {\bibinfo {author} {\bibfnamefont {H.-H.}\ \bibnamefont
  {Lu}}, \bibinfo {author} {\bibfnamefont {K.~V.}\ \bibnamefont {Myilswamy}},
  \bibinfo {author} {\bibfnamefont {R.~S.}\ \bibnamefont {Bennink}}, \bibinfo
  {author} {\bibfnamefont {S.}~\bibnamefont {Seshadri}}, \bibinfo {author}
  {\bibfnamefont {M.~S.}\ \bibnamefont {Alshaykh}}, \bibinfo {author}
  {\bibfnamefont {J.}~\bibnamefont {Liu}}, \bibinfo {author} {\bibfnamefont
  {T.~J.}\ \bibnamefont {Kippenberg}}, \bibinfo {author} {\bibfnamefont
  {D.~E.}\ \bibnamefont {Leaird}}, \bibinfo {author} {\bibfnamefont {A.~M.}\
  \bibnamefont {Weiner}},\ and\ \bibinfo {author} {\bibfnamefont {J.~M.}\
  \bibnamefont {Lukens}},\ }\bibfield  {title} {\bibinfo {title} {Bayesian
  tomography of high-dimensional on-chip biphoton frequency combs with
  randomized measurements},\ }\href
  {https://doi.org/10.1038/s41467-022-31639-z} {\bibfield  {journal} {\bibinfo
  {journal} {Nature Communications}\ }\textbf {\bibinfo {volume} {13}},\
  \bibinfo {pages} {4338} (\bibinfo {year} {2022})}\BibitemShut {NoStop}%
\bibitem [{\citenamefont {Chi}\ \emph {et~al.}(2022)\citenamefont {Chi},
  \citenamefont {Huang}, \citenamefont {Zhang}, \citenamefont {Mao},
  \citenamefont {Zhou}, \citenamefont {Chen}, \citenamefont {Zhai},
  \citenamefont {Bao}, \citenamefont {Dai}, \citenamefont {Yuan} \emph
  {et~al.}}]{chi2022programmable}%
  \BibitemOpen
  \bibfield  {author} {\bibinfo {author} {\bibfnamefont {Y.}~\bibnamefont
  {Chi}}, \bibinfo {author} {\bibfnamefont {J.}~\bibnamefont {Huang}}, \bibinfo
  {author} {\bibfnamefont {Z.}~\bibnamefont {Zhang}}, \bibinfo {author}
  {\bibfnamefont {J.}~\bibnamefont {Mao}}, \bibinfo {author} {\bibfnamefont
  {Z.}~\bibnamefont {Zhou}}, \bibinfo {author} {\bibfnamefont {X.}~\bibnamefont
  {Chen}}, \bibinfo {author} {\bibfnamefont {C.}~\bibnamefont {Zhai}}, \bibinfo
  {author} {\bibfnamefont {J.}~\bibnamefont {Bao}}, \bibinfo {author}
  {\bibfnamefont {T.}~\bibnamefont {Dai}}, \bibinfo {author} {\bibfnamefont
  {H.}~\bibnamefont {Yuan}}, \emph {et~al.},\ }\bibfield  {title} {\bibinfo
  {title} {A programmable qudit-based quantum processor},\ }\href
  {https://doi.org/10.1038/s41467-022-28767-x} {\bibfield  {journal} {\bibinfo
  {journal} {Nature Communications}\ }\textbf {\bibinfo {volume} {13}},\
  \bibinfo {pages} {1166} (\bibinfo {year} {2022})}\BibitemShut {NoStop}%
\bibitem [{\citenamefont {Neeley}\ \emph {et~al.}(2009)\citenamefont {Neeley},
  \citenamefont {Ansmann}, \citenamefont {Bialczak}, \citenamefont {Hofheinz},
  \citenamefont {Lucero}, \citenamefont {O'Connell}, \citenamefont {Sank},
  \citenamefont {Wang}, \citenamefont {Wenner}, \citenamefont {Cleland} \emph
  {et~al.}}]{neeley2009emulation}%
  \BibitemOpen
  \bibfield  {author} {\bibinfo {author} {\bibfnamefont {M.}~\bibnamefont
  {Neeley}}, \bibinfo {author} {\bibfnamefont {M.}~\bibnamefont {Ansmann}},
  \bibinfo {author} {\bibfnamefont {R.~C.}\ \bibnamefont {Bialczak}}, \bibinfo
  {author} {\bibfnamefont {M.}~\bibnamefont {Hofheinz}}, \bibinfo {author}
  {\bibfnamefont {E.}~\bibnamefont {Lucero}}, \bibinfo {author} {\bibfnamefont
  {A.~D.}\ \bibnamefont {O'Connell}}, \bibinfo {author} {\bibfnamefont
  {D.}~\bibnamefont {Sank}}, \bibinfo {author} {\bibfnamefont {H.}~\bibnamefont
  {Wang}}, \bibinfo {author} {\bibfnamefont {J.}~\bibnamefont {Wenner}},
  \bibinfo {author} {\bibfnamefont {A.~N.}\ \bibnamefont {Cleland}}, \emph
  {et~al.},\ }\bibfield  {title} {\bibinfo {title} {Emulation of a quantum spin
  with a superconducting phase qudit},\ }\href
  {https://doi.org/10.1126/science.1173440} {\bibfield  {journal} {\bibinfo
  {journal} {Science}\ }\textbf {\bibinfo {volume} {325}},\ \bibinfo {pages}
  {722} (\bibinfo {year} {2009})}\BibitemShut {NoStop}%
\bibitem [{\citenamefont {Yurtalan}\ \emph {et~al.}(2020)\citenamefont
  {Yurtalan}, \citenamefont {Shi}, \citenamefont {Kononenko}, \citenamefont
  {Lupascu},\ and\ \citenamefont {Ashhab}}]{yurtalan2020}%
  \BibitemOpen
  \bibfield  {author} {\bibinfo {author} {\bibfnamefont {M.~A.}\ \bibnamefont
  {Yurtalan}}, \bibinfo {author} {\bibfnamefont {J.}~\bibnamefont {Shi}},
  \bibinfo {author} {\bibfnamefont {M.}~\bibnamefont {Kononenko}}, \bibinfo
  {author} {\bibfnamefont {A.}~\bibnamefont {Lupascu}},\ and\ \bibinfo {author}
  {\bibfnamefont {S.}~\bibnamefont {Ashhab}},\ }\bibfield  {title} {\bibinfo
  {title} {{I}mplementation of a {W}alsh-{H}adamard {G}ate in a
  {S}uperconducting {Q}utrit},\ }\href
  {https://doi.org/10.1103/PhysRevLett.125.180504} {\bibfield  {journal}
  {\bibinfo  {journal} {Physical Review Letters}\ }\textbf {\bibinfo {volume}
  {125}},\ \bibinfo {pages} {180504} (\bibinfo {year} {2020})}\BibitemShut
  {NoStop}%
\bibitem [{\citenamefont {Goss}\ \emph {et~al.}(2022)\citenamefont {Goss},
  \citenamefont {Morvan}, \citenamefont {Marinelli}, \citenamefont {Mitchell},
  \citenamefont {Nguyen}, \citenamefont {Naik}, \citenamefont {Chen},
  \citenamefont {J{\"u}nger}, \citenamefont {Kreikebaum}, \citenamefont
  {Santiago} \emph {et~al.}}]{goss2022high}%
  \BibitemOpen
  \bibfield  {author} {\bibinfo {author} {\bibfnamefont {N.}~\bibnamefont
  {Goss}}, \bibinfo {author} {\bibfnamefont {A.}~\bibnamefont {Morvan}},
  \bibinfo {author} {\bibfnamefont {B.}~\bibnamefont {Marinelli}}, \bibinfo
  {author} {\bibfnamefont {B.~K.}\ \bibnamefont {Mitchell}}, \bibinfo {author}
  {\bibfnamefont {L.~B.}\ \bibnamefont {Nguyen}}, \bibinfo {author}
  {\bibfnamefont {R.~K.}\ \bibnamefont {Naik}}, \bibinfo {author}
  {\bibfnamefont {L.}~\bibnamefont {Chen}}, \bibinfo {author} {\bibfnamefont
  {C.}~\bibnamefont {J{\"u}nger}}, \bibinfo {author} {\bibfnamefont {J.~M.}\
  \bibnamefont {Kreikebaum}}, \bibinfo {author} {\bibfnamefont {D.~I.}\
  \bibnamefont {Santiago}}, \emph {et~al.},\ }\bibfield  {title} {\bibinfo
  {title} {High-fidelity qutrit entangling gates for superconducting
  circuits},\ }\href {https://doi.org/10.1038/s41467-022-34851-z} {\bibfield
  {journal} {\bibinfo  {journal} {Nature Communications}\ }\textbf {\bibinfo
  {volume} {13}},\ \bibinfo {pages} {7481} (\bibinfo {year}
  {2022})}\BibitemShut {NoStop}%
\bibitem [{\citenamefont {Ringbauer}\ \emph {et~al.}(2022)\citenamefont
  {Ringbauer}, \citenamefont {Meth}, \citenamefont {Postler}, \citenamefont
  {Stricker}, \citenamefont {Blatt}, \citenamefont {Schindler},\ and\
  \citenamefont {Monz}}]{ringbauer2022universal}%
  \BibitemOpen
  \bibfield  {author} {\bibinfo {author} {\bibfnamefont {M.}~\bibnamefont
  {Ringbauer}}, \bibinfo {author} {\bibfnamefont {M.}~\bibnamefont {Meth}},
  \bibinfo {author} {\bibfnamefont {L.}~\bibnamefont {Postler}}, \bibinfo
  {author} {\bibfnamefont {R.}~\bibnamefont {Stricker}}, \bibinfo {author}
  {\bibfnamefont {R.}~\bibnamefont {Blatt}}, \bibinfo {author} {\bibfnamefont
  {P.}~\bibnamefont {Schindler}},\ and\ \bibinfo {author} {\bibfnamefont
  {T.}~\bibnamefont {Monz}},\ }\bibfield  {title} {\bibinfo {title} {A
  universal qudit quantum processor with trapped ions},\ }\href
  {https://doi.org/10.1038/s41567-022-01658-0} {\bibfield  {journal} {\bibinfo
  {journal} {Nature Physics}\ }\textbf {\bibinfo {volume} {18}},\ \bibinfo
  {pages} {1053} (\bibinfo {year} {2022})}\BibitemShut {NoStop}%
\bibitem [{\citenamefont {Anderson}\ \emph {et~al.}(2015)\citenamefont
  {Anderson}, \citenamefont {Sosa-Martinez}, \citenamefont {Riofr\'{\i}o},
  \citenamefont {Deutsch},\ and\ \citenamefont {Jessen}}]{anderson2015}%
  \BibitemOpen
  \bibfield  {author} {\bibinfo {author} {\bibfnamefont {B.~E.}\ \bibnamefont
  {Anderson}}, \bibinfo {author} {\bibfnamefont {H.}~\bibnamefont
  {Sosa-Martinez}}, \bibinfo {author} {\bibfnamefont {C.~A.}\ \bibnamefont
  {Riofr\'{\i}o}}, \bibinfo {author} {\bibfnamefont {I.~H.}\ \bibnamefont
  {Deutsch}},\ and\ \bibinfo {author} {\bibfnamefont {P.~S.}\ \bibnamefont
  {Jessen}},\ }\bibfield  {title} {\bibinfo {title} {Accurate and robust
  unitary transformations of a {H}igh-{D}imensional {Q}uantum {S}ystem},\
  }\href {https://doi.org/10.1103/PhysRevLett.114.240401} {\bibfield  {journal}
  {\bibinfo  {journal} {Physical Review Letters}\ }\textbf {\bibinfo {volume}
  {114}},\ \bibinfo {pages} {240401} (\bibinfo {year} {2015})}\BibitemShut
  {NoStop}%
\bibitem [{\citenamefont {Godfrin}\ \emph {et~al.}(2018)\citenamefont
  {Godfrin}, \citenamefont {Ballou}, \citenamefont {Bonet}, \citenamefont
  {Ruben}, \citenamefont {Klyatskaya}, \citenamefont {Wernsdorfer},\ and\
  \citenamefont {Balestro}}]{godfrin2018generalized}%
  \BibitemOpen
  \bibfield  {author} {\bibinfo {author} {\bibfnamefont {C.}~\bibnamefont
  {Godfrin}}, \bibinfo {author} {\bibfnamefont {R.}~\bibnamefont {Ballou}},
  \bibinfo {author} {\bibfnamefont {E.}~\bibnamefont {Bonet}}, \bibinfo
  {author} {\bibfnamefont {M.}~\bibnamefont {Ruben}}, \bibinfo {author}
  {\bibfnamefont {S.}~\bibnamefont {Klyatskaya}}, \bibinfo {author}
  {\bibfnamefont {W.}~\bibnamefont {Wernsdorfer}},\ and\ \bibinfo {author}
  {\bibfnamefont {F.}~\bibnamefont {Balestro}},\ }\bibfield  {title} {\bibinfo
  {title} {Generalized {R}amsey interferometry explored with a single nuclear
  spin qudit},\ }\href {https://doi.org/10.1038/s41534-018-0101-3} {\bibfield
  {journal} {\bibinfo  {journal} {npj Quantum Information}\ }\textbf {\bibinfo
  {volume} {4}},\ \bibinfo {pages} {53} (\bibinfo {year} {2018})}\BibitemShut
  {NoStop}%
\bibitem [{\citenamefont {Zwanenburg}\ \emph {et~al.}(2013)\citenamefont
  {Zwanenburg}, \citenamefont {Dzurak}, \citenamefont {Morello}, \citenamefont
  {Simmons}, \citenamefont {Hollenberg}, \citenamefont {Klimeck}, \citenamefont
  {Rogge}, \citenamefont {Coppersmith},\ and\ \citenamefont
  {Eriksson}}]{zwanenburg2013silicon}%
  \BibitemOpen
  \bibfield  {author} {\bibinfo {author} {\bibfnamefont {F.~A.}\ \bibnamefont
  {Zwanenburg}}, \bibinfo {author} {\bibfnamefont {A.~S.}\ \bibnamefont
  {Dzurak}}, \bibinfo {author} {\bibfnamefont {A.}~\bibnamefont {Morello}},
  \bibinfo {author} {\bibfnamefont {M.~Y.}\ \bibnamefont {Simmons}}, \bibinfo
  {author} {\bibfnamefont {L.~C.}\ \bibnamefont {Hollenberg}}, \bibinfo
  {author} {\bibfnamefont {G.}~\bibnamefont {Klimeck}}, \bibinfo {author}
  {\bibfnamefont {S.}~\bibnamefont {Rogge}}, \bibinfo {author} {\bibfnamefont
  {S.~N.}\ \bibnamefont {Coppersmith}},\ and\ \bibinfo {author} {\bibfnamefont
  {M.~A.}\ \bibnamefont {Eriksson}},\ }\bibfield  {title} {\bibinfo {title}
  {Silicon quantum electronics},\ }\href
  {https://doi.org/10.1103/RevModPhys.85.961} {\bibfield  {journal} {\bibinfo
  {journal} {Reviews of Modern Physics}\ }\textbf {\bibinfo {volume} {85}},\
  \bibinfo {pages} {961} (\bibinfo {year} {2013})}\BibitemShut {NoStop}%
\bibitem [{\citenamefont {Veldhorst}\ \emph {et~al.}(2014)\citenamefont
  {Veldhorst}, \citenamefont {Hwang}, \citenamefont {Yang}, \citenamefont
  {Leenstra}, \citenamefont {Dehollain}, \citenamefont {Muhonen}, \citenamefont
  {Hudson}, \citenamefont {Morello} \emph {et~al.}}]{veldhorst2014addressable}%
  \BibitemOpen
  \bibfield  {author} {\bibinfo {author} {\bibfnamefont {M.}~\bibnamefont
  {Veldhorst}}, \bibinfo {author} {\bibfnamefont {J.}~\bibnamefont {Hwang}},
  \bibinfo {author} {\bibfnamefont {C.}~\bibnamefont {Yang}}, \bibinfo {author}
  {\bibfnamefont {A.}~\bibnamefont {Leenstra}}, \bibinfo {author}
  {\bibfnamefont {J.}~\bibnamefont {Dehollain}}, \bibinfo {author}
  {\bibfnamefont {J.}~\bibnamefont {Muhonen}}, \bibinfo {author} {\bibfnamefont
  {F.}~\bibnamefont {Hudson}}, \bibinfo {author} {\bibfnamefont
  {A.}~\bibnamefont {Morello}}, \emph {et~al.},\ }\bibfield  {title} {\bibinfo
  {title} {An addressable quantum dot qubit with fault-tolerant
  control-fidelity},\ }\href {https://doi.org/10.1038/nnano.2014.216}
  {\bibfield  {journal} {\bibinfo  {journal} {Nature Nanotechnology}\ }\textbf
  {\bibinfo {volume} {9}},\ \bibinfo {pages} {981} (\bibinfo {year}
  {2014})}\BibitemShut {NoStop}%
\bibitem [{\citenamefont {Muhonen}\ \emph {et~al.}(2014)\citenamefont
  {Muhonen}, \citenamefont {Dehollain}, \citenamefont {Laucht}, \citenamefont
  {Hudson}, \citenamefont {Kalra}, \citenamefont {Sekiguchi}, \citenamefont
  {Itoh}, \citenamefont {Jamieson}, \citenamefont {McCallum}, \citenamefont
  {Dzurak} \emph {et~al.}}]{muhonen2014storing}%
  \BibitemOpen
  \bibfield  {author} {\bibinfo {author} {\bibfnamefont {J.~T.}\ \bibnamefont
  {Muhonen}}, \bibinfo {author} {\bibfnamefont {J.~P.}\ \bibnamefont
  {Dehollain}}, \bibinfo {author} {\bibfnamefont {A.}~\bibnamefont {Laucht}},
  \bibinfo {author} {\bibfnamefont {F.~E.}\ \bibnamefont {Hudson}}, \bibinfo
  {author} {\bibfnamefont {R.}~\bibnamefont {Kalra}}, \bibinfo {author}
  {\bibfnamefont {T.}~\bibnamefont {Sekiguchi}}, \bibinfo {author}
  {\bibfnamefont {K.~M.}\ \bibnamefont {Itoh}}, \bibinfo {author}
  {\bibfnamefont {D.~N.}\ \bibnamefont {Jamieson}}, \bibinfo {author}
  {\bibfnamefont {J.~C.}\ \bibnamefont {McCallum}}, \bibinfo {author}
  {\bibfnamefont {A.~S.}\ \bibnamefont {Dzurak}}, \emph {et~al.},\ }\bibfield
  {title} {\bibinfo {title} {Storing quantum information for 30 seconds in a
  nanoelectronic device},\ }\href {https://doi.org/10.1038/nnano.2014.211}
  {\bibfield  {journal} {\bibinfo  {journal} {Nature Nanotechnology}\ }\textbf
  {\bibinfo {volume} {9}},\ \bibinfo {pages} {986} (\bibinfo {year}
  {2014})}\BibitemShut {NoStop}%
\bibitem [{\citenamefont {M{\k{a}}dzik}\ \emph {et~al.}(2022)\citenamefont
  {M{\k{a}}dzik}, \citenamefont {Asaad}, \citenamefont {Youssry}, \citenamefont
  {J\"ocker}, \citenamefont {Rudinger}, \citenamefont {Nielsen}, \citenamefont
  {Young}, \citenamefont {Proctor}, \citenamefont {Baczewski}, \citenamefont
  {Laucht} \emph {et~al.}}]{mkadzik2022precision}%
  \BibitemOpen
  \bibfield  {author} {\bibinfo {author} {\bibfnamefont {M.~T.}\ \bibnamefont
  {M{\k{a}}dzik}}, \bibinfo {author} {\bibfnamefont {S.}~\bibnamefont {Asaad}},
  \bibinfo {author} {\bibfnamefont {A.}~\bibnamefont {Youssry}}, \bibinfo
  {author} {\bibfnamefont {B.}~\bibnamefont {J\"ocker}}, \bibinfo {author}
  {\bibfnamefont {K.~M.}\ \bibnamefont {Rudinger}}, \bibinfo {author}
  {\bibfnamefont {E.}~\bibnamefont {Nielsen}}, \bibinfo {author} {\bibfnamefont
  {K.~C.}\ \bibnamefont {Young}}, \bibinfo {author} {\bibfnamefont {T.~J.}\
  \bibnamefont {Proctor}}, \bibinfo {author} {\bibfnamefont {A.~D.}\
  \bibnamefont {Baczewski}}, \bibinfo {author} {\bibfnamefont {A.}~\bibnamefont
  {Laucht}}, \emph {et~al.},\ }\bibfield  {title} {\bibinfo {title} {Precision
  tomography of a three-qubit donor quantum processor in silicon},\ }\href
  {https://doi.org/10.1038/s41586-021-04292-7} {\bibfield  {journal} {\bibinfo
  {journal} {Nature}\ }\textbf {\bibinfo {volume} {601}},\ \bibinfo {pages}
  {348} (\bibinfo {year} {2022})}\BibitemShut {NoStop}%
\bibitem [{\citenamefont {Noiri}\ \emph {et~al.}(2022)\citenamefont {Noiri},
  \citenamefont {Takeda}, \citenamefont {Nakajima}, \citenamefont {Kobayashi},
  \citenamefont {Sammak}, \citenamefont {Scappucci},\ and\ \citenamefont
  {Tarucha}}]{noiri2022fast}%
  \BibitemOpen
  \bibfield  {author} {\bibinfo {author} {\bibfnamefont {A.}~\bibnamefont
  {Noiri}}, \bibinfo {author} {\bibfnamefont {K.}~\bibnamefont {Takeda}},
  \bibinfo {author} {\bibfnamefont {T.}~\bibnamefont {Nakajima}}, \bibinfo
  {author} {\bibfnamefont {T.}~\bibnamefont {Kobayashi}}, \bibinfo {author}
  {\bibfnamefont {A.}~\bibnamefont {Sammak}}, \bibinfo {author} {\bibfnamefont
  {G.}~\bibnamefont {Scappucci}},\ and\ \bibinfo {author} {\bibfnamefont
  {S.}~\bibnamefont {Tarucha}},\ }\bibfield  {title} {\bibinfo {title} {Fast
  universal quantum gate above the fault-tolerance threshold in silicon},\
  }\href {https://doi.org/10.1038/s41586-021-04182-y} {\bibfield  {journal}
  {\bibinfo  {journal} {Nature}\ }\textbf {\bibinfo {volume} {601}},\ \bibinfo
  {pages} {338} (\bibinfo {year} {2022})}\BibitemShut {NoStop}%
\bibitem [{\citenamefont {Xue}\ \emph {et~al.}(2022)\citenamefont {Xue},
  \citenamefont {Russ}, \citenamefont {Samkharadze}, \citenamefont {Undseth},
  \citenamefont {Sammak}, \citenamefont {Scappucci},\ and\ \citenamefont
  {Vandersypen}}]{xue2022quantum}%
  \BibitemOpen
  \bibfield  {author} {\bibinfo {author} {\bibfnamefont {X.}~\bibnamefont
  {Xue}}, \bibinfo {author} {\bibfnamefont {M.}~\bibnamefont {Russ}}, \bibinfo
  {author} {\bibfnamefont {N.}~\bibnamefont {Samkharadze}}, \bibinfo {author}
  {\bibfnamefont {B.}~\bibnamefont {Undseth}}, \bibinfo {author} {\bibfnamefont
  {A.}~\bibnamefont {Sammak}}, \bibinfo {author} {\bibfnamefont
  {G.}~\bibnamefont {Scappucci}},\ and\ \bibinfo {author} {\bibfnamefont
  {L.~M.}\ \bibnamefont {Vandersypen}},\ }\bibfield  {title} {\bibinfo {title}
  {Quantum logic with spin qubits crossing the surface code threshold},\ }\href
  {https://doi.org/10.1038/s41586-021-04273-w} {\bibfield  {journal} {\bibinfo
  {journal} {Nature}\ }\textbf {\bibinfo {volume} {601}},\ \bibinfo {pages}
  {343} (\bibinfo {year} {2022})}\BibitemShut {NoStop}%
\bibitem [{\citenamefont {Mills}\ \emph {et~al.}(2022)\citenamefont {Mills},
  \citenamefont {Guinn}, \citenamefont {Gullans}, \citenamefont {Sigillito},
  \citenamefont {Feldman}, \citenamefont {Nielsen},\ and\ \citenamefont
  {Petta}}]{mills2022two}%
  \BibitemOpen
  \bibfield  {author} {\bibinfo {author} {\bibfnamefont {A.~R.}\ \bibnamefont
  {Mills}}, \bibinfo {author} {\bibfnamefont {C.~R.}\ \bibnamefont {Guinn}},
  \bibinfo {author} {\bibfnamefont {M.~J.}\ \bibnamefont {Gullans}}, \bibinfo
  {author} {\bibfnamefont {A.~J.}\ \bibnamefont {Sigillito}}, \bibinfo {author}
  {\bibfnamefont {M.~M.}\ \bibnamefont {Feldman}}, \bibinfo {author}
  {\bibfnamefont {E.}~\bibnamefont {Nielsen}},\ and\ \bibinfo {author}
  {\bibfnamefont {J.~R.}\ \bibnamefont {Petta}},\ }\bibfield  {title} {\bibinfo
  {title} {Two-qubit silicon quantum processor with operation fidelity
  exceeding 99\%},\ }\href {https://doi.org/10.1126/sciadv.abn5130} {\bibfield
  {journal} {\bibinfo  {journal} {Science Advances}\ }\textbf {\bibinfo
  {volume} {8}},\ \bibinfo {pages} {eabn5130} (\bibinfo {year}
  {2022})}\BibitemShut {NoStop}%
\bibitem [{\citenamefont {Zwerver}\ \emph {et~al.}(2022)\citenamefont
  {Zwerver}, \citenamefont {Kr{\"a}henmann}, \citenamefont {Watson},
  \citenamefont {Lampert}, \citenamefont {George}, \citenamefont
  {Pillarisetty}, \citenamefont {Bojarski}, \citenamefont {Amin}, \citenamefont
  {Amitonov}, \citenamefont {Boter} \emph {et~al.}}]{zwerver2022qubits}%
  \BibitemOpen
  \bibfield  {author} {\bibinfo {author} {\bibfnamefont {A.}~\bibnamefont
  {Zwerver}}, \bibinfo {author} {\bibfnamefont {T.}~\bibnamefont
  {Kr{\"a}henmann}}, \bibinfo {author} {\bibfnamefont {T.}~\bibnamefont
  {Watson}}, \bibinfo {author} {\bibfnamefont {L.}~\bibnamefont {Lampert}},
  \bibinfo {author} {\bibfnamefont {H.~C.}\ \bibnamefont {George}}, \bibinfo
  {author} {\bibfnamefont {R.}~\bibnamefont {Pillarisetty}}, \bibinfo {author}
  {\bibfnamefont {S.}~\bibnamefont {Bojarski}}, \bibinfo {author}
  {\bibfnamefont {P.}~\bibnamefont {Amin}}, \bibinfo {author} {\bibfnamefont
  {S.}~\bibnamefont {Amitonov}}, \bibinfo {author} {\bibfnamefont
  {J.}~\bibnamefont {Boter}}, \emph {et~al.},\ }\bibfield  {title} {\bibinfo
  {title} {Qubits made by advanced semiconductor manufacturing},\ }\href
  {https://doi.org/10.1038/s41928-022-00727-9} {\bibfield  {journal} {\bibinfo
  {journal} {Nature Electronics}\ }\textbf {\bibinfo {volume} {5}},\ \bibinfo
  {pages} {184} (\bibinfo {year} {2022})}\BibitemShut {NoStop}%
\bibitem [{\citenamefont {Pla}\ \emph {et~al.}(2012)\citenamefont {Pla},
  \citenamefont {Tan}, \citenamefont {Dehollain}, \citenamefont {Lim},
  \citenamefont {Morton}, \citenamefont {Jamieson}, \citenamefont {Dzurak},\
  and\ \citenamefont {Morello}}]{pla2012single}%
  \BibitemOpen
  \bibfield  {author} {\bibinfo {author} {\bibfnamefont {J.~J.}\ \bibnamefont
  {Pla}}, \bibinfo {author} {\bibfnamefont {K.~Y.}\ \bibnamefont {Tan}},
  \bibinfo {author} {\bibfnamefont {J.~P.}\ \bibnamefont {Dehollain}}, \bibinfo
  {author} {\bibfnamefont {W.~H.}\ \bibnamefont {Lim}}, \bibinfo {author}
  {\bibfnamefont {J.~J.}\ \bibnamefont {Morton}}, \bibinfo {author}
  {\bibfnamefont {D.~N.}\ \bibnamefont {Jamieson}}, \bibinfo {author}
  {\bibfnamefont {A.~S.}\ \bibnamefont {Dzurak}},\ and\ \bibinfo {author}
  {\bibfnamefont {A.}~\bibnamefont {Morello}},\ }\bibfield  {title} {\bibinfo
  {title} {A single-atom electron spin qubit in silicon},\ }\href
  {https://doi.org/10.1038/nature11449} {\bibfield  {journal} {\bibinfo
  {journal} {Nature}\ }\textbf {\bibinfo {volume} {489}},\ \bibinfo {pages}
  {541} (\bibinfo {year} {2012})}\BibitemShut {NoStop}%
\bibitem [{\citenamefont {Pla}\ \emph {et~al.}(2013)\citenamefont {Pla},
  \citenamefont {Tan}, \citenamefont {Dehollain}, \citenamefont {Lim},
  \citenamefont {Morton}, \citenamefont {Zwanenburg}, \citenamefont {Jamieson},
  \citenamefont {Dzurak},\ and\ \citenamefont {Morello}}]{pla2013high}%
  \BibitemOpen
  \bibfield  {author} {\bibinfo {author} {\bibfnamefont {J.~J.}\ \bibnamefont
  {Pla}}, \bibinfo {author} {\bibfnamefont {K.~Y.}\ \bibnamefont {Tan}},
  \bibinfo {author} {\bibfnamefont {J.~P.}\ \bibnamefont {Dehollain}}, \bibinfo
  {author} {\bibfnamefont {W.~H.}\ \bibnamefont {Lim}}, \bibinfo {author}
  {\bibfnamefont {J.~J.}\ \bibnamefont {Morton}}, \bibinfo {author}
  {\bibfnamefont {F.~A.}\ \bibnamefont {Zwanenburg}}, \bibinfo {author}
  {\bibfnamefont {D.~N.}\ \bibnamefont {Jamieson}}, \bibinfo {author}
  {\bibfnamefont {A.~S.}\ \bibnamefont {Dzurak}},\ and\ \bibinfo {author}
  {\bibfnamefont {A.}~\bibnamefont {Morello}},\ }\bibfield  {title} {\bibinfo
  {title} {High-fidelity readout and control of a nuclear spin qubit in
  silicon},\ }\href {https://doi.org/10.1038/nature12011} {\bibfield  {journal}
  {\bibinfo  {journal} {Nature}\ }\textbf {\bibinfo {volume} {496}},\ \bibinfo
  {pages} {334} (\bibinfo {year} {2013})}\BibitemShut {NoStop}%
\bibitem [{\citenamefont {Asaad}\ \emph {et~al.}(2020)\citenamefont {Asaad},
  \citenamefont {Mourik}, \citenamefont {J\"ocker}, \citenamefont {Johnson},
  \citenamefont {Baczewski}, \citenamefont {Firgau}, \citenamefont
  {M{\k{a}}dzik}, \citenamefont {Schmitt}, \citenamefont {Pla}, \citenamefont
  {Hudson} \emph {et~al.}}]{asaad2020coherent}%
  \BibitemOpen
  \bibfield  {author} {\bibinfo {author} {\bibfnamefont {S.}~\bibnamefont
  {Asaad}}, \bibinfo {author} {\bibfnamefont {V.}~\bibnamefont {Mourik}},
  \bibinfo {author} {\bibfnamefont {B.}~\bibnamefont {J\"ocker}}, \bibinfo
  {author} {\bibfnamefont {M.~A.}\ \bibnamefont {Johnson}}, \bibinfo {author}
  {\bibfnamefont {A.~D.}\ \bibnamefont {Baczewski}}, \bibinfo {author}
  {\bibfnamefont {H.~R.}\ \bibnamefont {Firgau}}, \bibinfo {author}
  {\bibfnamefont {M.~T.}\ \bibnamefont {M{\k{a}}dzik}}, \bibinfo {author}
  {\bibfnamefont {V.}~\bibnamefont {Schmitt}}, \bibinfo {author} {\bibfnamefont
  {J.~J.}\ \bibnamefont {Pla}}, \bibinfo {author} {\bibfnamefont {F.~E.}\
  \bibnamefont {Hudson}}, \emph {et~al.},\ }\bibfield  {title} {\bibinfo
  {title} {Coherent electrical control of a single high-spin nucleus in
  silicon},\ }\href {https://doi.org/10.1038/s41586-020-2057-7} {\bibfield
  {journal} {\bibinfo  {journal} {Nature}\ }\textbf {\bibinfo {volume} {579}},\
  \bibinfo {pages} {205} (\bibinfo {year} {2020})}\BibitemShut {NoStop}%
\bibitem [{\citenamefont {O'Neill}\ \emph {et~al.}(2021)\citenamefont
  {O'Neill}, \citenamefont {J\"ocker}, \citenamefont {Baczewski},\ and\
  \citenamefont {Morello}}]{o2021engineering}%
  \BibitemOpen
  \bibfield  {author} {\bibinfo {author} {\bibfnamefont {L.~A.}\ \bibnamefont
  {O'Neill}}, \bibinfo {author} {\bibfnamefont {B.}~\bibnamefont {J\"ocker}},
  \bibinfo {author} {\bibfnamefont {A.~D.}\ \bibnamefont {Baczewski}},\ and\
  \bibinfo {author} {\bibfnamefont {A.}~\bibnamefont {Morello}},\ }\bibfield
  {title} {\bibinfo {title} {Engineering local strain for single-atom nuclear
  acoustic resonance in silicon},\ }\href {https://doi.org/10.1063/5.0069305}
  {\bibfield  {journal} {\bibinfo  {journal} {Applied Physics Letters}\
  }\textbf {\bibinfo {volume} {119}},\ \bibinfo {pages} {174001} (\bibinfo
  {year} {2021})}\BibitemShut {NoStop}%
\bibitem [{\citenamefont {Laucht}\ \emph {et~al.}(2015)\citenamefont {Laucht},
  \citenamefont {Muhonen}, \citenamefont {Mohiyaddin}, \citenamefont {Kalra},
  \citenamefont {Dehollain}, \citenamefont {Freer}, \citenamefont {Hudson},
  \citenamefont {Veldhorst}, \citenamefont {Rahman}, \citenamefont {Klimeck}
  \emph {et~al.}}]{laucht2015electrically}%
  \BibitemOpen
  \bibfield  {author} {\bibinfo {author} {\bibfnamefont {A.}~\bibnamefont
  {Laucht}}, \bibinfo {author} {\bibfnamefont {J.~T.}\ \bibnamefont {Muhonen}},
  \bibinfo {author} {\bibfnamefont {F.~A.}\ \bibnamefont {Mohiyaddin}},
  \bibinfo {author} {\bibfnamefont {R.}~\bibnamefont {Kalra}}, \bibinfo
  {author} {\bibfnamefont {J.~P.}\ \bibnamefont {Dehollain}}, \bibinfo {author}
  {\bibfnamefont {S.}~\bibnamefont {Freer}}, \bibinfo {author} {\bibfnamefont
  {F.~E.}\ \bibnamefont {Hudson}}, \bibinfo {author} {\bibfnamefont
  {M.}~\bibnamefont {Veldhorst}}, \bibinfo {author} {\bibfnamefont
  {R.}~\bibnamefont {Rahman}}, \bibinfo {author} {\bibfnamefont
  {G.}~\bibnamefont {Klimeck}}, \emph {et~al.},\ }\bibfield  {title} {\bibinfo
  {title} {Electrically controlling single-spin qubits in a continuous
  microwave field},\ }\href {https://doi.org/10.1126/sciadv.1500022} {\bibfield
   {journal} {\bibinfo  {journal} {Science Advances}\ }\textbf {\bibinfo
  {volume} {1}},\ \bibinfo {pages} {e1500022} (\bibinfo {year}
  {2015})}\BibitemShut {NoStop}%
\bibitem [{\citenamefont {Slack-Smith}\ \emph {et~al.}(2022)\citenamefont
  {Slack-Smith}, \citenamefont {Hudson}, \citenamefont {Cifuentes} \emph
  {et~al.}}]{vahapoglu2022coherent}%
  \BibitemOpen
  \bibfield  {author} {\bibinfo {author} {\bibfnamefont {J.}~\bibnamefont
  {Slack-Smith}}, \bibinfo {author} {\bibfnamefont {F.}~\bibnamefont {Hudson}},
  \bibinfo {author} {\bibfnamefont {J.}~\bibnamefont {Cifuentes}}, \emph
  {et~al.},\ }\bibfield  {title} {\bibinfo {title} {Coherent control of
  electron spin qubits in silicon using a global field},\ }\href
  {https://doi.org/10.1038/s41534-022-00645-w} {\bibfield  {journal} {\bibinfo
  {journal} {npj Quantum Information}\ }\textbf {\bibinfo {volume} {8}},\
  \bibinfo {pages} {126} (\bibinfo {year} {2022})}\BibitemShut {NoStop}%
\bibitem [{\citenamefont {Mourik}\ \emph {et~al.}(2018)\citenamefont {Mourik},
  \citenamefont {Asaad}, \citenamefont {Firgau}, \citenamefont {Pla},
  \citenamefont {Holmes}, \citenamefont {Milburn}, \citenamefont {McCallum},\
  and\ \citenamefont {Morello}}]{mourik2018exploring}%
  \BibitemOpen
  \bibfield  {author} {\bibinfo {author} {\bibfnamefont {V.}~\bibnamefont
  {Mourik}}, \bibinfo {author} {\bibfnamefont {S.}~\bibnamefont {Asaad}},
  \bibinfo {author} {\bibfnamefont {H.}~\bibnamefont {Firgau}}, \bibinfo
  {author} {\bibfnamefont {J.~J.}\ \bibnamefont {Pla}}, \bibinfo {author}
  {\bibfnamefont {C.}~\bibnamefont {Holmes}}, \bibinfo {author} {\bibfnamefont
  {G.~J.}\ \bibnamefont {Milburn}}, \bibinfo {author} {\bibfnamefont {J.~C.}\
  \bibnamefont {McCallum}},\ and\ \bibinfo {author} {\bibfnamefont
  {A.}~\bibnamefont {Morello}},\ }\bibfield  {title} {\bibinfo {title}
  {Exploring quantum chaos with a single nuclear spin},\ }\href
  {https://doi.org/10.1103/PhysRevE.98.042206} {\bibfield  {journal} {\bibinfo
  {journal} {Physical Review E}\ }\textbf {\bibinfo {volume} {98}},\ \bibinfo
  {pages} {042206} (\bibinfo {year} {2018})}\BibitemShut {NoStop}%
\bibitem [{\citenamefont {Chiesa}\ \emph {et~al.}(2020)\citenamefont {Chiesa},
  \citenamefont {Macaluso}, \citenamefont {Petiziol}, \citenamefont
  {Wimberger}, \citenamefont {Santini},\ and\ \citenamefont
  {Carretta}}]{chiesa2020molecular}%
  \BibitemOpen
  \bibfield  {author} {\bibinfo {author} {\bibfnamefont {A.}~\bibnamefont
  {Chiesa}}, \bibinfo {author} {\bibfnamefont {E.}~\bibnamefont {Macaluso}},
  \bibinfo {author} {\bibfnamefont {F.}~\bibnamefont {Petiziol}}, \bibinfo
  {author} {\bibfnamefont {S.}~\bibnamefont {Wimberger}}, \bibinfo {author}
  {\bibfnamefont {P.}~\bibnamefont {Santini}},\ and\ \bibinfo {author}
  {\bibfnamefont {S.}~\bibnamefont {Carretta}},\ }\bibfield  {title} {\bibinfo
  {title} {Molecular nanomagnets as qubits with embedded quantum-error
  correction},\ }\href {https://doi.org/10.1021/acs.jpclett.0c02213} {\bibfield
   {journal} {\bibinfo  {journal} {The Journal of Physical Chemistry Letters}\
  }\textbf {\bibinfo {volume} {11}},\ \bibinfo {pages} {8610} (\bibinfo {year}
  {2020})}\BibitemShut {NoStop}%
\bibitem [{\citenamefont {Gross}(2021)}]{gross2021designing}%
  \BibitemOpen
  \bibfield  {author} {\bibinfo {author} {\bibfnamefont {J.~A.}\ \bibnamefont
  {Gross}},\ }\bibfield  {title} {\bibinfo {title} {Designing {C}odes around
  interactions: {T}he {C}ase of a {S}pin},\ }\href
  {https://doi.org/10.1103/PhysRevLett.127.010504} {\bibfield  {journal}
  {\bibinfo  {journal} {Physical Review Letters}\ }\textbf {\bibinfo {volume}
  {127}},\ \bibinfo {pages} {010504} (\bibinfo {year} {2021})}\BibitemShut
  {NoStop}%
\bibitem [{\citenamefont {Gross}\ \emph {et~al.}(2021)\citenamefont {Gross},
  \citenamefont {Godfrin}, \citenamefont {Blais},\ and\ \citenamefont
  {Dupont-Ferrier}}]{gross2021hardware}%
  \BibitemOpen
  \bibfield  {author} {\bibinfo {author} {\bibfnamefont {J.~A.}\ \bibnamefont
  {Gross}}, \bibinfo {author} {\bibfnamefont {C.}~\bibnamefont {Godfrin}},
  \bibinfo {author} {\bibfnamefont {A.}~\bibnamefont {Blais}},\ and\ \bibinfo
  {author} {\bibfnamefont {E.}~\bibnamefont {Dupont-Ferrier}},\ }\bibfield
  {title} {\bibinfo {title} {Hardware-efficient error-correcting codes for
  large nuclear spins},\ }\href {https://arxiv.org/abs/2103.08548} {\bibfield
  {journal} {\bibinfo  {journal} {arXiv preprint arXiv:2103.08548}\ } (\bibinfo
  {year} {2021})}\BibitemShut {NoStop}%
\bibitem [{\citenamefont {Franke}\ \emph {et~al.}(2015)\citenamefont {Franke},
  \citenamefont {Hrubesch}, \citenamefont {K{\"u}nzl}, \citenamefont {Becker},
  \citenamefont {Itoh}, \citenamefont {Stutzmann}, \citenamefont {Hoehne},
  \citenamefont {Dreher},\ and\ \citenamefont
  {Brandt}}]{franke2015interaction}%
  \BibitemOpen
  \bibfield  {author} {\bibinfo {author} {\bibfnamefont {D.~P.}\ \bibnamefont
  {Franke}}, \bibinfo {author} {\bibfnamefont {F.~M.}\ \bibnamefont
  {Hrubesch}}, \bibinfo {author} {\bibfnamefont {M.}~\bibnamefont {K{\"u}nzl}},
  \bibinfo {author} {\bibfnamefont {H.-W.}\ \bibnamefont {Becker}}, \bibinfo
  {author} {\bibfnamefont {K.~M.}\ \bibnamefont {Itoh}}, \bibinfo {author}
  {\bibfnamefont {M.}~\bibnamefont {Stutzmann}}, \bibinfo {author}
  {\bibfnamefont {F.}~\bibnamefont {Hoehne}}, \bibinfo {author} {\bibfnamefont
  {L.}~\bibnamefont {Dreher}},\ and\ \bibinfo {author} {\bibfnamefont {M.~S.}\
  \bibnamefont {Brandt}},\ }\bibfield  {title} {\bibinfo {title} {Interaction
  of strain and nuclear spins in silicon: {Q}uadrupolar effects on ionized
  donors},\ }\href {https://doi.org/10.1103/PhysRevLett.115.057601} {\bibfield
  {journal} {\bibinfo  {journal} {Physical Review Letters}\ }\textbf {\bibinfo
  {volume} {115}},\ \bibinfo {pages} {057601} (\bibinfo {year}
  {2015})}\BibitemShut {NoStop}%
\bibitem [{\citenamefont {Morley}\ \emph {et~al.}(2010)\citenamefont {Morley},
  \citenamefont {Warner}, \citenamefont {Stoneham}, \citenamefont {Greenland},
  \citenamefont {Van~Tol}, \citenamefont {Kay},\ and\ \citenamefont
  {Aeppli}}]{morley2010initialization}%
  \BibitemOpen
  \bibfield  {author} {\bibinfo {author} {\bibfnamefont {G.~W.}\ \bibnamefont
  {Morley}}, \bibinfo {author} {\bibfnamefont {M.}~\bibnamefont {Warner}},
  \bibinfo {author} {\bibfnamefont {A.~M.}\ \bibnamefont {Stoneham}}, \bibinfo
  {author} {\bibfnamefont {P.~T.}\ \bibnamefont {Greenland}}, \bibinfo {author}
  {\bibfnamefont {J.}~\bibnamefont {Van~Tol}}, \bibinfo {author} {\bibfnamefont
  {C.~W.}\ \bibnamefont {Kay}},\ and\ \bibinfo {author} {\bibfnamefont
  {G.}~\bibnamefont {Aeppli}},\ }\bibfield  {title} {\bibinfo {title} {The
  initialization and manipulation of quantum information stored in silicon by
  bismuth dopants},\ }\href {https://doi.org/10.1038/nmat2828} {\bibfield
  {journal} {\bibinfo  {journal} {Nature Materials}\ }\textbf {\bibinfo
  {volume} {9}},\ \bibinfo {pages} {725} (\bibinfo {year} {2010})}\BibitemShut
  {NoStop}%
\bibitem [{\citenamefont {Ono}\ \emph {et~al.}(2013)\citenamefont {Ono},
  \citenamefont {Ishihara}, \citenamefont {Sato}, \citenamefont {Ohno},\ and\
  \citenamefont {Ohno}}]{ono2013coherent}%
  \BibitemOpen
  \bibfield  {author} {\bibinfo {author} {\bibfnamefont {M.}~\bibnamefont
  {Ono}}, \bibinfo {author} {\bibfnamefont {J.}~\bibnamefont {Ishihara}},
  \bibinfo {author} {\bibfnamefont {G.}~\bibnamefont {Sato}}, \bibinfo {author}
  {\bibfnamefont {Y.}~\bibnamefont {Ohno}},\ and\ \bibinfo {author}
  {\bibfnamefont {H.}~\bibnamefont {Ohno}},\ }\bibfield  {title} {\bibinfo
  {title} {Coherent manipulation of nuclear spins in semiconductors with an
  electric field},\ }\href {https://doi.org/10.7567/APEX.6.033002} {\bibfield
  {journal} {\bibinfo  {journal} {Applied Physics Express}\ }\textbf {\bibinfo
  {volume} {6}},\ \bibinfo {pages} {033002} (\bibinfo {year}
  {2013})}\BibitemShut {NoStop}%
\bibitem [{\citenamefont {Savytskyy}\ \emph {et~al.}(2023)\citenamefont
  {Savytskyy}, \citenamefont {Botzem}, \citenamefont {Fern\'andez~de Fuentes},
  \citenamefont {J\"ocker}, \citenamefont {Pla}, \citenamefont {Hudson},
  \citenamefont {Itoh}, \citenamefont {Jakob}, \citenamefont {Johnson},
  \citenamefont {Jamieson} \emph {et~al.}}]{savytskyy2023electrically}%
  \BibitemOpen
  \bibfield  {author} {\bibinfo {author} {\bibfnamefont {R.}~\bibnamefont
  {Savytskyy}}, \bibinfo {author} {\bibfnamefont {T.}~\bibnamefont {Botzem}},
  \bibinfo {author} {\bibfnamefont {I.}~\bibnamefont {Fern\'andez~de Fuentes}},
  \bibinfo {author} {\bibfnamefont {B.}~\bibnamefont {J\"ocker}}, \bibinfo
  {author} {\bibfnamefont {J.~J.}\ \bibnamefont {Pla}}, \bibinfo {author}
  {\bibfnamefont {F.~E.}\ \bibnamefont {Hudson}}, \bibinfo {author}
  {\bibfnamefont {K.~M.}\ \bibnamefont {Itoh}}, \bibinfo {author}
  {\bibfnamefont {A.~M.}\ \bibnamefont {Jakob}}, \bibinfo {author}
  {\bibfnamefont {B.~C.}\ \bibnamefont {Johnson}}, \bibinfo {author}
  {\bibfnamefont {D.~N.}\ \bibnamefont {Jamieson}}, \emph {et~al.},\ }\bibfield
   {title} {\bibinfo {title} {An electrically driven single-atom
  “flip-flop” qubit},\ }\href {https://doi.org/10.1126/sciadv.add9408}
  {\bibfield  {journal} {\bibinfo  {journal} {Science Advances}\ }\textbf
  {\bibinfo {volume} {9}},\ \bibinfo {pages} {eadd9408} (\bibinfo {year}
  {2023})}\BibitemShut {NoStop}%
\bibitem [{\citenamefont {Pica}\ \emph {et~al.}(2014)\citenamefont {Pica},
  \citenamefont {Wolfowicz}, \citenamefont {Urdampilleta}, \citenamefont
  {Thewalt}, \citenamefont {Riemann}, \citenamefont {Abrosimov}, \citenamefont
  {Becker}, \citenamefont {Pohl}, \citenamefont {Morton}, \citenamefont
  {Bhatt}, \citenamefont {Lyon},\ and\ \citenamefont
  {Lovett}}]{pica2014hyperfine}%
  \BibitemOpen
  \bibfield  {author} {\bibinfo {author} {\bibfnamefont {G.}~\bibnamefont
  {Pica}}, \bibinfo {author} {\bibfnamefont {G.}~\bibnamefont {Wolfowicz}},
  \bibinfo {author} {\bibfnamefont {M.}~\bibnamefont {Urdampilleta}}, \bibinfo
  {author} {\bibfnamefont {M.~L.~W.}\ \bibnamefont {Thewalt}}, \bibinfo
  {author} {\bibfnamefont {H.}~\bibnamefont {Riemann}}, \bibinfo {author}
  {\bibfnamefont {N.~V.}\ \bibnamefont {Abrosimov}}, \bibinfo {author}
  {\bibfnamefont {P.}~\bibnamefont {Becker}}, \bibinfo {author} {\bibfnamefont
  {H.-J.}\ \bibnamefont {Pohl}}, \bibinfo {author} {\bibfnamefont {J.~J.~L.}\
  \bibnamefont {Morton}}, \bibinfo {author} {\bibfnamefont {R.~N.}\
  \bibnamefont {Bhatt}}, \bibinfo {author} {\bibfnamefont {S.~A.}\ \bibnamefont
  {Lyon}},\ and\ \bibinfo {author} {\bibfnamefont {B.~W.}\ \bibnamefont
  {Lovett}},\ }\bibfield  {title} {\bibinfo {title} {Hyperfine stark effect of
  shallow donors in silicon},\ }\href
  {https://doi.org/10.1103/PhysRevB.90.195204} {\bibfield  {journal} {\bibinfo
  {journal} {Physical Review B}\ }\textbf {\bibinfo {volume} {90}},\ \bibinfo
  {pages} {195204} (\bibinfo {year} {2014})}\BibitemShut {NoStop}%
\bibitem [{\citenamefont {Morello}\ \emph {et~al.}(2010)\citenamefont
  {Morello}, \citenamefont {Pla}, \citenamefont {Zwanenburg}, \citenamefont
  {Chan}, \citenamefont {Tan}, \citenamefont {Huebl}, \citenamefont
  {M{\"o}tt{\"o}nen}, \citenamefont {Nugroho}, \citenamefont {Yang},
  \citenamefont {Van~Donkelaar} \emph {et~al.}}]{morello2010single}%
  \BibitemOpen
  \bibfield  {author} {\bibinfo {author} {\bibfnamefont {A.}~\bibnamefont
  {Morello}}, \bibinfo {author} {\bibfnamefont {J.~J.}\ \bibnamefont {Pla}},
  \bibinfo {author} {\bibfnamefont {F.~A.}\ \bibnamefont {Zwanenburg}},
  \bibinfo {author} {\bibfnamefont {K.~W.}\ \bibnamefont {Chan}}, \bibinfo
  {author} {\bibfnamefont {K.~Y.}\ \bibnamefont {Tan}}, \bibinfo {author}
  {\bibfnamefont {H.}~\bibnamefont {Huebl}}, \bibinfo {author} {\bibfnamefont
  {M.}~\bibnamefont {M{\"o}tt{\"o}nen}}, \bibinfo {author} {\bibfnamefont
  {C.~D.}\ \bibnamefont {Nugroho}}, \bibinfo {author} {\bibfnamefont
  {C.}~\bibnamefont {Yang}}, \bibinfo {author} {\bibfnamefont {J.~A.}\
  \bibnamefont {Van~Donkelaar}}, \emph {et~al.},\ }\bibfield  {title} {\bibinfo
  {title} {Single-shot readout of an electron spin in silicon},\ }\href
  {https://doi.org/10.1038/nature09392} {\bibfield  {journal} {\bibinfo
  {journal} {Nature}\ }\textbf {\bibinfo {volume} {467}},\ \bibinfo {pages}
  {687} (\bibinfo {year} {2010})}\BibitemShut {NoStop}%
\bibitem [{\citenamefont {Thorbeck}\ and\ \citenamefont
  {Zimmerman}(2015)}]{thorbeck2015formation}%
  \BibitemOpen
  \bibfield  {author} {\bibinfo {author} {\bibfnamefont {T.}~\bibnamefont
  {Thorbeck}}\ and\ \bibinfo {author} {\bibfnamefont {N.~M.}\ \bibnamefont
  {Zimmerman}},\ }\bibfield  {title} {\bibinfo {title} {Formation of
  strain-induced quantum dots in gated semiconductor nanostructures},\ }\href
  {https://doi.org/10.1063/1.4928320} {\bibfield  {journal} {\bibinfo
  {journal} {AIP Advances}\ }\textbf {\bibinfo {volume} {5}},\ \bibinfo {pages}
  {087107} (\bibinfo {year} {2015})}\BibitemShut {NoStop}%
\bibitem [{\citenamefont {Adambukulam}\ \emph {et~al.}(2021)\citenamefont
  {Adambukulam}, \citenamefont {Sewani}, \citenamefont {Stemp}, \citenamefont
  {Asaad}, \citenamefont {M{\k{a}}dzik}, \citenamefont {Morello},\ and\
  \citenamefont {Laucht}}]{adambukulam2021ultra}%
  \BibitemOpen
  \bibfield  {author} {\bibinfo {author} {\bibfnamefont {C.}~\bibnamefont
  {Adambukulam}}, \bibinfo {author} {\bibfnamefont {V.}~\bibnamefont {Sewani}},
  \bibinfo {author} {\bibfnamefont {H.}~\bibnamefont {Stemp}}, \bibinfo
  {author} {\bibfnamefont {S.}~\bibnamefont {Asaad}}, \bibinfo {author}
  {\bibfnamefont {M.}~\bibnamefont {M{\k{a}}dzik}}, \bibinfo {author}
  {\bibfnamefont {A.}~\bibnamefont {Morello}},\ and\ \bibinfo {author}
  {\bibfnamefont {A.}~\bibnamefont {Laucht}},\ }\bibfield  {title} {\bibinfo
  {title} {An ultra-stable 1.5 {T} permanent magnet assembly for qubit
  experiments at cryogenic temperatures},\ }\href
  {https://doi.org/10.1063/5.0055318} {\bibfield  {journal} {\bibinfo
  {journal} {Review of Scientific Instruments}\ }\textbf {\bibinfo {volume}
  {92}},\ \bibinfo {pages} {085106} (\bibinfo {year} {2021})}\BibitemShut
  {NoStop}%
\bibitem [{\citenamefont {Franke}\ \emph {et~al.}(2016)\citenamefont {Franke},
  \citenamefont {Pfl{\"u}ger}, \citenamefont {Mortemousque}, \citenamefont
  {Itoh},\ and\ \citenamefont {Brandt}}]{franke2016quadrupolar}%
  \BibitemOpen
  \bibfield  {author} {\bibinfo {author} {\bibfnamefont {D.~P.}\ \bibnamefont
  {Franke}}, \bibinfo {author} {\bibfnamefont {M.~P.}\ \bibnamefont
  {Pfl{\"u}ger}}, \bibinfo {author} {\bibfnamefont {P.-A.}\ \bibnamefont
  {Mortemousque}}, \bibinfo {author} {\bibfnamefont {K.~M.}\ \bibnamefont
  {Itoh}},\ and\ \bibinfo {author} {\bibfnamefont {M.~S.}\ \bibnamefont
  {Brandt}},\ }\bibfield  {title} {\bibinfo {title} {Quadrupolar effects on
  nuclear spins of neutral arsenic donors in silicon},\ }\href
  {https://doi.org/10.1103/PhysRevB.93.161303} {\bibfield  {journal} {\bibinfo
  {journal} {Physical Review B}\ }\textbf {\bibinfo {volume} {93}},\ \bibinfo
  {pages} {161303} (\bibinfo {year} {2016})}\BibitemShut {NoStop}%
\bibitem [{\citenamefont {Sangtawesin}\ \emph {et~al.}(2016)\citenamefont
  {Sangtawesin}, \citenamefont {McLellan}, \citenamefont {Myers}, \citenamefont
  {Jayich}, \citenamefont {Awschalom},\ and\ \citenamefont
  {Petta}}]{sangtawesin2016hyperfine}%
  \BibitemOpen
  \bibfield  {author} {\bibinfo {author} {\bibfnamefont {S.}~\bibnamefont
  {Sangtawesin}}, \bibinfo {author} {\bibfnamefont {C.}~\bibnamefont
  {McLellan}}, \bibinfo {author} {\bibfnamefont {B.}~\bibnamefont {Myers}},
  \bibinfo {author} {\bibfnamefont {A.~B.}\ \bibnamefont {Jayich}}, \bibinfo
  {author} {\bibfnamefont {D.}~\bibnamefont {Awschalom}},\ and\ \bibinfo
  {author} {\bibfnamefont {J.~R.}\ \bibnamefont {Petta}},\ }\bibfield  {title}
  {\bibinfo {title} {Hyperfine-enhanced gyromagnetic ratio of a nuclear spin in
  diamond},\ }\href {https://doi.org/10.1088/1367-2630/18/8/083016} {\bibfield
  {journal} {\bibinfo  {journal} {New Journal of Physics}\ }\textbf {\bibinfo
  {volume} {18}},\ \bibinfo {pages} {083016} (\bibinfo {year}
  {2016})}\BibitemShut {NoStop}%
\bibitem [{\citenamefont {Tosi}\ \emph {et~al.}(2017)\citenamefont {Tosi},
  \citenamefont {Mohiyaddin}, \citenamefont {Schmitt}, \citenamefont {Tenberg},
  \citenamefont {Rahman}, \citenamefont {Klimeck},\ and\ \citenamefont
  {Morello}}]{tosi2017silicon}%
  \BibitemOpen
  \bibfield  {author} {\bibinfo {author} {\bibfnamefont {G.}~\bibnamefont
  {Tosi}}, \bibinfo {author} {\bibfnamefont {F.~A.}\ \bibnamefont
  {Mohiyaddin}}, \bibinfo {author} {\bibfnamefont {V.}~\bibnamefont {Schmitt}},
  \bibinfo {author} {\bibfnamefont {S.}~\bibnamefont {Tenberg}}, \bibinfo
  {author} {\bibfnamefont {R.}~\bibnamefont {Rahman}}, \bibinfo {author}
  {\bibfnamefont {G.}~\bibnamefont {Klimeck}},\ and\ \bibinfo {author}
  {\bibfnamefont {A.}~\bibnamefont {Morello}},\ }\bibfield  {title} {\bibinfo
  {title} {Silicon quantum processor with robust long-distance qubit
  couplings},\ }\href {https://doi.org/10.1038/s41467-017-00378-x} {\bibfield
  {journal} {\bibinfo  {journal} {Nature Communications}\ }\textbf {\bibinfo
  {volume} {8}},\ \bibinfo {pages} {1} (\bibinfo {year} {2017})}\BibitemShut
  {NoStop}%
\bibitem [{\citenamefont {Armstrong}\ \emph {et~al.}(1961)\citenamefont
  {Armstrong}, \citenamefont {Bloembergen},\ and\ \citenamefont
  {Gill}}]{armstrong1961linear}%
  \BibitemOpen
  \bibfield  {author} {\bibinfo {author} {\bibfnamefont {J.}~\bibnamefont
  {Armstrong}}, \bibinfo {author} {\bibfnamefont {N.}~\bibnamefont
  {Bloembergen}},\ and\ \bibinfo {author} {\bibfnamefont {D.}~\bibnamefont
  {Gill}},\ }\bibfield  {title} {\bibinfo {title} {Linear effect of applied
  electric field on nuclear quadrupole resonance},\ }\href
  {https://doi.org/10.1103/PhysRevLett.7.11} {\bibfield  {journal} {\bibinfo
  {journal} {Physical Review Letters}\ }\textbf {\bibinfo {volume} {7}},\
  \bibinfo {pages} {11} (\bibinfo {year} {1961})}\BibitemShut {NoStop}%
\bibitem [{\citenamefont {Franke}\ \emph {et~al.}(2017)\citenamefont {Franke},
  \citenamefont {Pfl{\"u}ger}, \citenamefont {Itoh},\ and\ \citenamefont
  {Brandt}}]{franke2017multiple}%
  \BibitemOpen
  \bibfield  {author} {\bibinfo {author} {\bibfnamefont {D.~P.}\ \bibnamefont
  {Franke}}, \bibinfo {author} {\bibfnamefont {M.~P.}\ \bibnamefont
  {Pfl{\"u}ger}}, \bibinfo {author} {\bibfnamefont {K.~M.}\ \bibnamefont
  {Itoh}},\ and\ \bibinfo {author} {\bibfnamefont {M.~S.}\ \bibnamefont
  {Brandt}},\ }\bibfield  {title} {\bibinfo {title} {Multiple-quantum
  transitions and charge-induced decoherence of donor nuclear spins in
  silicon},\ }\href {https://doi.org/10.1103/PhysRevLett.118.246401} {\bibfield
   {journal} {\bibinfo  {journal} {Physical Review Letters}\ }\textbf {\bibinfo
  {volume} {118}},\ \bibinfo {pages} {246401} (\bibinfo {year}
  {2017})}\BibitemShut {NoStop}%
\bibitem [{\citenamefont {Nielsen}\ \emph {et~al.}(2021)\citenamefont
  {Nielsen}, \citenamefont {Gamble}, \citenamefont {Rudinger}, \citenamefont
  {Scholten}, \citenamefont {Young},\ and\ \citenamefont
  {Blume-Kohout}}]{nielsen2021gate}%
  \BibitemOpen
  \bibfield  {author} {\bibinfo {author} {\bibfnamefont {E.}~\bibnamefont
  {Nielsen}}, \bibinfo {author} {\bibfnamefont {J.~K.}\ \bibnamefont {Gamble}},
  \bibinfo {author} {\bibfnamefont {K.}~\bibnamefont {Rudinger}}, \bibinfo
  {author} {\bibfnamefont {T.}~\bibnamefont {Scholten}}, \bibinfo {author}
  {\bibfnamefont {K.}~\bibnamefont {Young}},\ and\ \bibinfo {author}
  {\bibfnamefont {R.}~\bibnamefont {Blume-Kohout}},\ }\bibfield  {title}
  {\bibinfo {title} {Gate set tomography},\ }\href
  {https://doi.org/10.22331/q-2021-10-05-557} {\bibfield  {journal} {\bibinfo
  {journal} {Quantum}\ }\textbf {\bibinfo {volume} {5}},\ \bibinfo {pages}
  {557} (\bibinfo {year} {2021})}\BibitemShut {NoStop}%
\bibitem [{\citenamefont {Undseth}\ \emph {et~al.}(2023)\citenamefont
  {Undseth}, \citenamefont {Pietx-Casas}, \citenamefont {Raymenants},
  \citenamefont {Mehmandoost}, \citenamefont {Madzik}, \citenamefont {Philips},
  \citenamefont {de~Snoo}, \citenamefont {Michalak}, \citenamefont {Amitonov},
  \citenamefont {Tryputen} \emph {et~al.}}]{undseth2023hotter}%
  \BibitemOpen
  \bibfield  {author} {\bibinfo {author} {\bibfnamefont {B.}~\bibnamefont
  {Undseth}}, \bibinfo {author} {\bibfnamefont {O.}~\bibnamefont
  {Pietx-Casas}}, \bibinfo {author} {\bibfnamefont {E.}~\bibnamefont
  {Raymenants}}, \bibinfo {author} {\bibfnamefont {M.}~\bibnamefont
  {Mehmandoost}}, \bibinfo {author} {\bibfnamefont {M.~T.}\ \bibnamefont
  {Madzik}}, \bibinfo {author} {\bibfnamefont {S.~G.}\ \bibnamefont {Philips}},
  \bibinfo {author} {\bibfnamefont {S.~L.}\ \bibnamefont {de~Snoo}}, \bibinfo
  {author} {\bibfnamefont {D.~J.}\ \bibnamefont {Michalak}}, \bibinfo {author}
  {\bibfnamefont {S.~V.}\ \bibnamefont {Amitonov}}, \bibinfo {author}
  {\bibfnamefont {L.}~\bibnamefont {Tryputen}}, \emph {et~al.},\ }\bibfield
  {title} {\bibinfo {title} {Hotter is easier: unexpected temperature
  dependence of spin qubit frequencies},\ }\href
  {https://arxiv.org/abs/2304.12984} {\bibfield  {journal} {\bibinfo  {journal}
  {arXiv preprint arXiv:2304.12984}\ } (\bibinfo {year} {2023})}\BibitemShut
  {NoStop}%
\bibitem [{\citenamefont {Gupta}\ \emph {et~al.}(2023)\citenamefont {Gupta},
  \citenamefont {Vaartjes}, \citenamefont {Yu}, \citenamefont {Morello},\ and\
  \citenamefont {Sanders}}]{gupta2023robust}%
  \BibitemOpen
  \bibfield  {author} {\bibinfo {author} {\bibfnamefont {P.}~\bibnamefont
  {Gupta}}, \bibinfo {author} {\bibfnamefont {A.}~\bibnamefont {Vaartjes}},
  \bibinfo {author} {\bibfnamefont {X.}~\bibnamefont {Yu}}, \bibinfo {author}
  {\bibfnamefont {A.}~\bibnamefont {Morello}},\ and\ \bibinfo {author}
  {\bibfnamefont {B.~C.}\ \bibnamefont {Sanders}},\ }\bibfield  {title}
  {\bibinfo {title} {Robust macroscopic schr\"{o}dinger's cat on a nucleus},\
  }\href {https://arxiv.org/abs/2304.13813} {\bibfield  {journal} {\bibinfo
  {journal} {arXiv preprint arXiv:2304.13813}\ } (\bibinfo {year}
  {2023})}\BibitemShut {NoStop}%
\bibitem [{\citenamefont {Chalopin}\ \emph {et~al.}(2018)\citenamefont
  {Chalopin}, \citenamefont {Bouazza}, \citenamefont {Evrard}, \citenamefont
  {Makhalov}, \citenamefont {Dreon}, \citenamefont {Dalibard}, \citenamefont
  {Sidorenkov},\ and\ \citenamefont {Nascimbene}}]{chalopin2018quantum}%
  \BibitemOpen
  \bibfield  {author} {\bibinfo {author} {\bibfnamefont {T.}~\bibnamefont
  {Chalopin}}, \bibinfo {author} {\bibfnamefont {C.}~\bibnamefont {Bouazza}},
  \bibinfo {author} {\bibfnamefont {A.}~\bibnamefont {Evrard}}, \bibinfo
  {author} {\bibfnamefont {V.}~\bibnamefont {Makhalov}}, \bibinfo {author}
  {\bibfnamefont {D.}~\bibnamefont {Dreon}}, \bibinfo {author} {\bibfnamefont
  {J.}~\bibnamefont {Dalibard}}, \bibinfo {author} {\bibfnamefont {L.~A.}\
  \bibnamefont {Sidorenkov}},\ and\ \bibinfo {author} {\bibfnamefont
  {S.}~\bibnamefont {Nascimbene}},\ }\bibfield  {title} {\bibinfo {title}
  {Quantum-enhanced sensing using non-classical spin states of a highly
  magnetic atom},\ }\href {https://doi.org/10.1038/s41467-018-07433-1}
  {\bibfield  {journal} {\bibinfo  {journal} {Nature Communications}\ }\textbf
  {\bibinfo {volume} {9}},\ \bibinfo {pages} {4955} (\bibinfo {year}
  {2018})}\BibitemShut {NoStop}%
\bibitem [{\citenamefont {Zaw}\ \emph {et~al.}(2022)\citenamefont {Zaw},
  \citenamefont {Aw}, \citenamefont {Lasmar},\ and\ \citenamefont
  {Scarani}}]{zaw2022detecting}%
  \BibitemOpen
  \bibfield  {author} {\bibinfo {author} {\bibfnamefont {L.~H.}\ \bibnamefont
  {Zaw}}, \bibinfo {author} {\bibfnamefont {C.~C.}\ \bibnamefont {Aw}},
  \bibinfo {author} {\bibfnamefont {Z.}~\bibnamefont {Lasmar}},\ and\ \bibinfo
  {author} {\bibfnamefont {V.}~\bibnamefont {Scarani}},\ }\bibfield  {title}
  {\bibinfo {title} {Detecting quantumness in uniform precessions},\ }\href
  {https://doi.org/10.1103/PhysRevA.106.032222} {\bibfield  {journal} {\bibinfo
   {journal} {Physical Review A}\ }\textbf {\bibinfo {volume} {106}},\ \bibinfo
  {pages} {032222} (\bibinfo {year} {2022})}\BibitemShut {NoStop}%
\bibitem [{\citenamefont {Corley-Wiciak}\ \emph {et~al.}(2023)\citenamefont
  {Corley-Wiciak}, \citenamefont {Richter}, \citenamefont {Zoellner},
  \citenamefont {Zaitsev}, \citenamefont {Manganelli}, \citenamefont
  {Zatterin}, \citenamefont {Schülli}, \citenamefont {Corley-Wiciak},
  \citenamefont {Katzer}, \citenamefont {Reichmann}, \citenamefont {Klesse},
  \citenamefont {Hendrickx}, \citenamefont {Sammak}, \citenamefont {Veldhorst},
  \citenamefont {Scappucci}, \citenamefont {Virgilio},\ and\ \citenamefont
  {Capellini}}]{corley2023nanoscale}%
  \BibitemOpen
  \bibfield  {author} {\bibinfo {author} {\bibfnamefont {C.}~\bibnamefont
  {Corley-Wiciak}}, \bibinfo {author} {\bibfnamefont {C.}~\bibnamefont
  {Richter}}, \bibinfo {author} {\bibfnamefont {M.~H.}\ \bibnamefont
  {Zoellner}}, \bibinfo {author} {\bibfnamefont {I.}~\bibnamefont {Zaitsev}},
  \bibinfo {author} {\bibfnamefont {C.~L.}\ \bibnamefont {Manganelli}},
  \bibinfo {author} {\bibfnamefont {E.}~\bibnamefont {Zatterin}}, \bibinfo
  {author} {\bibfnamefont {T.~U.}\ \bibnamefont {Schülli}}, \bibinfo {author}
  {\bibfnamefont {A.~A.}\ \bibnamefont {Corley-Wiciak}}, \bibinfo {author}
  {\bibfnamefont {J.}~\bibnamefont {Katzer}}, \bibinfo {author} {\bibfnamefont
  {F.}~\bibnamefont {Reichmann}}, \bibinfo {author} {\bibfnamefont {W.~M.}\
  \bibnamefont {Klesse}}, \bibinfo {author} {\bibfnamefont {N.~W.}\
  \bibnamefont {Hendrickx}}, \bibinfo {author} {\bibfnamefont {A.}~\bibnamefont
  {Sammak}}, \bibinfo {author} {\bibfnamefont {M.}~\bibnamefont {Veldhorst}},
  \bibinfo {author} {\bibfnamefont {G.}~\bibnamefont {Scappucci}}, \bibinfo
  {author} {\bibfnamefont {M.}~\bibnamefont {Virgilio}},\ and\ \bibinfo
  {author} {\bibfnamefont {G.}~\bibnamefont {Capellini}},\ }\bibfield  {title}
  {\bibinfo {title} {Nanoscale mapping of the 3d strain tensor in a germanium
  quantum well hosting a functional spin qubit device},\ }\href
  {https://doi.org/10.1021/acsami.2c17395} {\bibfield  {journal} {\bibinfo
  {journal} {ACS Applied Materials \& Interfaces}\ }\textbf {\bibinfo {volume}
  {15}},\ \bibinfo {pages} {3119} (\bibinfo {year} {2023})}\BibitemShut
  {NoStop}%
\end{thebibliography}%


\begin{thebibliography}{20}%
\makeatletter
\providecommand \@ifxundefined [1]{%
 \@ifx{#1\undefined}
}%
\providecommand \@ifnum [1]{%
 \ifnum #1\expandafter \@firstoftwo
 \else \expandafter \@secondoftwo
 \fi
}%
\providecommand \@ifx [1]{%
 \ifx #1\expandafter \@firstoftwo
 \else \expandafter \@secondoftwo
 \fi
}%
\providecommand \natexlab [1]{#1}%
\providecommand \enquote  [1]{``#1''}%
\providecommand \bibnamefont  [1]{#1}%
\providecommand \bibfnamefont [1]{#1}%
\providecommand \citenamefont [1]{#1}%
\providecommand \href@noop [0]{\@secondoftwo}%
\providecommand \href [0]{\begingroup \@sanitize@url \@href}%
\providecommand \@href[1]{\@@startlink{#1}\@@href}%
\providecommand \@@href[1]{\endgroup#1\@@endlink}%
\providecommand \@sanitize@url [0]{\catcode `\\12\catcode `\$12\catcode
  `\&12\catcode `\#12\catcode `\^12\catcode `\_12\catcode `\%12\relax}%
\providecommand \@@startlink[1]{}%
\providecommand \@@endlink[0]{}%
\providecommand \url  [0]{\begingroup\@sanitize@url \@url }%
\providecommand \@url [1]{\endgroup\@href {#1}{\urlprefix }}%
\providecommand \urlprefix  [0]{URL }%
\providecommand \Eprint [0]{\href }%
\providecommand \doibase [0]{https://doi.org/}%
\providecommand \selectlanguage [0]{\@gobble}%
\providecommand \bibinfo  [0]{\@secondoftwo}%
\providecommand \bibfield  [0]{\@secondoftwo}%
\providecommand \translation [1]{[#1]}%
\providecommand \BibitemOpen [0]{}%
\providecommand \bibitemStop [0]{}%
\providecommand \bibitemNoStop [0]{.\EOS\space}%
\providecommand \EOS [0]{\spacefactor3000\relax}%
\providecommand \BibitemShut  [1]{\csname bibitem#1\endcsname}%
\let\auto@bib@innerbib\@empty
\bibitem [{\citenamefont {Savytskyy}\ \emph {et~al.}(2023)\citenamefont
  {Savytskyy}, \citenamefont {Botzem}, \citenamefont {Fernandez~de Fuentes},
  \citenamefont {Joecker}, \citenamefont {Pla}, \citenamefont {Hudson},
  \citenamefont {Itoh}, \citenamefont {Jakob}, \citenamefont {Johnson},
  \citenamefont {Jamieson} \emph {et~al.}}]{savytskyy2023electrically}%
  \BibitemOpen
  \bibfield  {author} {\bibinfo {author} {\bibfnamefont {R.}~\bibnamefont
  {Savytskyy}}, \bibinfo {author} {\bibfnamefont {T.}~\bibnamefont {Botzem}},
  \bibinfo {author} {\bibfnamefont {I.}~\bibnamefont {Fernandez~de Fuentes}},
  \bibinfo {author} {\bibfnamefont {B.}~\bibnamefont {Joecker}}, \bibinfo
  {author} {\bibfnamefont {J.~J.}\ \bibnamefont {Pla}}, \bibinfo {author}
  {\bibfnamefont {F.~E.}\ \bibnamefont {Hudson}}, \bibinfo {author}
  {\bibfnamefont {K.~M.}\ \bibnamefont {Itoh}}, \bibinfo {author}
  {\bibfnamefont {A.~M.}\ \bibnamefont {Jakob}}, \bibinfo {author}
  {\bibfnamefont {B.~C.}\ \bibnamefont {Johnson}}, \bibinfo {author}
  {\bibfnamefont {D.~N.}\ \bibnamefont {Jamieson}}, \emph {et~al.},\ }\bibfield
   {title} {\bibinfo {title} {An electrically driven single-atom
  “flip-flop” qubit},\ }\href@noop {} {\bibfield  {journal} {\bibinfo
  {journal} {Science {A}dvances}\ }\textbf {\bibinfo {volume} {9}},\ \bibinfo
  {pages} {eadd9408} (\bibinfo {year} {2023})}\BibitemShut {NoStop}%
\bibitem [{\citenamefont {Morello}\ \emph {et~al.}(2010)\citenamefont
  {Morello}, \citenamefont {Pla}, \citenamefont {Zwanenburg}, \citenamefont
  {Chan}, \citenamefont {Tan}, \citenamefont {Huebl}, \citenamefont
  {M{\"o}tt{\"o}nen}, \citenamefont {Nugroho}, \citenamefont {Yang},
  \citenamefont {Van~Donkelaar} \emph {et~al.}}]{morello2010single}%
  \BibitemOpen
  \bibfield  {author} {\bibinfo {author} {\bibfnamefont {A.}~\bibnamefont
  {Morello}}, \bibinfo {author} {\bibfnamefont {J.~J.}\ \bibnamefont {Pla}},
  \bibinfo {author} {\bibfnamefont {F.~A.}\ \bibnamefont {Zwanenburg}},
  \bibinfo {author} {\bibfnamefont {K.~W.}\ \bibnamefont {Chan}}, \bibinfo
  {author} {\bibfnamefont {K.~Y.}\ \bibnamefont {Tan}}, \bibinfo {author}
  {\bibfnamefont {H.}~\bibnamefont {Huebl}}, \bibinfo {author} {\bibfnamefont
  {M.}~\bibnamefont {M{\"o}tt{\"o}nen}}, \bibinfo {author} {\bibfnamefont
  {C.~D.}\ \bibnamefont {Nugroho}}, \bibinfo {author} {\bibfnamefont
  {C.}~\bibnamefont {Yang}}, \bibinfo {author} {\bibfnamefont {J.~A.}\
  \bibnamefont {Van~Donkelaar}}, \emph {et~al.},\ }\bibfield  {title} {\bibinfo
  {title} {Single-shot readout of an electron spin in silicon},\ }\href@noop {}
  {\bibfield  {journal} {\bibinfo  {journal} {Nature}\ }\textbf {\bibinfo
  {volume} {467}},\ \bibinfo {pages} {687} (\bibinfo {year}
  {2010})}\BibitemShut {NoStop}%
\bibitem [{\citenamefont {Braginsky}\ and\ \citenamefont
  {Khalili}(1996)}]{braginsky1996quantum}%
  \BibitemOpen
  \bibfield  {author} {\bibinfo {author} {\bibfnamefont {V.~B.}\ \bibnamefont
  {Braginsky}}\ and\ \bibinfo {author} {\bibfnamefont {F.~Y.}\ \bibnamefont
  {Khalili}},\ }\bibfield  {title} {\bibinfo {title} {Quantum nondemolition
  measurements: the route from toys to tools},\ }\href@noop {} {\bibfield
  {journal} {\bibinfo  {journal} {Reviews of {M}odern {P}hysics}\ }\textbf
  {\bibinfo {volume} {68}},\ \bibinfo {pages} {1} (\bibinfo {year}
  {1996})}\BibitemShut {NoStop}%
\bibitem [{\citenamefont {Pla}\ \emph {et~al.}(2013)\citenamefont {Pla},
  \citenamefont {Tan}, \citenamefont {Dehollain}, \citenamefont {Lim},
  \citenamefont {Morton}, \citenamefont {Zwanenburg}, \citenamefont {Jamieson},
  \citenamefont {Dzurak},\ and\ \citenamefont {Morello}}]{pla2013high}%
  \BibitemOpen
  \bibfield  {author} {\bibinfo {author} {\bibfnamefont {J.~J.}\ \bibnamefont
  {Pla}}, \bibinfo {author} {\bibfnamefont {K.~Y.}\ \bibnamefont {Tan}},
  \bibinfo {author} {\bibfnamefont {J.~P.}\ \bibnamefont {Dehollain}}, \bibinfo
  {author} {\bibfnamefont {W.~H.}\ \bibnamefont {Lim}}, \bibinfo {author}
  {\bibfnamefont {J.~J.}\ \bibnamefont {Morton}}, \bibinfo {author}
  {\bibfnamefont {F.~A.}\ \bibnamefont {Zwanenburg}}, \bibinfo {author}
  {\bibfnamefont {D.~N.}\ \bibnamefont {Jamieson}}, \bibinfo {author}
  {\bibfnamefont {A.~S.}\ \bibnamefont {Dzurak}},\ and\ \bibinfo {author}
  {\bibfnamefont {A.}~\bibnamefont {Morello}},\ }\bibfield  {title} {\bibinfo
  {title} {High-fidelity readout and control of a nuclear spin qubit in
  silicon},\ }\href@noop {} {\bibfield  {journal} {\bibinfo  {journal}
  {Nature}\ }\textbf {\bibinfo {volume} {496}},\ \bibinfo {pages} {334}
  (\bibinfo {year} {2013})}\BibitemShut {NoStop}%
\bibitem [{\citenamefont {Hile}\ \emph {et~al.}(2018)\citenamefont {Hile},
  \citenamefont {Fricke}, \citenamefont {House}, \citenamefont {Peretz},
  \citenamefont {Chen}, \citenamefont {Wang}, \citenamefont {Broome},
  \citenamefont {Gorman}, \citenamefont {Keizer}, \citenamefont {Rahman} \emph
  {et~al.}}]{hile2018addressable}%
  \BibitemOpen
  \bibfield  {author} {\bibinfo {author} {\bibfnamefont {S.~J.}\ \bibnamefont
  {Hile}}, \bibinfo {author} {\bibfnamefont {L.}~\bibnamefont {Fricke}},
  \bibinfo {author} {\bibfnamefont {M.~G.}\ \bibnamefont {House}}, \bibinfo
  {author} {\bibfnamefont {E.}~\bibnamefont {Peretz}}, \bibinfo {author}
  {\bibfnamefont {C.~Y.}\ \bibnamefont {Chen}}, \bibinfo {author}
  {\bibfnamefont {Y.}~\bibnamefont {Wang}}, \bibinfo {author} {\bibfnamefont
  {M.}~\bibnamefont {Broome}}, \bibinfo {author} {\bibfnamefont {S.~K.}\
  \bibnamefont {Gorman}}, \bibinfo {author} {\bibfnamefont {J.~G.}\
  \bibnamefont {Keizer}}, \bibinfo {author} {\bibfnamefont {R.}~\bibnamefont
  {Rahman}}, \emph {et~al.},\ }\bibfield  {title} {\bibinfo {title}
  {Addressable electron spin resonance using donors and donor molecules in
  silicon},\ }\href@noop {} {\bibfield  {journal} {\bibinfo  {journal} {Science
  {A}dvances}\ }\textbf {\bibinfo {volume} {4}},\ \bibinfo {pages} {eaaq1459}
  (\bibinfo {year} {2018})}\BibitemShut {NoStop}%
\bibitem [{\citenamefont {Franke}\ \emph {et~al.}(2016)\citenamefont {Franke},
  \citenamefont {Pfl{\"u}ger}, \citenamefont {Mortemousque}, \citenamefont
  {Itoh},\ and\ \citenamefont {Brandt}}]{franke2016quadrupolar}%
  \BibitemOpen
  \bibfield  {author} {\bibinfo {author} {\bibfnamefont {D.~P.}\ \bibnamefont
  {Franke}}, \bibinfo {author} {\bibfnamefont {M.~P.}\ \bibnamefont
  {Pfl{\"u}ger}}, \bibinfo {author} {\bibfnamefont {P.-A.}\ \bibnamefont
  {Mortemousque}}, \bibinfo {author} {\bibfnamefont {K.~M.}\ \bibnamefont
  {Itoh}},\ and\ \bibinfo {author} {\bibfnamefont {M.~S.}\ \bibnamefont
  {Brandt}},\ }\bibfield  {title} {\bibinfo {title} {Quadrupolar effects on
  nuclear spins of neutral arsenic donors in silicon},\ }\href@noop {}
  {\bibfield  {journal} {\bibinfo  {journal} {Physical {R}eview {B}}\ }\textbf
  {\bibinfo {volume} {93}},\ \bibinfo {pages} {161303} (\bibinfo {year}
  {2016})}\BibitemShut {NoStop}%
\bibitem [{\citenamefont {M{\k{a}}dzik}\ \emph {et~al.}(2022)\citenamefont
  {M{\k{a}}dzik}, \citenamefont {Asaad}, \citenamefont {Youssry}, \citenamefont
  {Joecker}, \citenamefont {Rudinger}, \citenamefont {Nielsen}, \citenamefont
  {Young}, \citenamefont {Proctor}, \citenamefont {Baczewski}, \citenamefont
  {Laucht} \emph {et~al.}}]{mkadzik2022precision}%
  \BibitemOpen
  \bibfield  {author} {\bibinfo {author} {\bibfnamefont {M.~T.}\ \bibnamefont
  {M{\k{a}}dzik}}, \bibinfo {author} {\bibfnamefont {S.}~\bibnamefont {Asaad}},
  \bibinfo {author} {\bibfnamefont {A.}~\bibnamefont {Youssry}}, \bibinfo
  {author} {\bibfnamefont {B.}~\bibnamefont {Joecker}}, \bibinfo {author}
  {\bibfnamefont {K.~M.}\ \bibnamefont {Rudinger}}, \bibinfo {author}
  {\bibfnamefont {E.}~\bibnamefont {Nielsen}}, \bibinfo {author} {\bibfnamefont
  {K.~C.}\ \bibnamefont {Young}}, \bibinfo {author} {\bibfnamefont {T.~J.}\
  \bibnamefont {Proctor}}, \bibinfo {author} {\bibfnamefont {A.~D.}\
  \bibnamefont {Baczewski}}, \bibinfo {author} {\bibfnamefont {A.}~\bibnamefont
  {Laucht}}, \emph {et~al.},\ }\bibfield  {title} {\bibinfo {title} {Precision
  tomography of a three-qubit donor quantum processor in silicon},\ }\href@noop
  {} {\bibfield  {journal} {\bibinfo  {journal} {Nature}\ }\textbf {\bibinfo
  {volume} {601}},\ \bibinfo {pages} {348} (\bibinfo {year}
  {2022})}\BibitemShut {NoStop}%
\bibitem [{\citenamefont {Asaad}\ \emph {et~al.}(2020)\citenamefont {Asaad},
  \citenamefont {Mourik}, \citenamefont {Joecker}, \citenamefont {Johnson},
  \citenamefont {Baczewski}, \citenamefont {Firgau}, \citenamefont
  {M{\k{a}}dzik}, \citenamefont {Schmitt}, \citenamefont {Pla}, \citenamefont
  {Hudson} \emph {et~al.}}]{asaad2020coherent}%
  \BibitemOpen
  \bibfield  {author} {\bibinfo {author} {\bibfnamefont {S.}~\bibnamefont
  {Asaad}}, \bibinfo {author} {\bibfnamefont {V.}~\bibnamefont {Mourik}},
  \bibinfo {author} {\bibfnamefont {B.}~\bibnamefont {Joecker}}, \bibinfo
  {author} {\bibfnamefont {M.~A.}\ \bibnamefont {Johnson}}, \bibinfo {author}
  {\bibfnamefont {A.~D.}\ \bibnamefont {Baczewski}}, \bibinfo {author}
  {\bibfnamefont {H.~R.}\ \bibnamefont {Firgau}}, \bibinfo {author}
  {\bibfnamefont {M.~T.}\ \bibnamefont {M{\k{a}}dzik}}, \bibinfo {author}
  {\bibfnamefont {V.}~\bibnamefont {Schmitt}}, \bibinfo {author} {\bibfnamefont
  {J.~J.}\ \bibnamefont {Pla}}, \bibinfo {author} {\bibfnamefont {F.~E.}\
  \bibnamefont {Hudson}}, \emph {et~al.},\ }\bibfield  {title} {\bibinfo
  {title} {Coherent electrical control of a single high-spin nucleus in
  silicon},\ }\href@noop {} {\bibfield  {journal} {\bibinfo  {journal}
  {Nature}\ }\textbf {\bibinfo {volume} {579}},\ \bibinfo {pages} {205}
  (\bibinfo {year} {2020})}\BibitemShut {NoStop}%
\bibitem [{\citenamefont {Dehollain}\ \emph {et~al.}(2012)\citenamefont
  {Dehollain}, \citenamefont {Pla}, \citenamefont {Siew}, \citenamefont {Tan},
  \citenamefont {Dzurak},\ and\ \citenamefont
  {Morello}}]{dehollain2012nanoscale}%
  \BibitemOpen
  \bibfield  {author} {\bibinfo {author} {\bibfnamefont {J.}~\bibnamefont
  {Dehollain}}, \bibinfo {author} {\bibfnamefont {J.}~\bibnamefont {Pla}},
  \bibinfo {author} {\bibfnamefont {E.}~\bibnamefont {Siew}}, \bibinfo {author}
  {\bibfnamefont {K.}~\bibnamefont {Tan}}, \bibinfo {author} {\bibfnamefont
  {A.}~\bibnamefont {Dzurak}},\ and\ \bibinfo {author} {\bibfnamefont
  {A.}~\bibnamefont {Morello}},\ }\bibfield  {title} {\bibinfo {title}
  {Nanoscale broadband transmission lines for spin qubit control},\ }\href@noop
  {} {\bibfield  {journal} {\bibinfo  {journal} {Nanotechnology}\ }\textbf
  {\bibinfo {volume} {24}},\ \bibinfo {pages} {015202} (\bibinfo {year}
  {2012})}\BibitemShut {NoStop}%
\bibitem [{\citenamefont {Adambukulam}\ \emph {et~al.}(2021)\citenamefont
  {Adambukulam}, \citenamefont {Sewani}, \citenamefont {Stemp}, \citenamefont
  {Asaad}, \citenamefont {M{\k{a}}dzik}, \citenamefont {Morello},\ and\
  \citenamefont {Laucht}}]{adambukulam2021ultra}%
  \BibitemOpen
  \bibfield  {author} {\bibinfo {author} {\bibfnamefont {C.}~\bibnamefont
  {Adambukulam}}, \bibinfo {author} {\bibfnamefont {V.}~\bibnamefont {Sewani}},
  \bibinfo {author} {\bibfnamefont {H.}~\bibnamefont {Stemp}}, \bibinfo
  {author} {\bibfnamefont {S.}~\bibnamefont {Asaad}}, \bibinfo {author}
  {\bibfnamefont {M.}~\bibnamefont {M{\k{a}}dzik}}, \bibinfo {author}
  {\bibfnamefont {A.}~\bibnamefont {Morello}},\ and\ \bibinfo {author}
  {\bibfnamefont {A.}~\bibnamefont {Laucht}},\ }\bibfield  {title} {\bibinfo
  {title} {An ultra-stable 1.5 {T} permanent magnet assembly for qubit
  experiments at cryogenic temperatures},\ }\href@noop {} {\bibfield  {journal}
  {\bibinfo  {journal} {Review of {S}cientific {I}nstruments}\ }\textbf
  {\bibinfo {volume} {92}},\ \bibinfo {pages} {085106} (\bibinfo {year}
  {2021})}\BibitemShut {NoStop}%
\bibitem [{\citenamefont {Kalra}\ \emph {et~al.}(2016)\citenamefont {Kalra},
  \citenamefont {Laucht}, \citenamefont {Dehollain}, \citenamefont {Bar},
  \citenamefont {Freer}, \citenamefont {Simmons}, \citenamefont {Muhonen},\
  and\ \citenamefont {Morello}}]{kalra2016vibration}%
  \BibitemOpen
  \bibfield  {author} {\bibinfo {author} {\bibfnamefont {R.}~\bibnamefont
  {Kalra}}, \bibinfo {author} {\bibfnamefont {A.}~\bibnamefont {Laucht}},
  \bibinfo {author} {\bibfnamefont {J.~P.}\ \bibnamefont {Dehollain}}, \bibinfo
  {author} {\bibfnamefont {D.}~\bibnamefont {Bar}}, \bibinfo {author}
  {\bibfnamefont {S.}~\bibnamefont {Freer}}, \bibinfo {author} {\bibfnamefont
  {S.}~\bibnamefont {Simmons}}, \bibinfo {author} {\bibfnamefont {J.~T.}\
  \bibnamefont {Muhonen}},\ and\ \bibinfo {author} {\bibfnamefont
  {A.}~\bibnamefont {Morello}},\ }\bibfield  {title} {\bibinfo {title}
  {Vibration-induced electrical noise in a cryogen-free dilution refrigerator:
  Characterization, mitigation, and impact on qubit coherence},\ }\href@noop {}
  {\bibfield  {journal} {\bibinfo  {journal} {Review of {S}cientific
  {I}nstruments}\ }\textbf {\bibinfo {volume} {87}},\ \bibinfo {pages} {073905}
  (\bibinfo {year} {2016})}\BibitemShut {NoStop}%
\bibitem [{\citenamefont {Asaad}\ and\ \citenamefont {Johnson}(2017)}]{silq}%
  \BibitemOpen
  \bibfield  {author} {\bibinfo {author} {\bibfnamefont {S.}~\bibnamefont
  {Asaad}}\ and\ \bibinfo {author} {\bibfnamefont {M.}~\bibnamefont
  {Johnson}},\ }\href {https://nulinspiratie.github.io/SilQ/index.html}
  {\bibinfo {title} {Silq measurement software}} (\bibinfo {year}
  {2017})\BibitemShut {NoStop}%
\bibitem [{\citenamefont {Nielsen}\ \emph {et~al.}(2019)\citenamefont
  {Nielsen}, \citenamefont {Nielsen}, \citenamefont {Astafev}, \citenamefont
  {alexcjohnson}, \citenamefont {Vogel}, \citenamefont {sohail chatoor},
  \citenamefont {Ungaretti}, \citenamefont {MerlinSmiles}, \citenamefont
  {Adriaan}, \citenamefont {Pauka}, \citenamefont {Eendebak}, \citenamefont
  {qSaevar}, \citenamefont {Eendebak}, \citenamefont {van Gulik}, \citenamefont
  {Pearson}, \citenamefont {damazter}, \citenamefont {Corna}, \citenamefont
  {Droege}, \citenamefont {damazter2}, \citenamefont {ThorvaldLarsen},
  \citenamefont {Geller}, \citenamefont {euchas}, \citenamefont {Hartong},
  \citenamefont {Asaad}, \citenamefont {Granade}, \citenamefont {Drmić},
  \citenamefont {Borghardt},\ and\ \citenamefont {mltls}}]{qcodes}%
  \BibitemOpen
  \bibfield  {author} {\bibinfo {author} {\bibfnamefont {J.~H.}\ \bibnamefont
  {Nielsen}}, \bibinfo {author} {\bibfnamefont {W.~H.}\ \bibnamefont
  {Nielsen}}, \bibinfo {author} {\bibfnamefont {M.}~\bibnamefont {Astafev}},
  \bibinfo {author} {\bibnamefont {alexcjohnson}}, \bibinfo {author}
  {\bibfnamefont {D.}~\bibnamefont {Vogel}}, \bibinfo {author} {\bibnamefont
  {sohail chatoor}}, \bibinfo {author} {\bibfnamefont {G.}~\bibnamefont
  {Ungaretti}}, \bibinfo {author} {\bibnamefont {MerlinSmiles}}, \bibinfo
  {author} {\bibnamefont {Adriaan}}, \bibinfo {author} {\bibfnamefont
  {S.}~\bibnamefont {Pauka}}, \bibinfo {author} {\bibfnamefont
  {P.}~\bibnamefont {Eendebak}}, \bibinfo {author} {\bibnamefont {qSaevar}},
  \bibinfo {author} {\bibfnamefont {P.}~\bibnamefont {Eendebak}}, \bibinfo
  {author} {\bibfnamefont {R.}~\bibnamefont {van Gulik}}, \bibinfo {author}
  {\bibfnamefont {N.}~\bibnamefont {Pearson}}, \bibinfo {author} {\bibnamefont
  {damazter}}, \bibinfo {author} {\bibfnamefont {A.}~\bibnamefont {Corna}},
  \bibinfo {author} {\bibfnamefont {S.}~\bibnamefont {Droege}}, \bibinfo
  {author} {\bibnamefont {damazter2}}, \bibinfo {author} {\bibnamefont
  {ThorvaldLarsen}}, \bibinfo {author} {\bibfnamefont {A.}~\bibnamefont
  {Geller}}, \bibinfo {author} {\bibnamefont {euchas}}, \bibinfo {author}
  {\bibfnamefont {V.}~\bibnamefont {Hartong}}, \bibinfo {author} {\bibfnamefont
  {S.}~\bibnamefont {Asaad}}, \bibinfo {author} {\bibfnamefont
  {C.}~\bibnamefont {Granade}}, \bibinfo {author} {\bibfnamefont
  {L.}~\bibnamefont {Drmić}}, \bibinfo {author} {\bibfnamefont
  {S.}~\bibnamefont {Borghardt}},\ and\ \bibinfo {author} {\bibnamefont
  {mltls}},\ }\href {https://doi.org/10.5281/zenodo.2649884} {\bibinfo {title}
  {Qcodes/qcodes: Qcodes 0.2.1}} (\bibinfo {year} {2019})\BibitemShut {NoStop}%
\bibitem [{\citenamefont {Kushida}\ and\ \citenamefont
  {Saiki}(1961)}]{kushida1961shift}%
  \BibitemOpen
  \bibfield  {author} {\bibinfo {author} {\bibfnamefont {T.}~\bibnamefont
  {Kushida}}\ and\ \bibinfo {author} {\bibfnamefont {K.}~\bibnamefont
  {Saiki}},\ }\bibfield  {title} {\bibinfo {title} {Shift of nuclear quadrupole
  resonance frequency by electric field},\ }\href@noop {} {\bibfield  {journal}
  {\bibinfo  {journal} {Physical {R}eview {L}etters}\ }\textbf {\bibinfo
  {volume} {7}},\ \bibinfo {pages} {9} (\bibinfo {year} {1961})}\BibitemShut
  {NoStop}%
\bibitem [{\citenamefont {Armstrong}\ \emph {et~al.}(1961)\citenamefont
  {Armstrong}, \citenamefont {Bloembergen},\ and\ \citenamefont
  {Gill}}]{armstrong1961linear}%
  \BibitemOpen
  \bibfield  {author} {\bibinfo {author} {\bibfnamefont {J.}~\bibnamefont
  {Armstrong}}, \bibinfo {author} {\bibfnamefont {N.}~\bibnamefont
  {Bloembergen}},\ and\ \bibinfo {author} {\bibfnamefont {D.}~\bibnamefont
  {Gill}},\ }\bibfield  {title} {\bibinfo {title} {Linear effect of applied
  electric field on nuclear quadrupole resonance},\ }\href@noop {} {\bibfield
  {journal} {\bibinfo  {journal} {Physical {R}eview {L}etters}\ }\textbf
  {\bibinfo {volume} {7}},\ \bibinfo {pages} {11} (\bibinfo {year}
  {1961})}\BibitemShut {NoStop}%
\bibitem [{\citenamefont {Dixon}\ and\ \citenamefont
  {Bloembergen}(1964)}]{dixon1964linear}%
  \BibitemOpen
  \bibfield  {author} {\bibinfo {author} {\bibfnamefont {R.}~\bibnamefont
  {Dixon}}\ and\ \bibinfo {author} {\bibfnamefont {N.}~\bibnamefont
  {Bloembergen}},\ }\bibfield  {title} {\bibinfo {title} {Linear {E}lectric
  {S}hifts in the {N}uclear {Q}uadrupole interaction in al$_2$o$_3$},\
  }\href@noop {} {\bibfield  {journal} {\bibinfo  {journal} {Physical
  {R}eview}\ }\textbf {\bibinfo {volume} {135}},\ \bibinfo {pages} {A1669}
  (\bibinfo {year} {1964})}\BibitemShut {NoStop}%
\bibitem [{\citenamefont {Rahman}\ \emph {et~al.}(2007)\citenamefont {Rahman},
  \citenamefont {Wellard}, \citenamefont {Bradbury}, \citenamefont {Prada},
  \citenamefont {Cole}, \citenamefont {Klimeck},\ and\ \citenamefont
  {Hollenberg}}]{rahman2007high}%
  \BibitemOpen
  \bibfield  {author} {\bibinfo {author} {\bibfnamefont {R.}~\bibnamefont
  {Rahman}}, \bibinfo {author} {\bibfnamefont {C.~J.}\ \bibnamefont {Wellard}},
  \bibinfo {author} {\bibfnamefont {F.~R.}\ \bibnamefont {Bradbury}}, \bibinfo
  {author} {\bibfnamefont {M.}~\bibnamefont {Prada}}, \bibinfo {author}
  {\bibfnamefont {J.~H.}\ \bibnamefont {Cole}}, \bibinfo {author}
  {\bibfnamefont {G.}~\bibnamefont {Klimeck}},\ and\ \bibinfo {author}
  {\bibfnamefont {L.~C.}\ \bibnamefont {Hollenberg}},\ }\bibfield  {title}
  {\bibinfo {title} {High precision quantum control of single donor spins in
  silicon},\ }\href@noop {} {\bibfield  {journal} {\bibinfo  {journal}
  {Physical {R}eview {L}etters}\ }\textbf {\bibinfo {volume} {99}},\ \bibinfo
  {pages} {036403} (\bibinfo {year} {2007})}\BibitemShut {NoStop}%
\bibitem [{\citenamefont {Bradbury}\ \emph {et~al.}(2006)\citenamefont
  {Bradbury}, \citenamefont {Tyryshkin}, \citenamefont {Sabouret},
  \citenamefont {Bokor}, \citenamefont {Schenkel},\ and\ \citenamefont
  {Lyon}}]{bradbury2006stark}%
  \BibitemOpen
  \bibfield  {author} {\bibinfo {author} {\bibfnamefont {F.~R.}\ \bibnamefont
  {Bradbury}}, \bibinfo {author} {\bibfnamefont {A.~M.}\ \bibnamefont
  {Tyryshkin}}, \bibinfo {author} {\bibfnamefont {G.}~\bibnamefont {Sabouret}},
  \bibinfo {author} {\bibfnamefont {J.}~\bibnamefont {Bokor}}, \bibinfo
  {author} {\bibfnamefont {T.}~\bibnamefont {Schenkel}},\ and\ \bibinfo
  {author} {\bibfnamefont {S.~A.}\ \bibnamefont {Lyon}},\ }\bibfield  {title}
  {\bibinfo {title} {Stark tuning of donor electron spins in silicon},\
  }\href@noop {} {\bibfield  {journal} {\bibinfo  {journal} {Physical {R}eview
  {L}etters}\ }\textbf {\bibinfo {volume} {97}},\ \bibinfo {pages} {176404}
  (\bibinfo {year} {2006})}\BibitemShut {NoStop}%
\bibitem [{\citenamefont {Nielsen}\ \emph {et~al.}(2020)\citenamefont
  {Nielsen}, \citenamefont {Rudinger}, \citenamefont {Proctor}, \citenamefont
  {Russo}, \citenamefont {Young},\ and\ \citenamefont
  {Blume-Kohout}}]{nielsen2020probing}%
  \BibitemOpen
  \bibfield  {author} {\bibinfo {author} {\bibfnamefont {E.}~\bibnamefont
  {Nielsen}}, \bibinfo {author} {\bibfnamefont {K.}~\bibnamefont {Rudinger}},
  \bibinfo {author} {\bibfnamefont {T.}~\bibnamefont {Proctor}}, \bibinfo
  {author} {\bibfnamefont {A.}~\bibnamefont {Russo}}, \bibinfo {author}
  {\bibfnamefont {K.}~\bibnamefont {Young}},\ and\ \bibinfo {author}
  {\bibfnamefont {R.}~\bibnamefont {Blume-Kohout}},\ }\bibfield  {title}
  {\bibinfo {title} {Probing quantum processor performance with py{GST}i},\
  }\href@noop {} {\bibfield  {journal} {\bibinfo  {journal} {Quantum {S}cience
  and {T}echnology}\ }\textbf {\bibinfo {volume} {5}},\ \bibinfo {pages}
  {044002} (\bibinfo {year} {2020})}\BibitemShut {NoStop}%
\bibitem [{\citenamefont {Tenberg}\ \emph {et~al.}(2019)\citenamefont
  {Tenberg}, \citenamefont {Asaad}, \citenamefont {M{\k{a}}dzik}, \citenamefont
  {Johnson}, \citenamefont {Joecker}, \citenamefont {Laucht}, \citenamefont
  {Hudson}, \citenamefont {Itoh}, \citenamefont {Jakob}, \citenamefont
  {Johnson} \emph {et~al.}}]{tenberg2019electron}%
  \BibitemOpen
  \bibfield  {author} {\bibinfo {author} {\bibfnamefont {S.~B.}\ \bibnamefont
  {Tenberg}}, \bibinfo {author} {\bibfnamefont {S.}~\bibnamefont {Asaad}},
  \bibinfo {author} {\bibfnamefont {M.~T.}\ \bibnamefont {M{\k{a}}dzik}},
  \bibinfo {author} {\bibfnamefont {M.~A.}\ \bibnamefont {Johnson}}, \bibinfo
  {author} {\bibfnamefont {B.}~\bibnamefont {Joecker}}, \bibinfo {author}
  {\bibfnamefont {A.}~\bibnamefont {Laucht}}, \bibinfo {author} {\bibfnamefont
  {F.~E.}\ \bibnamefont {Hudson}}, \bibinfo {author} {\bibfnamefont {K.~M.}\
  \bibnamefont {Itoh}}, \bibinfo {author} {\bibfnamefont {A.~M.}\ \bibnamefont
  {Jakob}}, \bibinfo {author} {\bibfnamefont {B.~C.}\ \bibnamefont {Johnson}},
  \emph {et~al.},\ }\bibfield  {title} {\bibinfo {title} {Electron spin
  relaxation of single phosphorus donors in metal-oxide-semiconductor nanoscale
  devices},\ }\href@noop {} {\bibfield  {journal} {\bibinfo  {journal}
  {Physical Review B}\ }\textbf {\bibinfo {volume} {99}},\ \bibinfo {pages}
  {205306} (\bibinfo {year} {2019})}\BibitemShut {NoStop}%
\end{thebibliography}%

\end{document}


\title{Supplementary Materials: Navigating the 16-dimensional Hilbert space of a high-spin donor qudit with electric and magnetic fields}

\author{Irene Fern\'andez de Fuentes}
\author{Tim Botzem }
\author{Mark A. I. Johnson}
\author{Arjen Vaartjes }
\author{S. Asaad }
\author{V. Mourik }
\author{F. E. Hudson}
\affiliation{ARC Centre for Quantum Computation and Communication Technology (CQC2T), School of Electrical Engineering and Telecommunication, University of New South Wales, Sydney, NSW, Australia}
\author{K. M. Itoh}
\affiliation{School of Fundamental Science and Technology, Keio University, Minato City, Yokohama, Japan}
\author{B. C. Johnson}
\author{A. M. Jakob}
\author{J. C. McCallum}
\author{D. N. Jamieson}
\affiliation{School of Physics, University of Melbourne, Melbourne, Victoria, Australia}
\author{A. S. Dzurak}
\author{A. Morello}
\affiliation{ARC Centre for Quantum Computation and Communication Technology (CQC2T), School of Electrical Engineering and Telecommunication, University of New South Wales, Sydney, NSW, Australia}


\maketitle
\tableofcontents
\pagebreak
\section*{S1: Random nuclear spin flips / ionisation shock}
\label{Supp:Nuclear_flips}
The nuclear spin lattice relaxation time $T_{1\mathrm{n}}$ is known to be exceptionally long\,\cite{savytskyy2023electrically}, which makes nuclear relaxation processes entirely irrelevant within our measurements timescales (and beyond). However, during our experiments, we have observed that the \Sb nucleus undergoes frequent random spin flips. These nuclear flips must thus originate from the readout process which, relies on the electron acting as an ancilla qubit.

 In a hyperfine coupled donor system, the spin state of the nucleus (observable) can be measured in the $\hat{I}_z$ basis by first mapping its spin configuration onto the electron spin (ancilla) state by inverting the electron state conditional on one of the nuclear spin projections, i.e. driving the electron at one of the resonant peaks in Fig.\,1f. The electron spin is then read out using spin-dependent tunneling into the SET island \,\cite{morello2010single}. An $\ket{\uparrow}$ electron tunnels out to the SET island and is replaced by a $\ket{\downarrow}$. During the time the donor is ionized as a result of this process, the SET current exhibits a spike that can be detected with high fidelity.
		\begin{figure}[H]
		\centering
		\includegraphics[width =\textwidth]{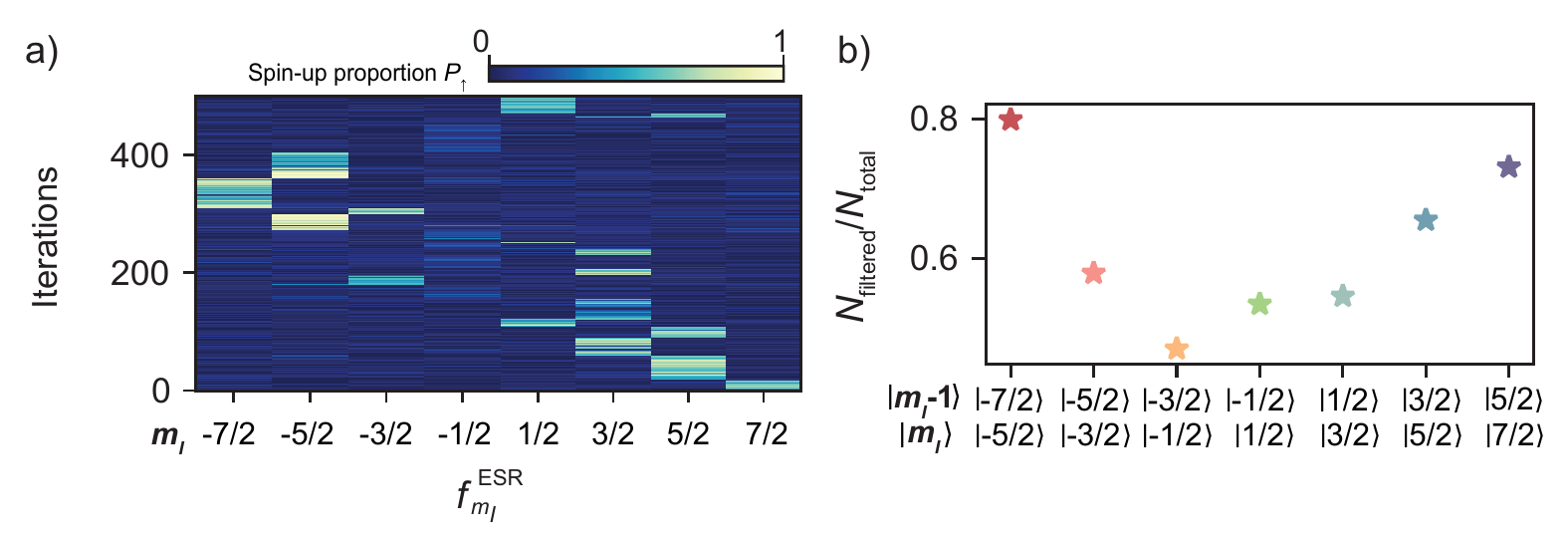}
		\caption[Statistics on the nuclear random flips for the \Sb donor.]{ \textbf{a)} Projective measurements of the nuclear state as a function of time (iterations), measured through electron single-shot readout after applying a microwave stimulus at a frequency $f_{m_I}^{\rm ESR}$. 
  \textbf{b)} Ratio between filtered $N_{\mathrm{filtered}}$ and total $N_{\mathrm{total}}$ points (see main text for definition), collected from various nuclear \Sb measurements in this work.}
		\label{COH-fig:Ionization_shock_Sb}
	\end{figure}
	If this is a quantum non demolition (QND)\,\cite{braginsky1996quantum} process, the nucleus should remain in the projected state after the ionization event, allowing for repeated ($n$) measurements to improve the readout accuracy. The condition for QND readout is that the Hamiltonian of the observable $\gamma_{\rm n}B_0\hat{I}_z$ commutes with the interaction Hamiltonian with the ancilla $\hat{H}_{\mathrm{in}} = A\boldsymbol{\hat{S}}\cdot\boldsymbol{\hat{I}}$, i.e. $[\hat{H}_{\mathrm{int}},\hat{I}_z] = 0$. This is true if the interaction Hamiltonian is only given by the secular component of the hyperfine interaction $\hat{H}_{\mathrm{in}} = A\hat{S}_z\hat{I}_z$\,.
	However, an accurate description of single-shot readout in a hyperfine-coupled system, requires accounting for the terms in the Hamiltonian that do not commute with $\hat{I}_z$. These arise from both the isotropic\,\cite{pla2013high} and anisotropic\,\cite{hile2018addressable} parts of the hyperfine interaction that precede the operators $\hat{S}_\alpha,\hat{I}_\beta$ with $\alpha\neq z$ and $\alpha,\beta \neq z,z$ . Moreover, in the case of \Sb, the removal of the electron could cause a change in the electric field gradient $\mathcal{V}_{\alpha\beta}$\,\cite{franke2016quadrupolar}, thus modifying the eigenstates between the ionized/neutral atom. These changes in eigenbasis result in a finite probability of flipping the nucleus during the readout process. 
	The phenomenon above is termed  \textit{ionization shock}, and in the extreme case where the hyperfine/quadrupolar interaction are highly anisotropic/highly dependent on the charge state, the nuclear states are randomized after each ionization event\,\cite{hile2018addressable}. 

Previous work on ion-implanted $^{31}$P donors showed that the rate of ionisation shock, i.e. the probability of the nuclear spin flipping as a result of a change in charge state of the donor, can be as low as $10^{-7}$~\cite{mkadzik2022precision,pla2013high}. However, much faster nuclear flipping rates have been observed in STM-fabricated donor clusters\,\cite{hile2018addressable} (with only hyperfine interaction) and for a \Sb donor\,\cite{asaad2020coherent} (with both hyperfine and quadrupolar interaction).

  Figure\,\ref{COH-fig:Ionization_shock_Sb} summarizes some of the observations made on \Sb in the present device.
	A first experiment tracks the nuclear spin state after a random initialization of the nucleus. After loading an electron $\ket{\downarrow}$ onto the donor, we apply an adiabatic ESR inversion pulse followed by an electron readout. We repeat this sequence, toggling the frequency of the microwave source between the eight possible ESR resonance frequencies $f_{m_I}^{\mathrm{ESR}}$. A high electron spin-up proportion $P_{\uparrow}$ after the adiabatic inversion flags the nuclear spin state where the donor is found. The results of these experiments are presented in Fig.\,\ref{COH-fig:Ionization_shock_Sb}\,a, where each iteration represents the population of the eight possible nuclear spin configurations. We observe  regular flips between spin configurations, visible from the meandering high spin-up proportion as a function of repetitions. A crude estimate of the nuclear spin flip probability \emph{per shot} can be obtained from Fig.\,\ref{COH-fig:Ionization_shock_Sb}\,a as follows. Each of the 8 data points (one for each nuclear spin orientation) at each iteration is obtained by repeating the nuclear readout process 30 times (shots). However, this does not cause $30 \times 8$ ionisation events, because only $\ket{\uparrow}$ electrons leave the donor, plus some `dark counts' ($\ket{\downarrow}$  electrons escaping the donors accidentally), so we estimate $\approx 50$ ionisation events for every iteration. The data shows that the nuclear spin typically flips after $\approx 20-50$ iterations. Therefore, the ionisation rate per shot is of order $10^{-3}$.

	An additional benchmarking parameter is given in Fig.\,\ref{COH-fig:Ionization_shock_Sb}\,b. Here, we post-analyze the experiments that include nuclear readout (nuclear Rabi oscillations, spectra, Ramsey sequences, etc.) to extract statistics on the nuclear spin flips. Using the outcome of a measurement that involves reading out the nuclear spin\,\cite{pla2013high}, we can calculate the fraction of successfully finding the nuclear state in the subspace under investigation (which defines  $N_{\mathrm{filtered}}$), after one measurement sequence which typically includes 200-300 electron shots (which defines $N_\mathrm{total}$) before the state is reinitialized. Note that this number also captures errors associated with the initialization sequence. We can thus define a probability of `success' as $N_{\mathrm{filtered}}/N_\mathrm{total}$, i.e. the probability that, after performing an experiment intended to measure a property of the system while the nuclear spin is in the $m_I$ projection, we find the nuclear spin still in $m_I$ at the end of the experiment. Unsuccessful experiments are subsequently filtered out of the averages.  \ref{COH-fig:Ionization_shock_Sb}\,b Shows that the probability of success is between 0.4 and 0.8, depending on the nuclear state.

\section*{S2: Device fabrication and operation}
This qubit device was fabricated on a natural silicon wafer with a 900\,nm thick epitaxial layer of isotopically enriched $^{28}$Si with 800\,ppm residual concentration of $^{29}$Si. We use electron beam lithography to pattern metallic aluminum structures on top of a thin ($\approx 8$\,nm) SiO$_2$ oxide layer, to control and readout the donor spins.

The device integrates a broadband microwave antenna \cite{dehollain2012nanoscale}, which is an on-chip 50\,$\Omega$ matched coplanar waveguide, terminated by a short circuit. The presence of stray electric fields in the GHz range allows us to use the waveguide to deliver oscillating electric fields.  

The device is wire-bonded to a high-frequency printed circuit board mounted within a copper enclosure, which is then bolted to a box where the qubit device sits in the air gap of a Halbach array of permanent magnets\,\cite{adambukulam2021ultra}. The assembly is then anchored to the mixing chamber plate of a Bluefors BF-LD400 dilution refrigerator, where it gets cooled down to $\lessapprox$20~mK. The static magnetic field $B_{0}$ (\SI{\approx{1}}{\tesla}) produced by the permanent magnets is applied along the short-circuit termination of the magnetic (ESR, NMR) antenna and parallel to the [110] plane of the Si substrate).

\begin{figure}
	\centering
	\includegraphics{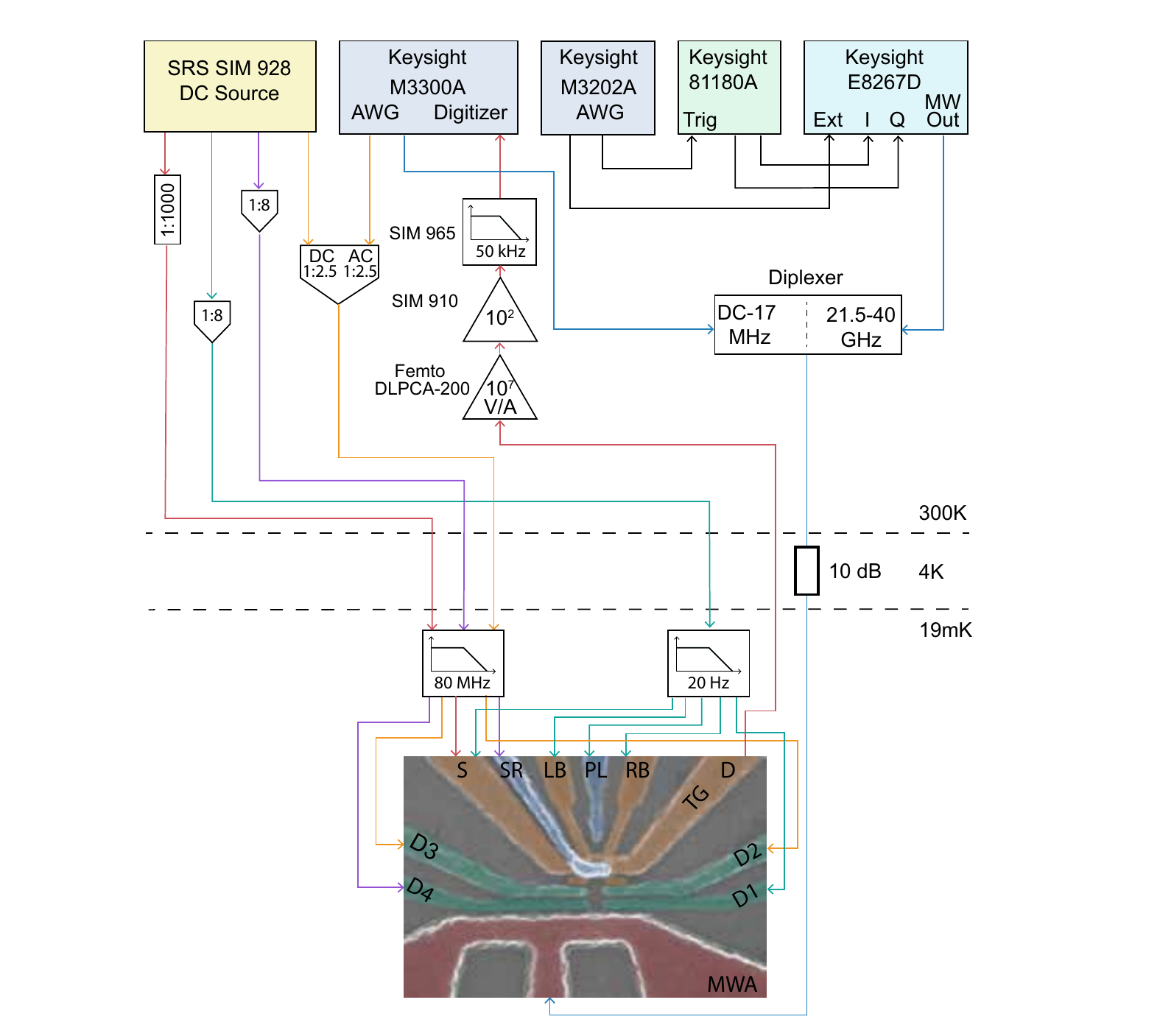}
	\caption[Experimental setup.]{ Schematics the cryogenic setup used to operate the silicon qubit device in this work. The green lines correspond to DC lines (\SI{20}{\hertz} cutoff frequency), purple and orange to AC lines (\SI{80}{\mega\hertz} cutoff frequency), and the blue line (no filtering) corresponds to the high-frequency line connected to the antenna. All the elements of the diagram, as well as the thermal connections of the cabling, are explained in the main text.}
	\label{ET-fig:setup}
\end{figure}

We use flexible copper cables to connect the enclosure to a filter box, attached to the mixing chamber plate which contains two types of low pass filters: second-order low-pass RC filters with a 20~Hz cut-off frequency used for the gates that provide a constant voltage bias to the device (TG, RB, LB, PL), and seventh-order low-pass filters with 80 MHz cut-off frequency, connected to the gates that we use for pulsing (typically donor gates DG and SR) or to measure conductance in the SET (source S and drain D). The gate layout can be found in Fig.\,\ref{ET-fig:setup}. 
Above the filter boxes, the DC lines consist of Constantan looms, which we thermalize by wrapping them around copper rods at various temperature stages. For the fast lines, we use coaxial cables with a graphite coating on the dielectric to minimize the triboelectric effect caused by mechanical vibrations from the pulse tube\,\cite{kalra2016vibration}. 

For the high-frequency lines, like the on-chip coplanar waveguides, we use silver-plated copper-nickel coaxial cables, with 2.92 mm coaxial connectors, and add a 10~dB attenuator at 4K to thermalize the line.
The qubit gates are DC-biased using battery-powered and opto-isolated SRS SIM 928 voltage sources. To increase the voltage resolution we use homemade resistive voltage dividers, with a division of 1:8. The AC signals are generated by arbitrary waveform generators (Keysight M3300A and M3202A), and combined with the DC signals using impedance-matched combiners with a voltage division 1:2.5. 

The microwave signals needed for ESR and EDSR are generated by a Keysight E8267D microwave vector source (100~kHz-44~GHz), which we IQ-modulate using the channels from a Keysight 81180A AWG. The NER and NMR control signals are synthesized directly by Keysight M3300A and M3202A AWG cards, which provide bandwidth of up to 500\,MS/s and 1\,GS/s, respectively. To combine the signals at radio and microwave frequencies to be delivered to the microwave antenna, we use a commercially available diplexer Marki Microwave DPX-1721 at room temperature.

The current from the SET is converted to a voltage using a Femto DLPCA-200 transimpedance amplifier. Typical currents are on the order of 1~nA, thus we use an amplification of 10$^7$\,V/A, to which corresponds an amplifier bandwidth of 50-~kHz. The signal is further amplified by an SRS SIM910 JFET amplifier, where we use 100\,V/V gain, and filtered by an SRS SIM965 analog \SI{50}{\kilo\hertz} low-pass Bessel filter. The converted signal is recorded using the digitizer in the the Keysight M3300A.

The last step is to interface our instruments at the software level, which is done using a Python-based in-house software called SilQ\,\cite{silq}, which uses the Python-based QCoDeS data acquisition framework\,\cite{qcodes}.

\section*{S3: Nuclear state preparation for \texorpdfstring{\Sb}{Sb}}
\label{Supp:Nuclear_pre}
\begin{figure}
 	\centering
 	\includegraphics[width = \textwidth]{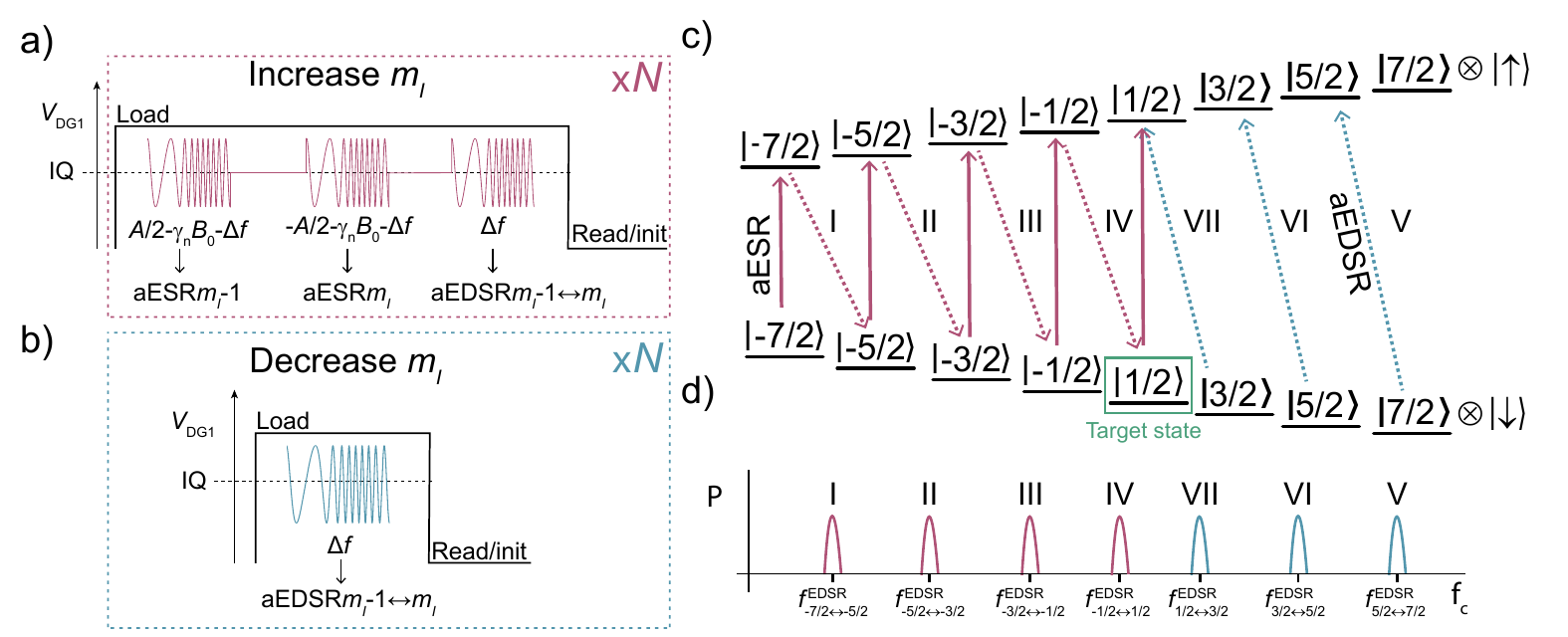}
 	\caption[Nuclear flip-flop initialization sequence for antimony.]{ \textbf{a)} Pulse sequence sent to the IQ inputs of the microwave source, used to increase the nuclear spin number $\ket{m_I-1}\rightarrow\ket{ m_{I}}$, consisting of two adiabatic ESR pulses and one adiabatic EDSR pulse. \textbf{b)} Pulse sequence sent to the IQ inputs of the microwave source,  used to decrease the nuclear spin number  $\ket{m_{I}}\rightarrow \ket{m_I-1}$, consisting of one adiabatic EDSR pulse. The frequencies for the pulses in \textbf{a)} and \textbf{b)} can be generalized by writing them in terms of the value of the hyperfine $A$, the Zeeman energy $\gamman B_{0}$, and the frequency modulation depth of the adiabatic pulse $\Delta f$ (see text for details). \textbf{c)} Schematics of the transitions involved in the initialization sequence, with the order of the operations indicated with roman numbers. \textbf{d)} Depiction of the frequency for the microwave local oscillator, which is set at the EDSR resonance frequency $f_{m_I-1\leftrightarrow m_I}^{\mathrm{EDSR}}$ of each subspace. }
 	\label{ET-fig:state_agnostic_prep}
 \end{figure}
 By leveraging the flip-flop drive, we can initialize the atom in an arbitrary nuclear subspace without relying on NMR pulses. This is achieved by concatenating sequences that include adiabatic flip-flop and ESR drive, as shown in Fig.\,\ref{ET-fig:state_agnostic_prep}. This technique enables high-fidelity nuclear state preparation with a compact instrument footprint, and was previously used by Asaad et al.\cite{asaad2020coherent}.
 
Using flip-flop initialization requires uploading only two different waveforms (Fig.\,\ref{ET-fig:state_agnostic_prep}\,a and Fig.\,\ref{ET-fig:state_agnostic_prep}\,b) to the AWG, depending on whether we want to increase or decrease the nuclear spin projection. \\
To increase the nuclear spin number ($\ket{m_I-1}\rightarrow \ket{m_{I}}$), we use the sequence presented in Fig.\ref{ET-fig:state_agnostic_prep}\,a. We first load an electron $\ket{\downarrow}$ from the SET reservoir, and apply two adiabatic ESR pulses which flip the state of the electron unconditionally of the state of the nucleus. A subsequent adiabatic EDSR pulse populates the $\ket{m_{I}}$ subspace only when the electron is found in the $\ket{\uparrow}$ state, therefore decreasing the nuclear spin projection. Conversely, to move from $\ket{m_{I}}\rightarrow \ket{m_I-1}$, we use the sequence shown in Fig.\ref{ET-fig:state_agnostic_prep}\,b consisting of a single adiabatic EDSR pulse, after an electron in the $\ket{\downarrow}$ state has been loaded onto the donor.\\
The high-frequency pulses  for this initialization technique are generated by performing IQ modulation on the microwave source (MWS). We set the carrier frequency of the MWS to be at the  flip-flop frequency $f_{\mathrm{c}} = f^{\mathrm{aEDSR}}_{m_I-1\leftrightarrow m}$, and write the frequency of the IQ pulses in terms of the hyperfine interaction $A$, the Zeeman energy $\gamman B_{0}$ and the frequency deviation  of the chirped pulses $\Delta f$. For equal amplitudes of the I and Q signals, this modulation should ideally be a single-sideband, i.e. a single tone at the frequency $f_{\mathrm{c}}+f_{\mathrm{IQ}}$ where $f_{\mathrm{c}}$ is the frequency of the microwave source carrier and $f_{\mathrm{IQ}}$  is the frequency of the IQ tones. We have 
\begin{align}
f^{\mathrm{aESR,IQ}}_{m_I-1} = - A/2-\gamman B_{0}-\Delta f,
\end{align}
\begin{align}
	f^{\mathrm{aESR,IQ}}_{m_{I}} = A/2-\gamman B_{0}-\Delta f,
\end{align}
and 
\begin{align}
	f^{\mathrm{aEDSR,IQ}}_{m_{I}} =\Delta f,
\end{align} 
In this way, we can upload fixed waveforms to the AWG the outputs into the IQ inputs of the microwave source (either the waveform in Fig.\,\ref{ET-fig:state_agnostic_prep}\,a or the waveform in Fig.\,\ref{ET-fig:state_agnostic_prep}\,b and independently switch the local oscillator frequency of the microwave source ($\sim\SI{10}{\milli\second}$), to target specific nuclear subspaces. We can repeat the previous sequence multiple times, and enhance the state preparation without running into instrument memory limitations or slow upload of long waveforms onto the AWG.\\
This protocol is exemplified in Figures\,\ref{ET-fig:state_agnostic_prep}\,c-d.
 For a target state $m_I = \ket{1/2}$, we load the pulse sequence in Fig.\,\ref{ET-fig:state_agnostic_prep}\,a to the AWG, and continuously play the waveform while switching the carrier $f_{\mathrm{c}}$ of the microwave source with a software instruction, between the frequencies I,II, III, and IV every $ t = t_{\mathrm{ENDOR}}$, where $t_{\mathrm{ENDOR}}$ is the duration of the Electron Nuclear Double Resonance (ENDOR) pulse sequence from Fig.\,\ref{ET-fig:state_agnostic_prep}\,a. This should bring any state from the left of $\ket{1/2}$ to the target state. After $N\approx20$ repetitions of this whole sequence, we upload the pulse sequence in Fig.\,\ref{ET-fig:state_agnostic_prep}\,b and repeat the procedure described above, this time switching the carrier frequency between V,VI,VII. This should bring any initial state that falls on the right-hand side of $\ket{1/2}$ to the target state.
 \section*{S4: Calculation of quadrupolar splitting for  \texorpdfstring{NMR$_{\pm 1}^{0}$}{NMR10}}
\label{Supp:Quadrupole_neutral}
The resonance frequencies for neutral NMR are mathematically obtained by taking the difference in expectation values of the Hamiltonian operator $\hat{H}_{D^{0}}$ (Eq.\,2) between neighbouring nuclear states with a change in nuclear spin projection number $\Delta m = \pm 1$ and $\Delta s= \pm 0$.\\
The explicit expression for the resonance frequencies is given by
\begin{align}
    f_{m_I-1\leftrightarrow m_{I}}^{\mathrm{NMR^0}} & =  \langle \downarrow\uparrow, m_I-1
		|\hat{H}_{D^{0}} |\downarrow\uparrow, m_I-1 \rangle-\langle \downarrow\uparrow, m_{I} |\hat{H}_{D^{0}}|\downarrow\uparrow m_{I} \rangle \nonumber \\
  & =\gamma_{\mathrm{n}}B_{\mathrm{0}} +\underbrace{\left(m_I-\frac{1}{2}\right)f_{\rm q} \pm \frac{A}{2}}_{f^1} \pm \underbrace{g_{m_I-1\leftrightarrow m_I}\frac{A^2}{\gammae B_0}}_{f^2} ,
\label{eq:NMR0_freq}
\end{align}
where the first term corresponds to the Zeeman splitting at magnetic field $B_0$ with $\gamma_{\rm n} = 5.55$\,MHz, the second and third terms are the first-order ($f^{1}$) contributions of the quadrupolar and isotropic hyperfine interactions, with quadrupole splitting $f_{\rm q}^{0}$ and hyperfine constant $A$, and the fourth term corresponds to the second-order ($f^{2}$) contribution of the hyperfine interaction to the energy splitting, where $g_{m_I-1\leftrightarrow m_I}$ is a coefficient that depends on the nuclear spin projection quantum number.\\
These coefficients are calculated by considering that each eigenfrequency is corrected to second order by
\begin{align}\label{FC-eq:second_order_hyperfine}
	f_{s, m_I}^{2} = & \sum_{{m'_{s},m_I'} \neq {m_{s},m_I}} \dfrac{|\langle m'_{s} , m_I' | \hat{H}_{\mathrm{A}} | m_{s}, m_I \rangle |^{2}}{f_{m_{s},m_I}^{0}-f_{m'_{s},m'_I}^{0}} ,\nonumber\\
	= & \frac{A^2}{4} \sum_{{m'_{s},m_I'} \neq {m_{s},m_I}} \dfrac{|\langle m'_{s} , m_I' | \left(\hat{S}_+\hat{I}_- + \hat{S}_-\hat{I}_+\right)| m_{s}, m_I \rangle|^2}{f_{m_{s},m_I}^{0}-f_{m'_{s},m'_I}^{0}}\nonumber\\
 = &\frac{A^2}{4} \sum_{{m'_{s},m_I'} \neq {m_{s},m_I}} \dfrac{\left(I\left(I+1\right)-m'_Im_I\right)\left(\delta_{m'_{s},m_{s}+1}\delta_{m_I'+1,m_I}+\delta_{m'_{s}+1,m_{s}\delta_{m_I',m_I+1}}\right)}{f_{m_{s},m_I}^{0}-f_{m'_{s},m'_I}^{0}},
\end{align}
where $\hat{H}_A = A\boldsymbol{\hat{S}}\cdot\boldsymbol{\hat{I}}$ is the hyperfine interaction Hamiltonian, which represents the perturbation to the Zeeman eigenbasis, and where $f_{m_{s},m_I}^{0} = \bra{m_{s},m_I}H_\mathrm{Z}\ket{m_{s},m_I}$ are the Zeeman eigenenergies (in units of Hz). The subscripts $\pm$ assigned to the spin operators $\hat{S}$ and $\hat{I}$ signify whether they represent creation or annihilation operators.\\
 As expected, the Kronecker delta conditions $\delta$ in Eq.\,\ref{FC-eq:second_order_hyperfine} reveal that corrections from the hyperfine interaction to  second order only arise from the associated antiparallel spin states, which are the ones that \ensuremath{\hat{H}_A} couples. 
 The nuclear-dependent coefficients $g_{m_I-1\leftrightarrow m_I}$ from Eq.\,\ref{eq:NMR0_freq} are thus calculated using
\begin{equation}
    g_{m_I-1\leftrightarrow m_I} = (f_{\pm 1/2, m_I-1}^{2} -f_{\pm 1/2, m_I}^{2})\frac{\gamma_{\rm{e}}B_0}{A^2} 
    \label{eq:2nd_order_coeff}
\end{equation}
Since the spectrum presented in Fig.\,1e    corresponds to the subspace where the electron is in the $\ket{\downarrow}$ state, our results consider the case where $m_s = -1/2$. This yields $g_{m_I-1\leftrightarrow m_I}$ = [1.75, 1.25, 0.75, 0.25, -0.25,-0.75, -1.25] for $m_I\in \{-5/2,...,7/2\}$. \\
To obtain the quadrupolar splitting from the experimental NMR$^{0}_{\pm 1}$ resonance frequencies, we first need to estimate the value of the hyperfine interaction strength $A$. Conveniently, we can do this without considering the quadrupolar splitting $f_q$, as its contribution to the transition frequencies is symmetric with respect to the nuclear spin number. This means that we can find an expression that only depends on the unknown value $A$ by adding the transition frequencies with same absolute value of the nuclear spin projection,
\begin{flalign}
 f_{m_{I}-1\leftrightarrow m_{I}}^{\mathrm{NMR^0}}+f_{m_{-I}\leftrightarrow m_{-I}+1}^{\mathrm{NMR^0}} =  2B_0+ A+\frac{A^2}{2\gammae B_0} ,
    \label{eq:A_neutral_nmr}
\end{flalign}
where $B_0 = 999.5(5)$~mT was obtained from the ionised NMR spectrum in Fig.\,1d and where $f^{\mathrm{NMR^0}}_{mI-1\leftrightarrow m_I}$ correspond to the experimental resonance frequencies for the neutral nucleus in  Fig.\,1e. The quadratic equation \,\ref{eq:A_neutral_nmr} can be solved exactly and the value of the hyperfine is calculated for the case where $m_I\in\{-\frac{1}{2},-\frac{3}{2},-\frac{5}{2}\}$. Using these three solutions we obtain an average value of the hyperfine interaction of $A = 96.584(2)$~MHz. By substituying this quantity in Eq.\,\ref{eq:NMR0_freq}, we calculate the quadrupolar splitting from the distance between transitions with $\Delta m = 1$, resulting in $f_q = -52.5(5)$~kHz for the neutral atom.
\section*{S5: Calculation of hyperfine interaction from ESR spectrum}
\label{Supp:Hypefine_ESR}
To obtain the resonance frequencies for the electron spin, we mathematically calculate the difference in the expectation values of the Hamiltonian operator $\hat{H}_{D^{0}}$ (as defined in Eq.\,2) when the nuclear spin projection number is fixed at $\Delta m = 0$ and there is a change in the electron spin projection number, $\Delta s =  1$:
\begin{equation}
	\label{FC-eq:ESR_first_order}
	f^{\mathrm{ESR}}_{m_{I}} = \langle \uparrow  m_{I} | \hat{H}_{\mathrm{D^0}} | \uparrow m_{I} \rangle-\langle \downarrow  m_{I} |\hat{H}_{\mathrm{D^0}} | \downarrow m_{I} \rangle =  \gamma_{\mathrm{e}}B_{0}+\underbrace{m_{I}A}_{f^1}+\underbrace{b_{m_{I}}\dfrac{A^2}{\gammae B_0}}_{f^2},
\end{equation}
where we noted the first- ($f^1$) and second- ($f^2$) order corrections resulting from the hyperfine interaction. The latter arise from the transversal components of the hyperfine interaction and correct the electron eigenenergies dependent on the state of the nuclear spin (whereas the  quadrupolar interaction term leaves them unaffected).\\
Using Eq.\,\ref{FC-eq:second_order_hyperfine}, we obtain that each ESR resonance peak shown in Fig.\,1f is corrected to second order by
\begin{align}\label{ESR-second-order-2}
	f_{m_I}^{2,\mathrm{ESR}} = & f_{\frac{1}{2},m_I}^{2}-f_{-\frac{1}{2}, m_I}^{2}.
\end{align}
Taking the difference between adjacent nuclear states shows that the distance between ESR features a second order contribution from $A$ that depends on the nuclear spin projection
\begin{align}
	\label{FC-eq:second_order_isolated}
	f_{m_I}^{\mathrm{ESR}}-f_{m_I-1}^{\mathrm{ESR}} = A+(f_{m_I}^{2,\mathrm{ESR}}-f_{m_I-1}^{2,\mathrm{ESR}}) = A\left(1 + c_{m_I-1\leftrightarrow m_I}\frac{A}{\left(\gammae+\gamman\right)B_{0}}\right),
\end{align}
where the values for $c_{m_I-1\leftrightarrow m_I}$ are given in Table.\,\ref{tab:second_order_ESR} 
\begin{table}[H]
\renewcommand*{\arraystretch}{2}
	\centering
	\begin{tabular}{|c|c|c|c|c|c|c|c|}
 \hline
		$\left(m_I-1,m_{I}\right)$ & $(-\frac{7}{2},-\frac{5}{2})$ & $\left(-\frac{5}{2},-\frac{3}{2}\right)$ & $\left(-\frac{3}{2},-\frac{1}{2}\right)$ & $\left(-\frac{1}{2},\frac{1}{2}\right)$ & $\left(\frac{1}{2},\frac{3}{2}\right)$ & $\left(\frac{3}{2},\frac{5}{2}\right)$ & $\left(\frac{5}{2},\frac{7}{2}\right)$ \\
  \hline

		$c_{m_I-1\leftrightarrow m_I}$ & $3$ & $2$ & $1$ & $0$& $-1$ & $-2$& $-3$ \\	\hline

	\end{tabular}
\caption{Coefficients for second order corrections to the hyperfine interaction between pairs of nuclear states}
	\label{tab:second_order_ESR}	
\end{table}
Since $c_{-1/2\leftrightarrow 1/2} = 0$ in this case, the value of $A$ can be obtained by calculating the difference between $f_{1/2}^{\mathrm{ESR}}$ and $f_{-1/2}^{\mathrm{ESR}}$, which yields $A = \SI{95.442(22)}{\mega\hertz}$. The disparities observed in the values of $A$ obtained from the NMR and ESR spectra may arise due to varying electrostatic configurations between the experiments.

\section*{S6: Rabi oscillations}
\label{Supp:Rabi_oscillations}
In this section we collate the raw data on the Rabi oscillations for the electron and nuclear spin transitions. In these experiments, a single frequency tone modulated by a baseband pulse is applied at each of the frequencies shown in Fig.1\,d-i, with a varying pulse duration. By varying the duration of the control pulse, Rabi oscillations are observed (Fig.\,S3-S5), demonstrating coherent control on each pair of sublevels. Fitting the data in Fig.\,S3-S5, we extract the values of $f_{\rm Rabi}$ reported in Fig.~2.
 
 \begin{figure}[hbt!]
	\includegraphics[width = 0.85\textwidth]{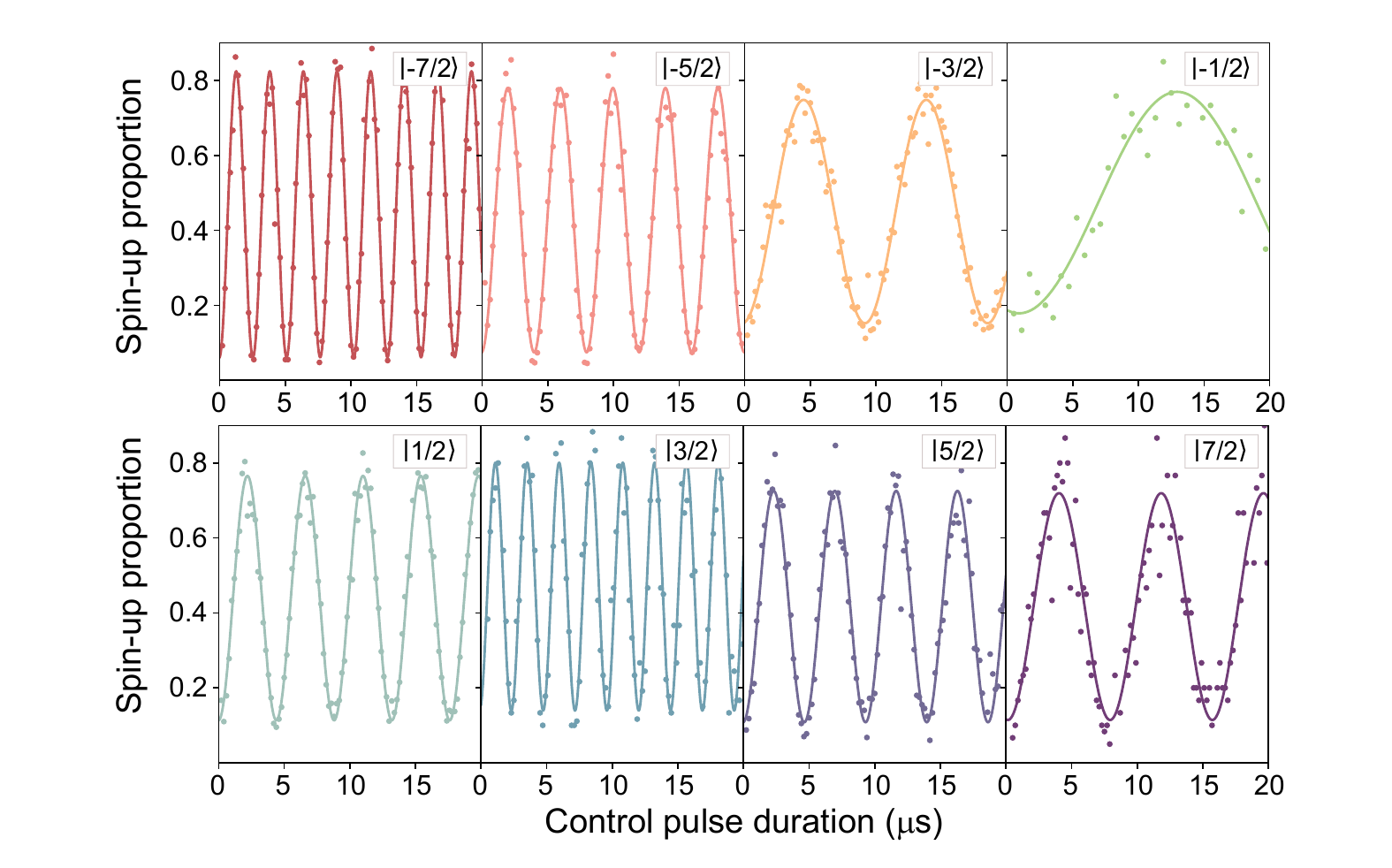}
	\caption{\textbf{ESR Rabi oscillations of a \Sb donor.} Electron spin Rabi oscillations on all eight ESR frequencies, where the state of the nuclear spin $\ket{m_I}$ for each panel is indicated in the legend. The ESR pulse amplitude at the source ($V^{\mathrm{pp}}_{\mathrm{MW}} = \SI{300}{\milli\volt}$) in constant in all measurement. The different Rabi frequencies are attributed to a frequency-dependent transmission of the microwave antenna.}
	\label{add_fig:ESR_Rabi_oscillations}
	\end{figure}
	\begin{figure}[hbt!]
	\centering
	\includegraphics[width = 0.75\textwidth]{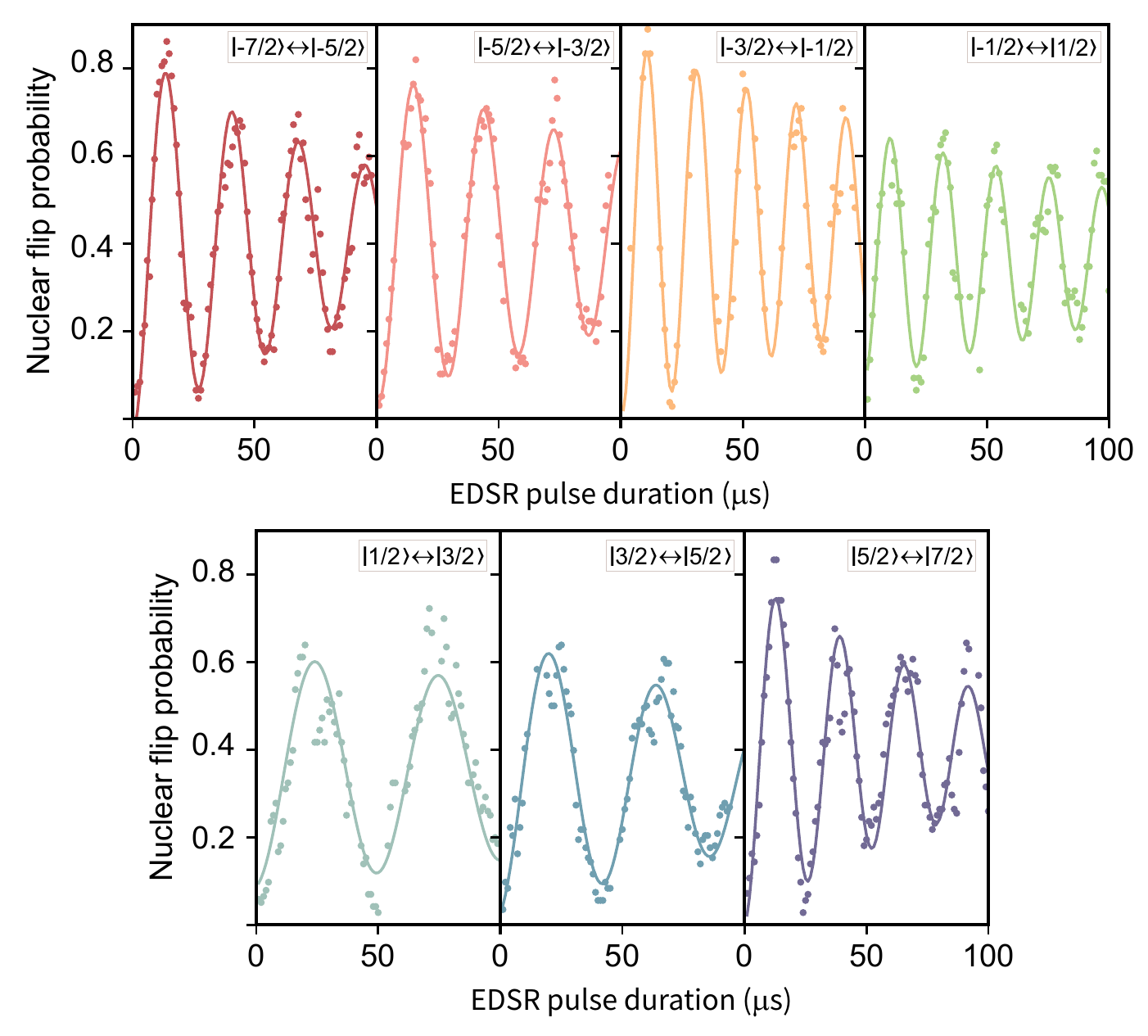}
		\caption{\textbf{EDSR Rabi oscillations of the \Sb donor.} Electric drive of the flip-flop transitions using the high-frequency microwave antenna, with a pulse amplitude of $V^{\mathrm{pp}}_{\mathrm{MW}} = \SI{300}{\milli\volt}$ at the source. The insets indicate the nuclear subspace. The data points have been smoothened using a  Savitzky–Golay filter, with a window length of 3 and a degree 1 polynomial to soothe an oversampling of the curves along the $x$-axis.}
	\label{add_fig:EDSR_Rabi_oscillations}
\end{figure}
 	\begin{figure}[hbt!]
 		\centering
 		\includegraphics[width = 0.9\textwidth]{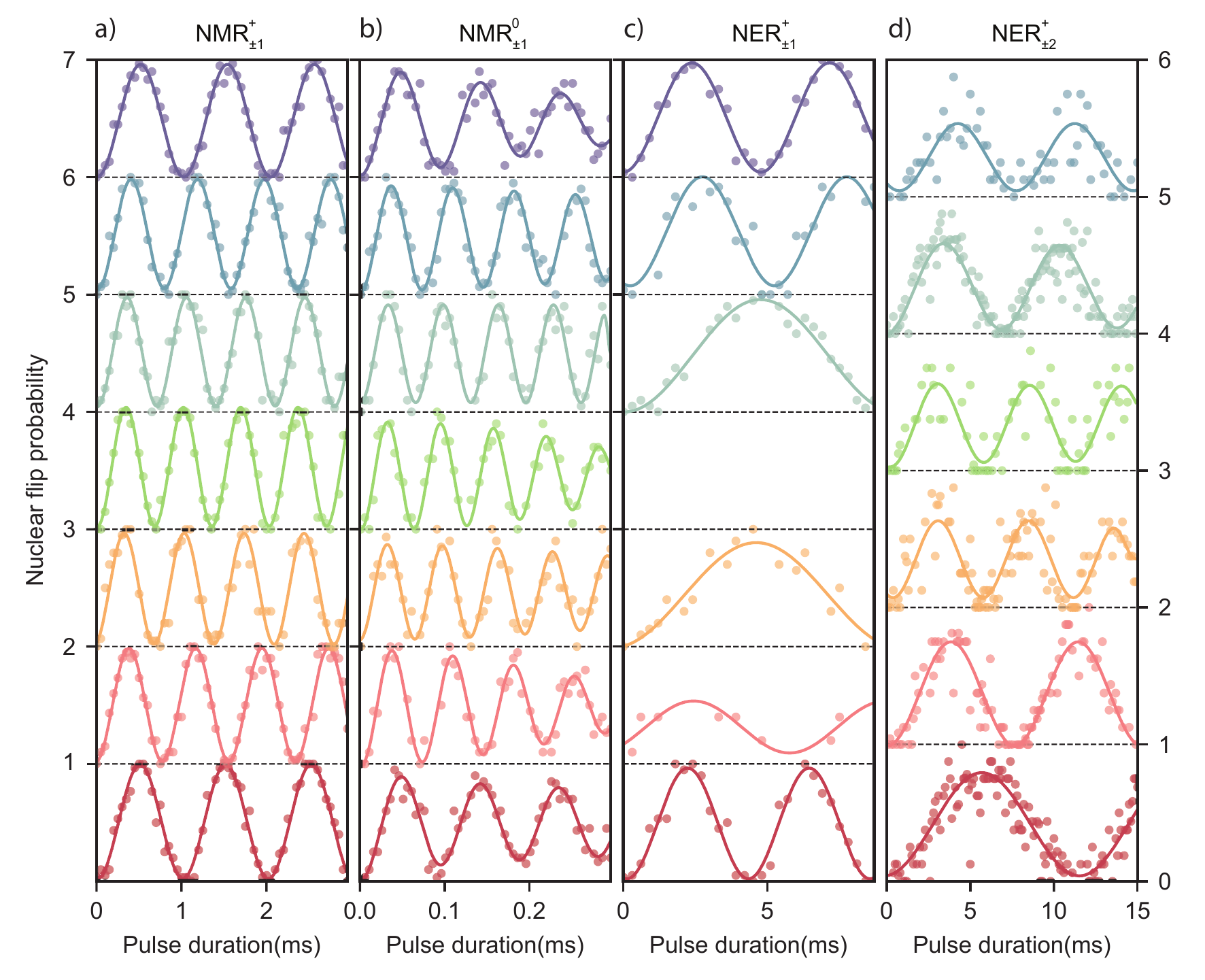}
 		\caption{ \textbf{Coherent nuclear spin drive of an \Sb donor.} \textbf{a), b)} Magnetically driven Rabi oscillations on the \textbf{a)} ionised (NMR$_{\pm 1}^{+}$), and \textbf{b)} neutral (NMR$_{\pm 1}^{0}$) nucleus in the  $\ket{\downarrow}$ electron spin configuration. \textbf{c), d)} Coherent electric drive on the ionised nuclear spin for \textbf{c)} $\Delta m_{I} = 1$ (NER$_{\pm 1}^{+}$), where the fourth row is left blank as driving $\ket{-1/2}\leftrightarrow\ket{1/2}$ is not allowed with NER drive (See Tab.\,1), and \textbf{d)} $\Delta m_{I} = 2$ (NER$_{\pm 2}^{+}$).
 	 } 		
   \label{add_fig:NMR_Rabi_oscillations}
 	\end{figure}
  
\clearpage
  
\section*{S7: Calculation of the Stark effect}
 \label{Supp:Stark_effect}
To calculate the Stark shift of the spectral lines, we examine the change in resonance frequencies as a function of bias voltage, presented experimentally in Fig.\,3a,b of the main text. These can be derived analytically by calculating the derivative of the electron ($f_{m_I}^{\mathrm{ESR}}$) and nuclear ($f_{m_I-1\leftrightarrow m_I}^{\mathrm{NMR}^{0}}$) resonance frequencies in Table.\,1 as a function of electric field (which in our devices is represented by a change in voltage $V$). 
\subsection{Electron spin}
Starting with the electron we have
\begin{flalign}
    f^{\mathrm{ESR}}_{m_{I}} = \gammae(E)B_{0} +m_{I}A(E)+b_{m_{I}} \frac{A(E)^{2}}{B_{0}\gammae(E)},
    \label{EC-eq:ESR_freqs}
\end{flalign}
where we have indicated the terms that are sensitive to static electric fields. The nuclear-state dependent coefficients $b_{m_I}$ are obtained from Eq.\,\ref{FC-eq:second_order_isolated}.
Taking the derivative yields
\begin{flalign}
		\label{EC-eq:ESR_freqs_derivative}
  \frac{\partial	f^{\mathrm{ESR}}_{m_I}}{ \partial V} =\frac{\partial	\gammae B_{0}}{\partial V}  + m_{I} \frac{\partial	A}{ \partial V}+ 2b_{m_{I}}\frac{A}{B_{0}\gammae}\frac{\partial A}{\partial V}.
\end{flalign}  
We note that $\gamma_{\rm{e}}$ in the denominator of $b_{m_I}A^2/B_{0}\gamma_{\rm e}$ should in principle be included in the derivative. However, we can safely ignore its contribution as it scales as $A^2/\gamma_{\rm e}^2$.
The symmetry of Eq.\,\ref{EC-eq:ESR_freqs_derivative} with respect to the nuclear spin number allows us to extract the slopes $\partial A/\partial V$ by substracting the expressions with opposite nuclear spin number
 \begin{flalign}
 	\label{eq:derivatives_A_stark_effect_ESR}
 	\frac{\partial f^{\mathrm{ESR}}_{m_{I}}}{ \partial V} -\frac{\partial f^{\mathrm{ESR}}_{m_{-I}}}{ \partial V} = 2m_{I}\frac{\partial A}{\partial V}. 
 \end{flalign}
 After this calculation, we can obtain the g-factor (gyromagnetic ratio) Stark effect, if we instead sum the derivatives with opposite nuclear spin projection
  \begin{flalign}
  \label{eq:derivatives_gyro_stark_effect_ESR}
 	\frac{\partial f^{\mathrm{ESR}}_{m_{I}}}{ \partial V} +\frac{\partial f^{\mathrm{ESR}}_{m_{-I}}}{ \partial V} = 2\frac{\partial\gamma_{\mathrm{e}}B_{0}}{\partial V}  +4b_{m_I}\frac{A}{B_{0}\gammae}\frac{\partial A}{\partial V}.
 \end{flalign}
 We extract the experimental slopes $\partial f^{\mathrm{ESR}}_{m_{I}}/\partial V$ by performing linear fits to the data in Fig.\,3\,a and use Eq.\,\ref{eq:derivatives_A_stark_effect_ESR} and Eq.\,\ref{eq:derivatives_gyro_stark_effect_ESR} to obtain the Stark effect on $A$ and $\gammae$ for each subspace $\ket{m_{I}}$. We calculate an average value of $\partial A/\partial V= 9.8(4)$~MHz/V and $\partial\gamma_{\mathrm{e}}B_{0}/\partial V= -1.4(6)$~MHz/V, where the errobars represent the standard error.\\
 Using the extracted slopes, we numerically obtain the Stark effect on the ESR resonance lines by calculating the eigenenergies using the Hamiltonian from Eq.\,2, as a function of voltage amplitude. The results are presented with the solid lines in Fig.\,3b of the main text.
 
Note that we have excluded the data for \ensuremath{m_{I} = 1/2} in the analysis since we observe a large deviation from the expected trend. We attribute this deviation from the expected trend to a sudden jump in the ESR frequency during the experiment, caused by a charge rearrangement, or a hyperfine-coupled \ensuremath{^{29}}Si, which erroneously attributes a change of gate voltage $\Delta V_{\mathrm{DC}}^{\mathrm{DG1}}$ to a change in resonance frequency $\Delta f_{\mathrm{ESR}}$.
  
	\subsection{Nuclear spin}
In the neutral antimony atom (\Sb$^0$), both the  hyperfine coupling and the quadrupolar interaction are sensitive to changes in the local electric field\,\cite{kushida1961shift,armstrong1961linear,dixon1964linear,asaad2020coherent}. The explicit formula for the resonance frequencies of the neutral \Sb atom, denoting the electric field dependency on the Hamiltonian parameters, is given by
 \begin{align}
		\label{EC-eq:NMR_frequencies_secondorder}
        f^{\mathrm{NMR^0}}_{m_{I}-1\leftrightarrow m_{I}} =  & \gamma_{\mathrm{n}}B_{\mathrm{0}} +\left(m_I-\frac{1}{2}\right)f_{\rm q}(E) \pm \frac{A(E)}{2} \pm g_{m_I-1\leftrightarrow m_I}\frac{A(E)^2}{\gammae B_0},
	\end{align}
	where $f_{\rm q}(E)$ is the $E$-field dependent value for the quadrupolar splitting, and $g_{m_I-1\leftrightarrow m_I}$ are the nuclear spin-dependent coefficients that scale the second order corrections from $A$, and are derived in Supplementary Information S4. 
 Calculating the derivative, we find
	\begin{align}
		\label{EC-eq:NMR_frequencies_stark}
		\frac{\partial f^{\mathrm{NMR^0}}_{m_{I}-1\leftrightarrow m_{I}} }{\partial V} =  &  \left(m_I-\frac{1}{2}\right)\frac{f_{\rm q}}{\partial V} \pm \frac{1}{2}\frac{\partial A}{\partial V} \pm g_{m_I-1\leftrightarrow m_I}\frac{2A}{\gammae B_0}\frac{\partial A}{\partial V}.
		\end{align}
	In this derivation, we are also ignoring the electric-field-dependence of $\gamma_{\rm e}$, as we did  for the electron in section S7A. We can see from Eq.\,\ref{EC-eq:NMR_frequencies_stark} that a change in the hyperfine coupling results in the same spectral shift for all nuclear spin projections to first order ($\frac{\partial A}{2\partial V}$), contrary to what we saw for ESR. However, different slopes arise as we evaluate the second order contributions from the hyperfine interaction, as well as the changes in the quadrupolar splitting.\\
    Using the same approach as we did for ESR (Eq.\,\ref{eq:derivatives_A_stark_effect_ESR} and Eq.\,\ref{eq:derivatives_gyro_stark_effect_ESR}) we now obtain the following relations
	\begin{align}
		\label{EC-eq:derivatives_quadrupole_stark_effect_NMR}
		\frac{\partial f^{\mathrm{NMR^0}}_{m_{I}-1\leftrightarrow m_I}}{ \partial V} -\frac{\partial f^{\mathrm{NMR^0}}_{m_{-I}\leftrightarrow m_{-I}+1}}{ \partial V} = & 2 (m_{I}-\frac{1}{2})\frac{\partial f_q}{\partial V} + \\ \nonumber &+(g_{m_{I}-1\leftrightarrow m_I}-g_{m_{-I}\leftrightarrow m_{-1}+1})\frac{2A}{\gammae B_{0}}\frac{\partial A}{\partial V},
	\end{align}
	and
	\begin{align}
		\label{EC-eq:derivatives_A_stark_effect_NMR}
		\frac{\partial f^{\mathrm{NMR^0}}_{m_{I}-1\leftrightarrow m_I}}{ \partial V} +\frac{\partial f^{\mathrm{NMR^0}}_{m_{-I}\leftrightarrow m_{-I}+1}}{ \partial V} = & \frac{\partial A}{\partial V} \left(1 +(g_{m_{I}-1\leftrightarrow m_I}+g_{m_{-I}\leftrightarrow m_{-I}+1})\frac{2A}{\gammae B_{0}}\right),
	\end{align}
where $m_{I}\in\{-\frac{1}{2},-\frac{3}{2},-\frac{5}{2}\}$ .
After performing linear fits to the data in Fig.\,3b, we use Eq.\,\ref{EC-eq:derivatives_A_stark_effect_NMR} to extract $\partial A/\partial V$ for each nuclear subspace, using $A = 96.584(2)$~MHz, and $B_0 = 999.5(5)$~mT from the ionized \Sb donor spectrum. We obtain an average value of $\partial A/\partial V= 11.6(5)\rm{MHz/V}$, where the error bars are expressed as the standard error. This result can be used to retrieve the contribution from the quadrupole Stark effect using  Eq.\,\ref{EC-eq:derivatives_quadrupole_stark_effect_NMR}. In this case, it is important that we consider the second order terms from $A$, as they might be on the order of the quadrupolar shifts. We obtain an average value of $\partial f_q/\partial V\approx \SI{-300(56)}{\kilo\hertz\volt^{-1}}$. As we did for the electron, we use these results to numerically calculate the Stark effect on the nuclear resonance frequencies, by solving the Hamiltonian in Eq.\,2 as a function of voltage gate $V_{\mathrm{DC}}^{\mathrm{DG1}}$ (see main text). The calculated values are presented as solid lines in Fig.\,3a of the main text.

\section*{S8: Quadratic nuclear Stark effect}
\label{Supp:Quadratic_Stark_effect}

Understanding the hyperfine shift caused by electric fields requires knowledge of the donor's distance to the interface (depth). For donors in bulk silicon, the hyperfine Stark shift is quadratic in the electric field. \cite{rahman2007high}. However, as the donor depth decreases, the change in hyperfine coupling becomes linearly dependent on the electric field. The total hyperfine shift can be expressed as follows: 
\begin{equation}
\label{EC-eq:Quadratic hyperfine_change_second}
\Delta A(\boldsymbol{E}) = A(0)\left(\eta_1E+\eta_2E^2\right),
\end{equation}
where \ensuremath{A(0) = |\psi(0,r_0)|^2} represents the hyperfine value in the absence of an electric field, and $\eta_{1}(\SI{}{\micro\meter/\volt})$ and $\eta_{2}(\SI{}{\micro\meter^2/\volt^2})$ are parameters that define the strength of the linear and quadratic Stark effects, respectively.

\begin{figure}[H]
		\centering
		\includegraphics[width =0.85\textwidth]{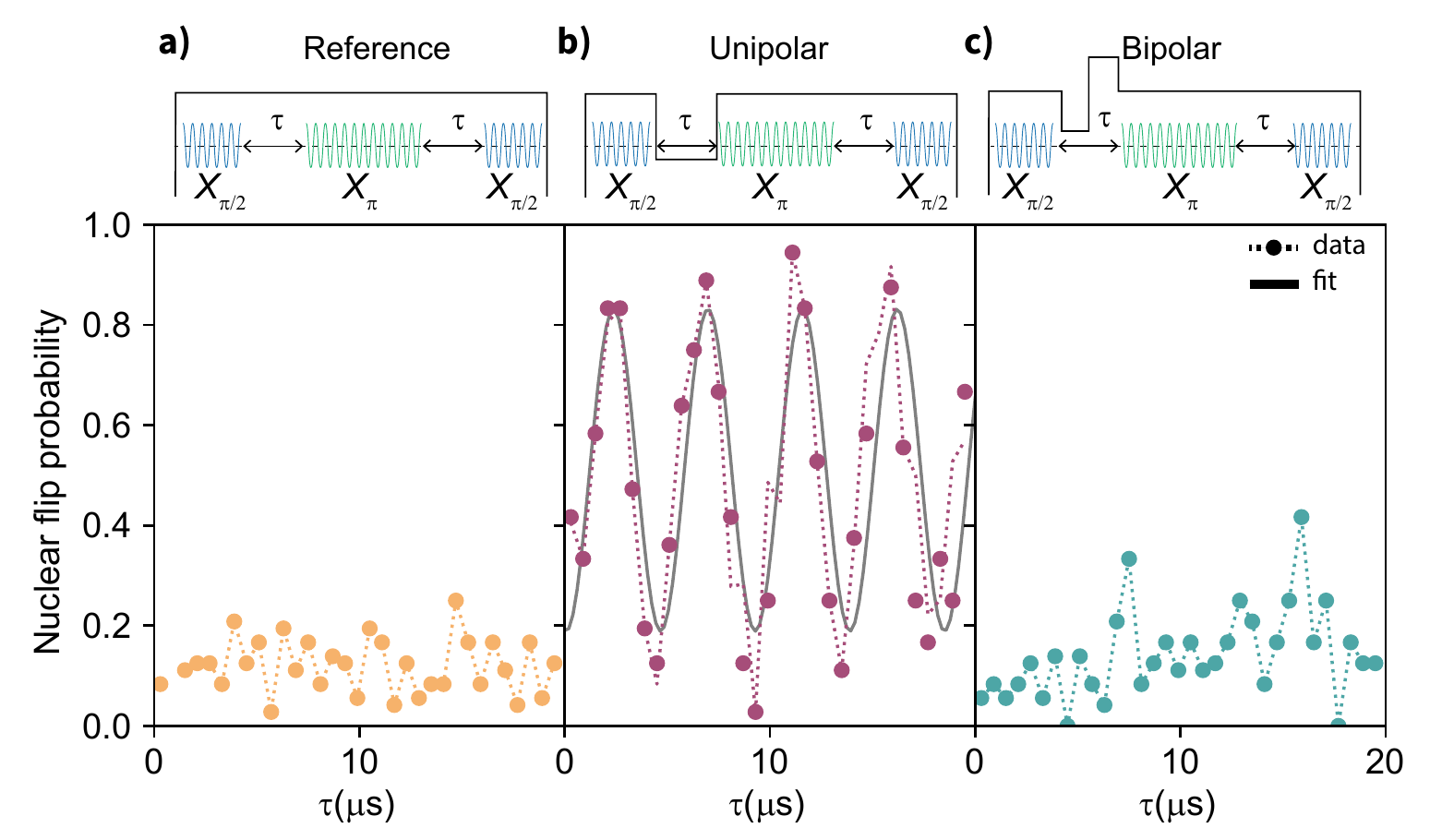}  \caption{\textbf{Linear and quadratic hyperfine Stark effect on a neutral \Sb$^0$ donor.} \textbf{a)} Hahn echo experiment on the neutral nucleus, without unipolar/bipolar pulses, used as a reference. As expected, the absence of electric pulses during the free precession causes no phase accumulation. \textbf{b)} Hahn echo experiment with a unipolar voltage pulse of amplitude \ensuremath{V_{\mathrm{DC}} = \SI{40}{\milli\volt}} during the first free precession time \ensuremath{\tau}. An accumulated phase as a function of $\tau$ is resolved in the oscillatory return probability. The unipolar pulse causes the spin to accumulate a phase given by Eq.\,\ref{EC-eq:phase_accumulation_quadratic}. 
  \textbf{c)} Hahn echo experiment with a bipolar pulse with amplitude \ensuremath{\pm V_{\mathrm{DC}} = \SI{40}{\milli\volt}} and duration \ensuremath{\tau/2}. The lack of oscillations indicates that the quadratic contribution is small and its effect cannot be resolved for \ensuremath{\tau<20\mathrm{\mu s}}.}
		\label{EC-fig:NMR_quadratic_Stark_shift}
	\end{figure}

 In order to characterize the quadratic contributions of electric fields $\eta_2E^2$  from Eq.\,\ref{EC-eq:Quadratic hyperfine_change_second}, we choose an echo refocusing pulse scheme like the one outlined in Ref.\,\cite{bradbury2006stark}. We concentrate our analysis on the subspace \ensuremath{\ket{\downarrow,7/2}\leftrightarrow \ket{\downarrow,5/2}}, and use magnetic control (NMR) to drive the nuclear spin.
	The modified Hahn echo works as follows: We start by bringing the nuclear spin to the \ensuremath{xy}-plane with a \ensuremath{X_{\pi/2}} pulse, where we let the spin precess for a time \ensuremath{\tau}. During this time, we choose to (i) wait (Fig.\,\ref{EC-fig:NMR_quadratic_Stark_shift}\,a), (ii) apply a `unipolar' voltage pulse with amplitude \ensuremath{V_\mathrm{DC}} (Fig.\,\ref{EC-fig:NMR_quadratic_Stark_shift}\,b) or (iii) apply a `bipolar' voltage pulse with amplitude \ensuremath{\pm V_\mathrm{DC}} (Fig.\,\ref{EC-fig:NMR_quadratic_Stark_shift}\,c) to one of the donor gates. We then invert the spin with a refocusing \ensuremath{X_{\pi}} pulse, let it precess for another time \ensuremath{\tau} and project it back to the \ensuremath{z-}axis with a final \ensuremath{X_{\pi/2}} pulse. \\
	The Hahn echo sequence cancels out dephasing during the free precession time $\tau$ if the phase accumulated is constant during both $\tau$ periods. However, the bipolar and unipolar pulse make the spin accumulate a phase $\theta_{m_{I}-1\leftrightarrow m_{I}}(\tau)$ during the first interval $\tau$:
	\begin{align}
		\label{EC-eq:phase_accumulation_quadratic}
		\theta_{m_{I}-1\leftrightarrow m_{I}}(\tau) =2\pi\Delta 	f^{\mathrm{NMR^0}}_{m_{I}-1\leftrightarrow m_{I}} \tau,
	\end{align}
	where 
	\begin{align}
		\label{EC-eq:phase_accumulation_nmr}
		\Delta f^{\mathrm{NMR^0}}_{m_{I}-1\leftrightarrow m_{I}} & = \frac{\partial f^{\mathrm{NMR^0}}_{m_{I}-1\leftrightarrow m_{I}}}{\partial V}\Delta V,\\
& = \frac{\partial f^{\mathrm{NMR^0}}_{m_{I}-1\leftrightarrow m_{I}}}{\partial V}V_{\mathrm{DC}},
	\end{align}
	is the change in resonance frequency caused  by the pulse with an amplitude $\Delta V = V_{\mathrm{DC}}$. In both cases, the phase accumulated during both free precession periods is different and the echo pulse $X_{\pi}$ does not refocus the spin. From Eq.\,\ref{EC-eq:Quadratic hyperfine_change_second} and Eq.\,\ref{EC-eq:phase_accumulation_quadratic} we see that in the absence of a quadratic contribution from the electric field, the bipolar pulse compensates the accumulated phase during the free precession time by accumulating a positive phase \ensuremath{\theta_{m_{I}-1\leftrightarrow m_{I}}(\tau/2)}, and a negative phase \ensuremath{-\ensuremath{\theta_{m_{I}-1\leftrightarrow m_{I}}(\tau/2)}} of equal magnitude. In the presence of a quadratic contribution, the sign of \ensuremath{\ensuremath{\theta_{m_{I}-1\leftrightarrow m_{I}}(\tau/2)}} does not change and thus gives a net accumulated phase of \ensuremath{2|\ensuremath{\theta_{m_{I}-1\leftrightarrow m_{I}}(\tau/2)}|}. For a unipolar pulse, the spin accumulates a phase during the first  time $\tau$ from both the linear \ensuremath{\eta_{1}E} and quadratic \ensuremath{\eta_{2}E^2} contributions of the electric field.\\
	The data for these experiments is presented in Fig.\,\ref{EC-fig:NMR_quadratic_Stark_shift}\,a-c. We confirm that the absence of an electric pulse causes no phase accumulation in the nuclear spin, and the return probability stays flat (Fig.\,\ref{EC-fig:NMR_quadratic_Stark_shift}\,a).\\
	 When a unipolar pulse of amplitude $V_{\mathrm{DC}} = \SI{40}{mV}$ is applied, we observe that the return probability oscillates. The oscillation frequency is obtained by fitting the data to a sinusoidal function $P\sin{(2\pi f t+\phi)}+P_{\mathrm{offset}}$, and using Eq.\,\ref{EC-eq:phase_accumulation_quadratic} and Eq.\,\ref{EC-eq:phase_accumulation_nmr} with $\theta_{-7/2\leftrightarrow-5/2} = \pi$ and $\tau = \SI{2.3}{\micro\second}$ we calculate a $\Delta f^{\mathrm{NMR^0}}_{m_{I}-1\leftrightarrow m_{I}} = \SI{217(2)}{\kilo\hertz}$. Using the the estimated parameters for the Stark effect obtained from directly measuring the spectrum as a function of gate voltage (Fig.\,4.\,b), we find $\Delta f^{\mathrm{NMR^0}}_{m_{I}-1\leftrightarrow m_{I}} = \SI{235(3)}{\kilo\hertz}$,showing an excellent agreement between the two methods.

	The experiment with the bipolar pulse, shown in Fig.\,\ref{EC-fig:NMR_quadratic_Stark_shift}\,c, displays no visible oscillation within the chosen evolution time, i.e. no detectable quadratic Stark effect. This is consistent with the expected behaviour of a donor close to an interface, and subjected to a strong static electric field.


\section*{S9: Voltage and frequency fluctuations with laboratory temperature}
	\label{Supp:Temperature_frequency_shifts}
	\begin{figure}[H]
		\centering
		\includegraphics[width =0.8\textwidth]{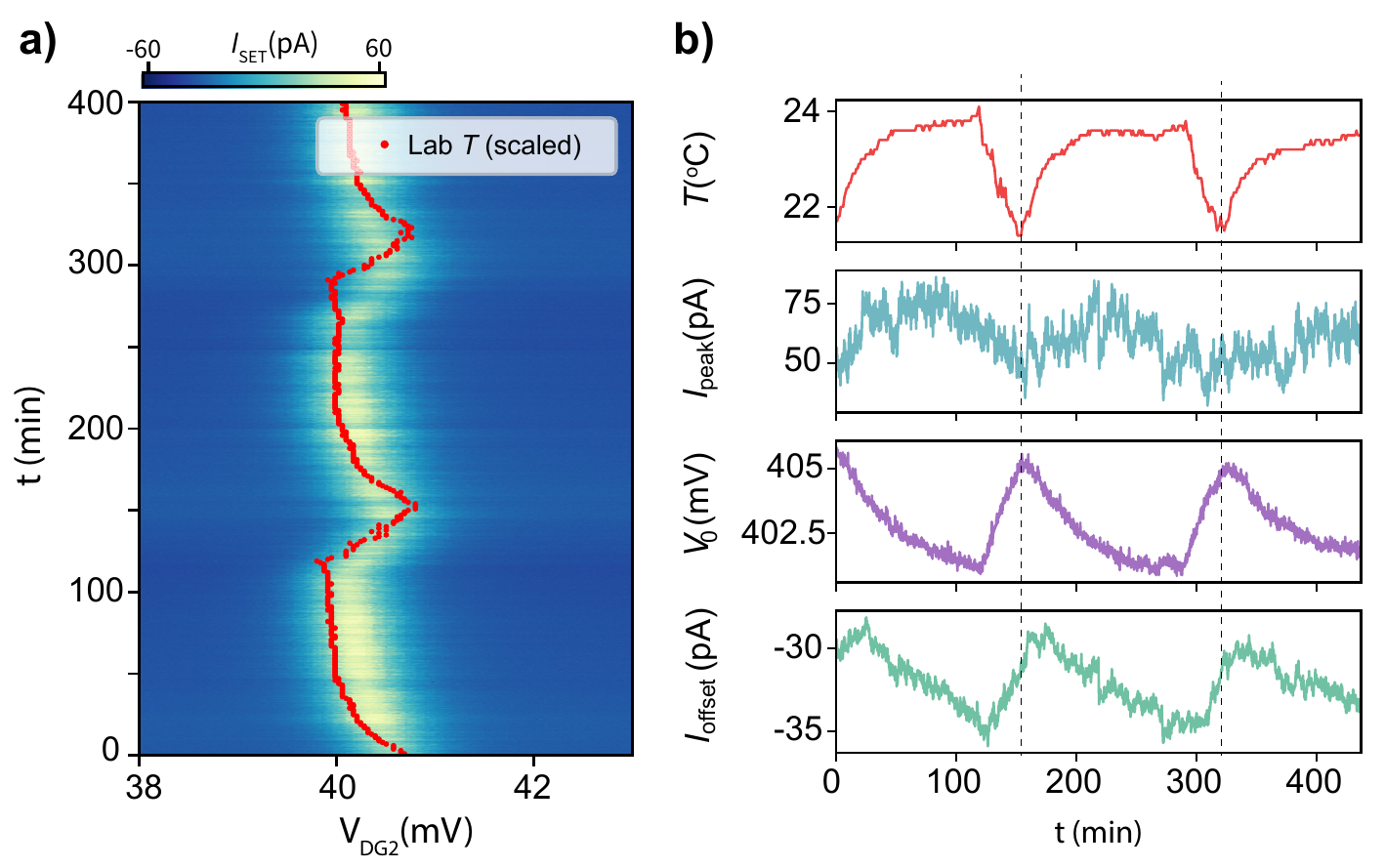}
		\caption{\textbf{Temperature-correlated voltage drifts sensed by the SET.} \textbf{a)} Line scans across a Coulomb peak (CP) over the course of 6 hours. We sweep the voltage of a donor gate $V_{\mathrm{DG2}}$ to track a CP over time and superimpose the lab temperature (red dots). The values of the temperature are rescaled to make the correlation between the temperature and CP drifts, visible to the reader. \textbf{b)} Individual SET current traces are fitted using a Gaussian function (explained in main text), and the resulting fitting parameters reveal oscillations in $I_{\mathrm{peak}}, V_{0}$ and $I_{\mathrm{offset}}$ which match the period of oscillations in the temperature T, as highlighted by the black dashed lines.}
		\label{EC-fig:Temperature_correalted_SET}
	\end{figure} 
 The temperature in our laboratory fluctuates by up to \SI{2}{\celsius} as a consequence of a controller that switches on/off the air conditioning unit with a typical period of two hours. This temperature fluctuation affects several experimental observables, such as the SET current.
	To investigate this relationship, we tracked the current of the SET by scanning over a Coulomb peak with one of the donor gates (DG2) over the course of \ensuremath{\approx 6} hours, while simultaneously measuring the temperature near the measurement equipment, using a EL-GFX-DTP data logger with a Thermistor Probe.	 Figure\,\ref{EC-fig:Temperature_correalted_SET}\,a shows the periodic drift of the Coulomb peak over time, following the periodically fluctuating lab temperature, plotted with the red dots. Fitting individual traces with a Gaussian function $I_{\mathrm{peak}}\exp{(-(V_{\mathrm{DG2}}-V_{0})^2/2\sigma^2)}+I_{\mathrm{offset}}$ reveals  that these fluctuations can be resolved in the free parameters of the fit, including the height of the Coulomb peak $I_{\mathrm{peak}}$, the offset current $I_{\mathrm{offset}}$, and the center of the Coulomb peak $V_{0}$, as shown in  Figure\,\ref{EC-fig:Temperature_correalted_SET}\,b.
 
\begin{figure}
	\centering
	\includegraphics[width =0.85\textwidth]{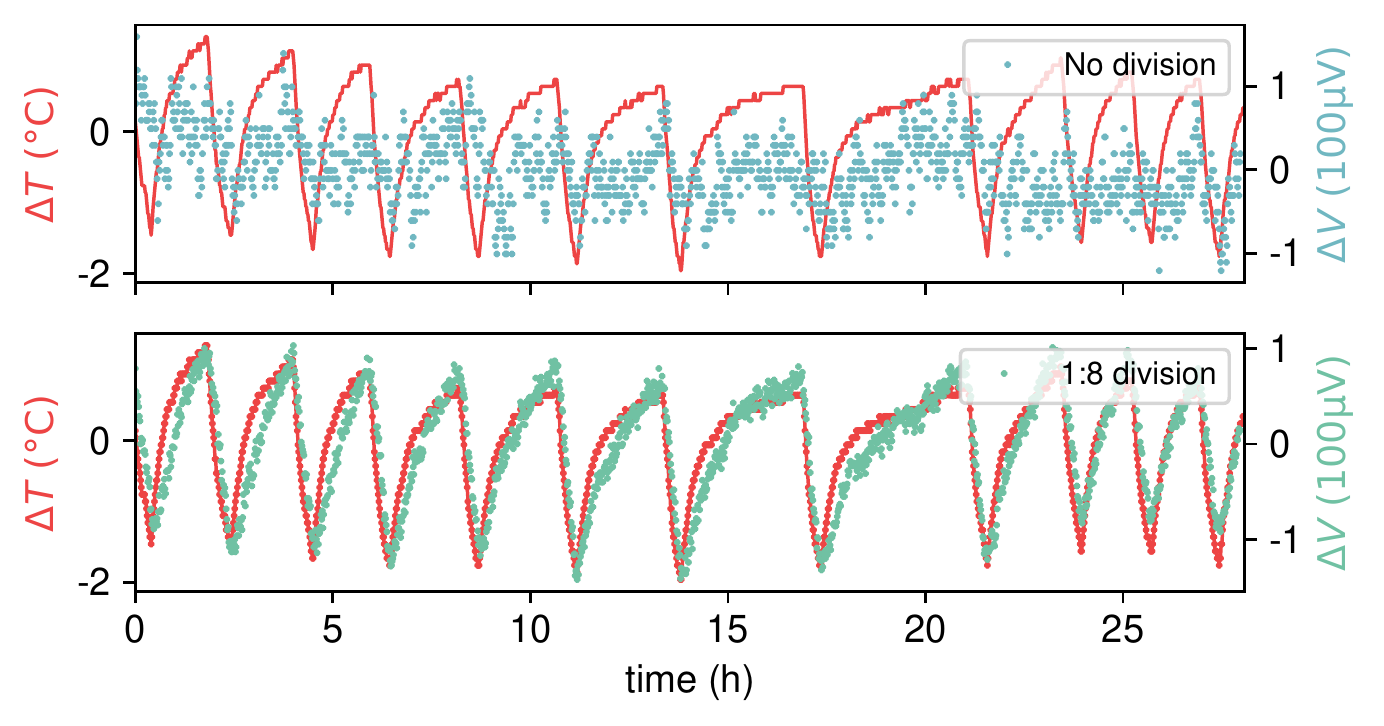}
	\caption{\textbf{Correlation between lab temperature and DC voltages.} We track the output voltage of a SRS SIM928 voltage source \textbf{a)} before and \textbf{b)} after a resistive voltage divider (1:8 division) over time, and plot it against the temperature T in the lab measured by a thermistor probe. In both \textbf{a)-b)} the voltage source from the SIM module was set to output \SI{10}{\volt}. We plot the deviation for the temperature $\Delta T = T-\overline{T}$ and voltage $\Delta V = V(t)-\overline{V}$ where $\overline{T}$ and $\overline{V}$ are the mean values for T and V, respectively.}
	\label{EC-fig:SIM_voltage_vs_time}
	\end{figure}

	Drifts in the Coulomb peaks and SET current can be explained by changes in voltages applied to the device, which modify the electrochemical potential of the SET island. We investigated whether the temperature affects directly the DC voltage source (SRS SIM928), or the 1:8 resistive voltage dividers used between the source and the device. In both cases, we set the source to output \SI{10}{\volt}. Figure\,\ref{EC-fig:SIM_voltage_vs_time}\,a shows the voltage fluctuations directly at the DC source output; Fig.\,\ref{EC-fig:SIM_voltage_vs_time}\,b for the second experiment. For both situations, we plot the deviation for the temperature $\Delta T = T(t)-\overline{T}$ and voltage $\Delta V = V(t)-\overline{V}$ where $\overline{T}$ and $\overline{V}$ are the mean values for the $T$ and $V$ datasets, respectively. 
 
We observe that the oscillations between voltage and temperature are present for both situations, indicating that the fluctuations in the SET from Fig.\,\ref{EC-fig:Temperature_correalted_SET} are likely caused by fluctuations in the voltage supplied to the gates. In the absence of temperature dependence in the resistors of the voltage divider, we would expect a decrease in the oscillations by a factor of 8 compared to the case with no division. Interestingly, we find that the standard deviation $\sigma(1:8) = \SI{58}{\micro\volt} $
and  $\sigma(1:1) = \SI{40}{\micro\volt}$, which is a measure of the oscillations amplitude, is similar for both. This suggests that the voltage dividers give a significant contribution to the temperature-dependent voltage fluctuations. 
 
\begin{figure}[hbt!]
		\centering
		\includegraphics[width =0.85\textwidth]{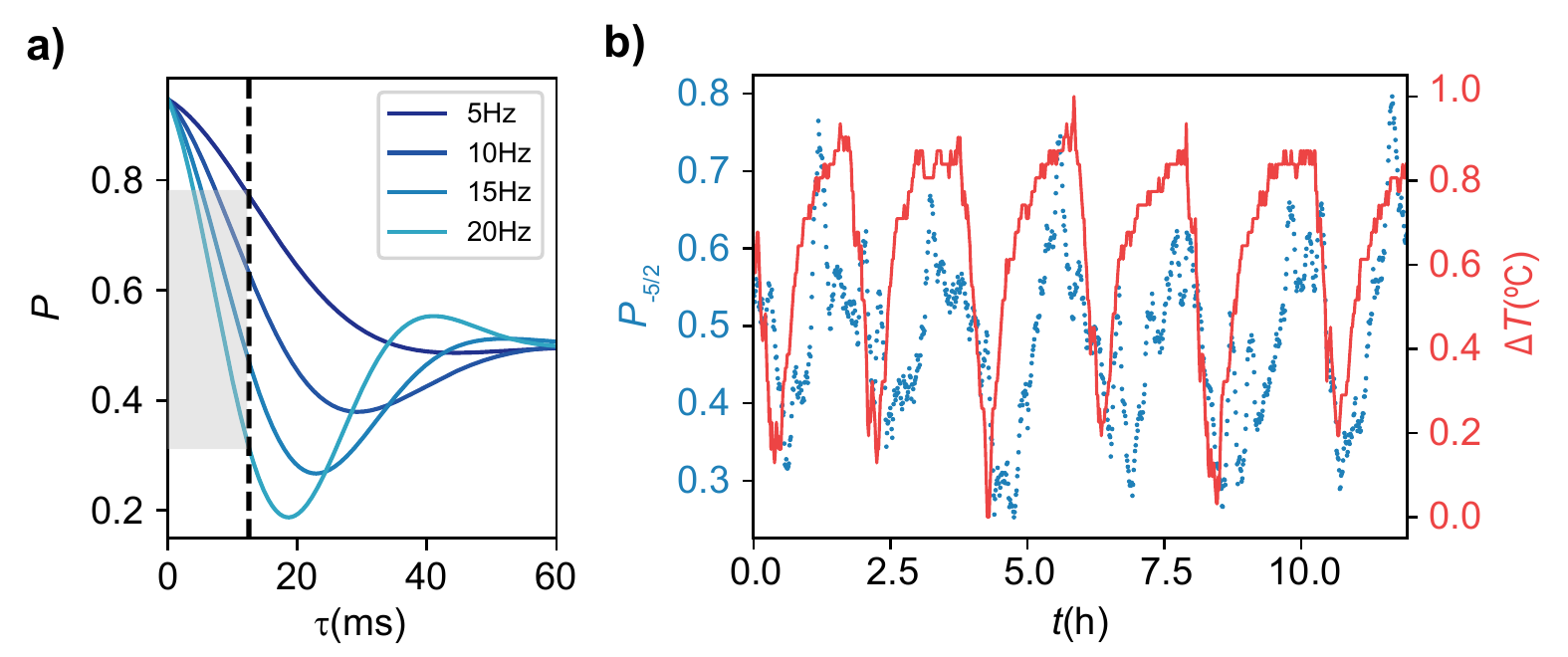}
		\caption{\textbf{Temperature-correlated frequency shifts on the ionized \Sb nucleus.} \textbf{a)} Depiction of a Ramsey experiment for different values of detuning $\Delta f$ (see main text for definition). The decay in the oscillations mimics the decoherence of the spin with a free induction decay time $T_{2\mathrm{n+}}^{*} = \SI{29}{\milli\second}$ for the $\ket{5/2}\leftrightarrow \ket{7/2}$ in this device. Taking a line cut at a fixed value for the free precession time $\tau$ highlights the variation in the nuclear state probability $P$ for the different values of detuning. \textbf{b)} We track the nuclear state probability $\ket{-5/2}$ as a function of time for a fixed $\tau  = \SI{12.5}{\milli\second}$, and plot it against the normalized lab temperature deviation $\Delta T = T(t)-\overline{T}$, where $\overline{T}$ is the mean value of the temperature dataset. We can use the results from Fig.\,\ref{EC-fig:SIM_voltage_vs_time}, to correlate the fluctuations from the gate voltages, with changes in the resonance frequency of the ionized donor. }
		\label{EC-fig:Fixed_tau_vs_temperature}
	\end{figure}

Finally, we investigate the impact of these voltage fluctuations on the frequency of the ionized nucleus, since the quadrupolar interaction is sensitive to changes in the electric field applied to the nucleus via the LQSE. We use a Ramsey interferometry scheme, where we perform consecutive Ramsey experiments on the ionized nucleus with a fixed free-precession time. The pulse sequence for a Ramsey experiment is given $X_{\pi/2}-\tau-X_{\pi/2}$, where $X_{\pi/2}$ are $\pi/2$ rotations around the $x$-axis and $\tau$ is the time the spin is left free to precess around the $xy$-plane.
 When the frequency of the control pulse $f_{\pi/2}$ is detuned from the resonance frequency $f_{0}$,  the nuclear state probability $P$ oscillates at a frequency $f_{\mathrm{fringe}} = \Delta f = f_{\pi/2}-f_{0}$, as illustrated in Fig.\,\ref{EC-fig:Fixed_tau_vs_temperature}\,a. Fixing the value for the free precession time $\tau$, thus allows detecting changes in resonance frequency, as they translate into changes in $P$, as depicted with the grey shaded area in Fig.\,\ref{EC-fig:Fixed_tau_vs_temperature}\,a.

We choose $\tau = \SI{12.5}{\milli\second}$ and track the nuclear state probability for $m_I = -5/2$ ($P_{-5/2}$) and the lab temperature T over time, which results in the data presented in Fig.\,\ref{EC-fig:Fixed_tau_vs_temperature}\,b, showing correlated oscillations for both. This experiments shows that the nuclear spin is capable of detecting temperature fluctuations in the lab, via their effect on the voltage applied to the gates in the device.

\clearpage
\section*{S10: GST Experiments}
\label{Supp:GST_experiments}
\begin{figure}[H]
    \centering
    \includegraphics[width = \textwidth]{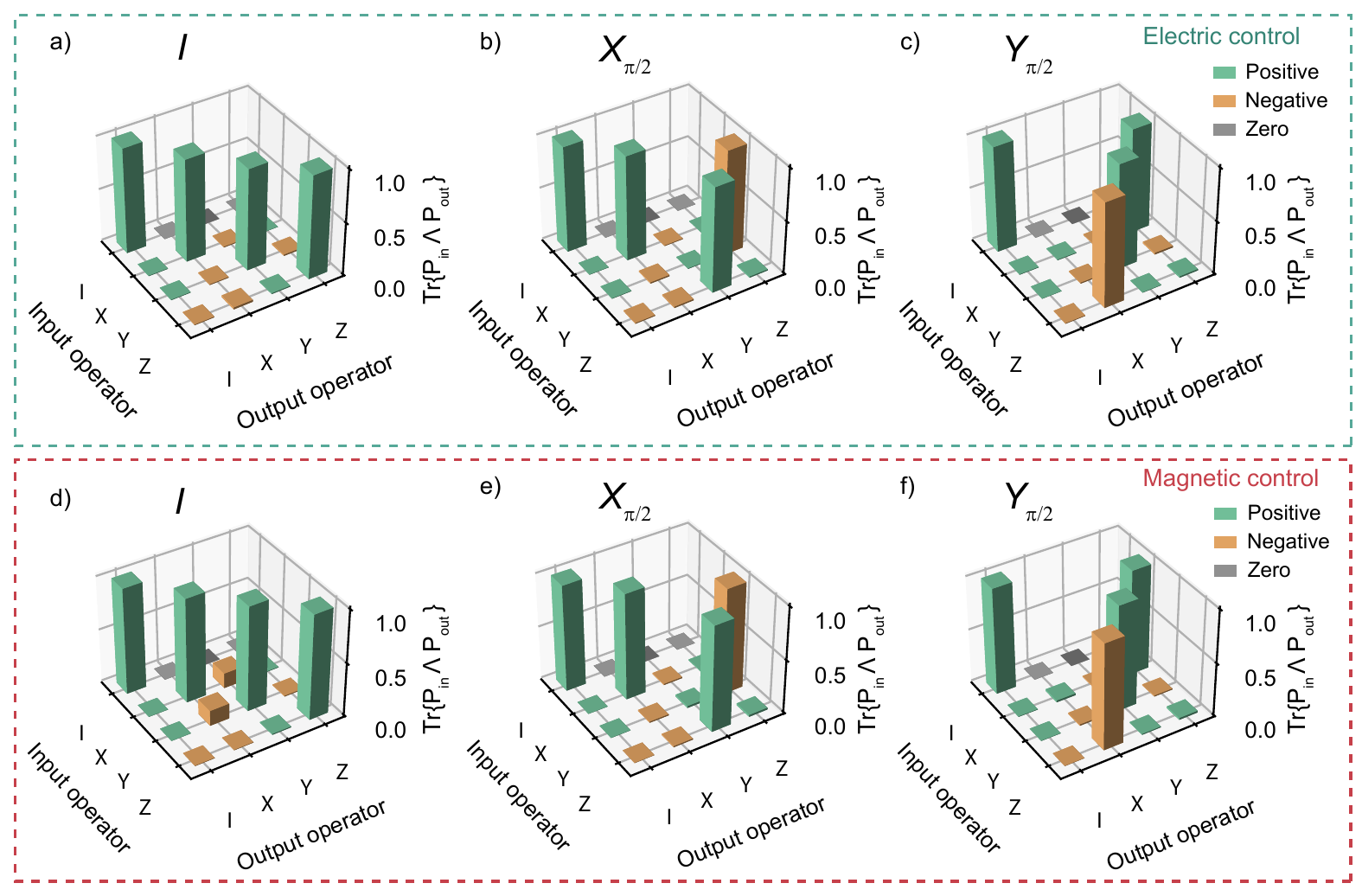}
    \caption{ \textbf{Process matrices for ionized nuclear 1Q-GST}. \textbf{a)-c)} Estimated process matrices for the qubit gates a) $\mathds{I}$, b) $X_{\pi/2}$ and c) $Y_{\pi/2}$, obtained using NER drive. \textbf{d)-e)} Estimated process matrices for the qubit gates d) $\mathds{I}$, e) $X_{\pi/2}$ and f) $Y_{\pi/2}$, obtained using NMR drive. In both cases, a circuit depth of L = 8 was used, corresponding to 448 circuits.}
\label{add_fig:GST_process_matrices}
\end{figure}
The one-qubit GST experiment aims to investigate the performance of the gates $\mathds{I}$ (idle gate), $X_{\pi/2}$ (a $\pi/2$ rotation around the $x$-axis), and $Y_{\pi/2}$ (a $\pi/2$ rotation around the $y$-axis). In our experiments, we apply NMR/NER on-resonance pulses to the magnetic antenna/donor gate for the $X_{\pi/2}$ and $Y_{\pi/2}$ gates, while the idle gate $\mathds{I}$ was an off-resonant NMR/NER pulse detuned by \SI{1}{\mega\hertz} from the resonance frequency.

To create the gate set, we used six fiducial sequences, namely, ($\mathds{I}$, $X_{\pi/2}$, $Y_{\pi/2}$, $X_{\pi/2}X_{\pi/2}X_{\pi/2}$, $Y_{\pi/2}Y_{\pi/2}Y_{\pi/2}$, $X_{\pi/2}X_{\pi/2}$), where $X_{\pi/2}$ ($Y_{\pi/2}$) are noisy $\pi/2$ rotations around $x$ ($y$). These fiducials map the qubit density matrix $\rho$ to the six Pauli eigenstates, defining an informationally complete experimental reference frame. We then selected the smallest set of germs ($\mathds{I}$, $X_{\pi/2}$, $Y_{\pi/2}$, $X_{\pi/2}Y_{\pi/2}$, $X_{\pi/2}X_{\pi/2}Y_{\pi/2}$) that amplified errors and included them in the gate set.
Using these sets of fiducials and germs, we created the circuit list using the open-source software pyGSTi\,\cite{nielsen2020probing}.

We set the circuit depth to $L = 8$, which limits the maximum number of gates in each circuit to 8 and results in a list of 448 different circuits. 
We chose $\ket{-5/2} = \ket{0}$ and $\ket{-7/2} = \ket{1}$ as the computational basis. 
After a measurement, statistics for each circuit are gathered as a binary count of $\ket{0}$ and $\ket{1}$.  We use the software pyGSTi to create a report that returns a detailed description of the gate errors. From this report we can extract the process matrices, the contribution of each type of error in the gates and the average gate fidelities.

The error bars for the average gate fidelities are expected to scale as $O(1/L\sqrt{N_{\mathrm{reps}}})$, where $L$ is the depth of the circuits (maximum number of gates on each circuit) and $N_{\mathrm{reps}}$ is the number of total repetitions for each circuit. For the chosen values of $L = 8$ and $N_{\mathrm{reps}} \approx 100$, the expected uncertainties are $\approx 1\%$.
The GST experiments with magnetic drive ran for 16 hours, and  24 hours for electric drive. During that time we did not perform any recalibration protocol (frequency or readout retuning). The model violation remained relatively low in both cases $\sigma_{\mathrm{NMR}} = 10.35$ and $\sigma_{\mathrm{NER}} = 35.94$.

Coherent errors dominate the total error in both magnetic and electric drive cases, accounting for more than 95\% of the total error. These errors result from over/under rotations of the qubit and can be corrected by adjusting the duration of specific pulses ($\mathds{I}$, $X_{\pi/2}$, and $Y_{\pi/2}$). Stochastic errors remained minimal, constituting less than 5\% of the total error for both NMR and NER drive. This outcome is expected since the maximum gate set length (L = 8) ensured that the longest pulse sequences were approximately 2\,ms for NMR and 10\,ms for NER. These durations are well below the dephasing times ($T^{*}_{2n+}\approx29$ ms). Affine errors were the smallest of the errors in both cases, representing < 2\% of the total error.

\section*{S11: Electron $T_1$ time}
 	\begin{figure}[ht]
 		\centering
 		\includegraphics{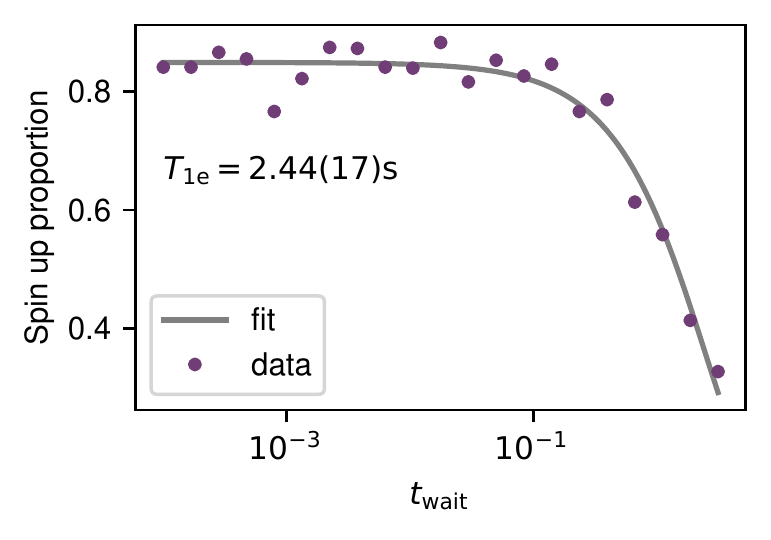}
 		\caption{ \textbf{Electron spin relaxation time $T_{1\mathrm{e}}$} Electron spin relaxation for \Sb measured in the nuclear state $m_{I} = 7/2$. The data is fitted using an exponentially decaying function  $P_0\exp{(-t_{\mathrm{wait}}/T_{1\mathrm{e}})}+P_{\mathrm{offset}}$, revealing a relaxation time of the electron into the $\ket{\downarrow}$ state of $T_{1\mathrm{e}} = 2.44(17)$\,s. 
 	 } 		
   \label{add_fig:T1_electron}
 	\end{figure}
We measure the relaxation time for the electron in the lowest electron-nuclear energy state, free of flip-flop transition. After initializing the nuclear state using the method outlined in Supplementary Section~3,
we load an electron in the $\ket{\downarrow}$ state, and invert it to the $\ket{\uparrow}$ state using an adiabatic ESR pulse. While monitoring the spin-up population as a function of $t_{\rm wait}$  (which equals the time between excitation to $\ket{\uparrow}$ and electron readout), we can measure the decay in spin-up fraction, due to the relaxation of the electron spin to the $\ket{\downarrow}$ state. At $B_0\approx 1T$ we measure $T_{1e} = \SI{2.44(17)}{\second}$ 
(Fig.\,\ref{add_fig:T1_electron}), which is similar to the typical values found in $^{31}$P donor electrons near a SiO$_2$ interface\,\cite{tenberg2019electron}.\\

\section*{S12: Coherence times} 
\begin{figure}[ht]
    \centering
    \includegraphics[width = \textwidth]{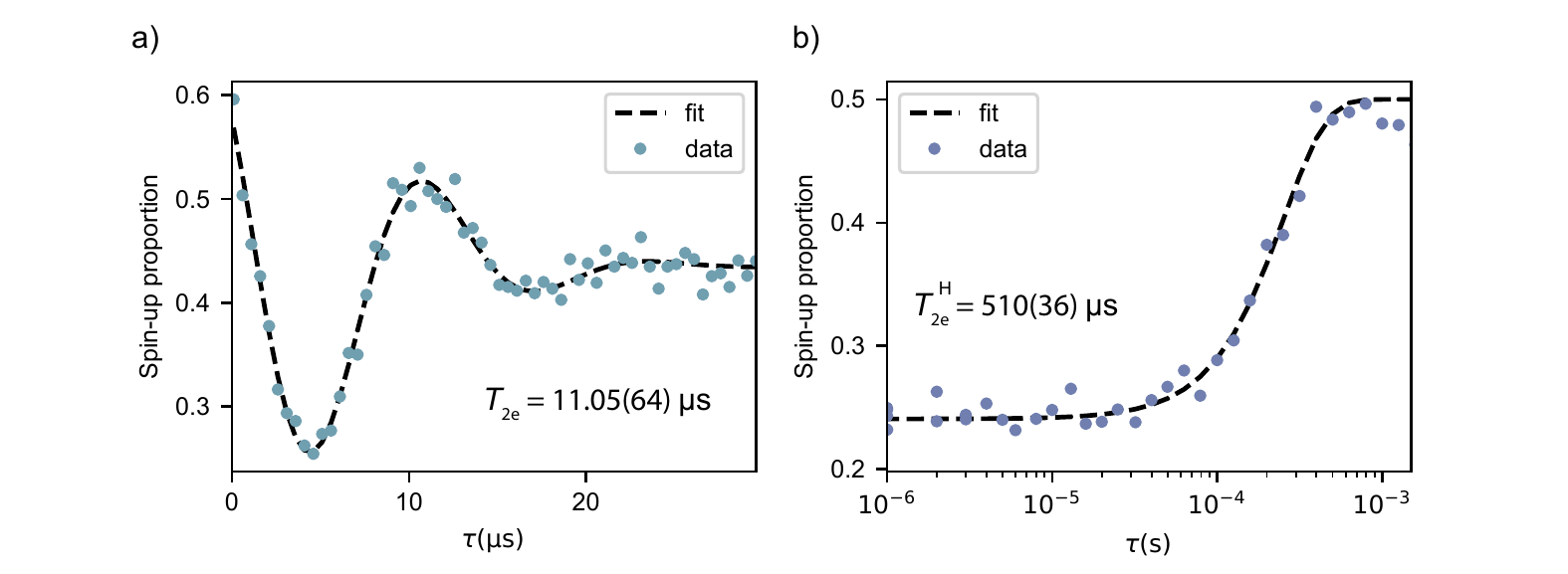}
    		\caption{\textbf{\Sb electron coherence times.} \textbf{a)} Off-resonant Ramsey experiment ($X_{\pi/2}-\tau-X_{\pi/2}$) on the electron of a \Sb donor for the transition with $m_{I} = 7/2$. The mean of 11 repetitions is fitted to a Gaussian decaying sinusoid (see text) and reveals a $T_{2\mathrm{e}}^{*} = \SI{11.06(64)}{\micro\second}$.  \textbf{b)} Hahn echo experiment ($X_{\pi/2}-\tau-X_{\pi}-\tau-X_{\pi/2}$) on the electron for the transition $m_I = 7/2$ . The decaying curve is fitted with an exponentially decaying function (see main text), and reveals a $T_{2\mathrm{e}}^{\mathrm{H}} = \SI{510(36)}{\micro\second}$ and a $\beta_{\mathrm{e}}^{\mathrm{H}} = 1.67(25)$.}
\label{add_fig:T2_ESR_Sb}	
\end{figure}

We measured the coherence times for the electron spin of the \Sb atom, in the nuclear state $m_I = \ket{7/2}$.  Figure\,\ref{add_fig:T2_ESR_Sb}\,a shows a Ramsey experiment performed on the electron. By fitting the curve to a Gaussian decaying sinusoid $P\exp{(-\tau/T_{2\rm e}^{*})}^{2} \sin{(2\pi f \tau+\phi)}+P_{\rm{offset}}$, we obtain $T_{2\rm e}^{*} = 11.05(64)~\mu\rm s$. Note that for this Ramsey experiment, we detune the $X_{\pi/2}$ pulse from the resonance frequency $f^{\rm{ESR}}_{7/2}$, resulting in oscillations the spin-up proportion as a function of $\tau$, known as the Ramsey fringes.  Also, note that the duration of the $X_{\pi/2}$ pulse is calibrated to give a maximum return probability (spin-up proportion) for $\tau = 0$ for a $X_{\pi/2}$ pulse on resonance with $f_{7/2}^{\rm{ESR}}$.

 We further characterize the coherence times of the \Sb electron with a Hahn echo experiment. The decoupling sequence extends the coherence times of the electron, as shown in Fig.\,\ref{add_fig:T2_ESR_Sb}\,b where we measure $T_{2\rm e}^{\mathrm{H}} = 510(36)~\mu\rm s$ obtained by fitting the decaying curve to  $P\exp{(-\tau/T_{2\rm e}^{\mathrm{H}})}^{\beta_{\rm e}^{\mathrm{H}}}+P_{\mathrm{offset}}$ where $\beta_{\rm e}^{\mathrm{H}} = 1.67(25)$.
 
\begin{figure}[ht]
    \centering
    \includegraphics{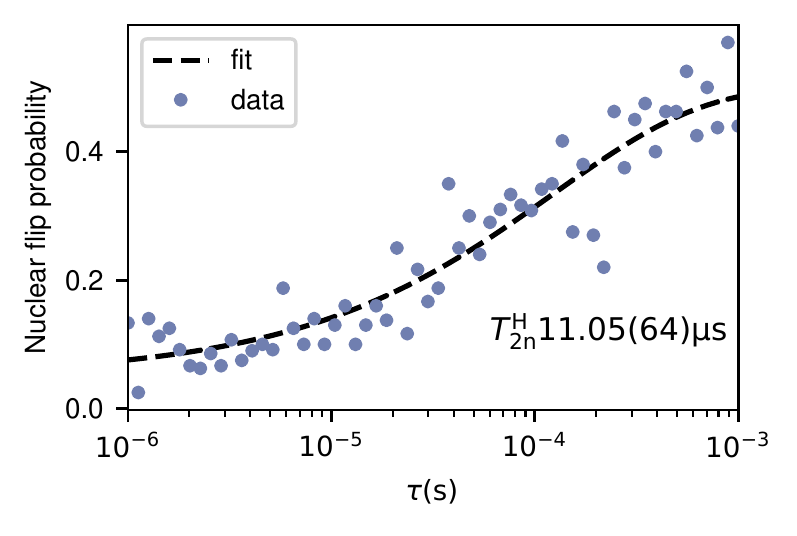}
    		\caption{\textbf{\Sb nuclear coherence times.} Hahn echo experiment on the transition $\ket{5/2}\leftrightarrow \ket{7/2}$. The decay is fitted with an exponentially decaying curve $P\exp{(\tau/T_{2\rm{n0}}^{\rm{H}})}^{\beta_{\rm{n0}}^{\mathrm{H}}}+P_{\mathrm{offset}}$. The fitting reveals a $T_{2\rm{n0}}^{\rm{H}} = \SI{247(11)}{\micro\second}$.}
\label{add_fig:T2H_NMR_Sb}	
\end{figure}

We next measured the dephasing times on the neutral nucleus using a Hahn echo sequence. The mean of five repetitions (Fig.\,\ref{add_fig:T2H_NMR_Sb}\,c) is fitted to an exponential decay $P\exp{(\tau/T_{2\mathrm{n0}}^{\mathrm{H}})}^{\beta_{\mathrm{n0}}^{\mathrm{H}}}+P_{\mathrm{offset}}$, from which we extract a coherence time $T_{2\rm{n0}}^{\rm{H}} = 247(41)~\mu\rm s$.  
We note that this value sets a lower bound to the Rabi rates needed to observe coherent drive on the neutral atom. This potentially explains our inability to achieve electrical control over the neutral nucleus in this work, since NER is a slow process, with typical Rabi periods on the order of milliseconds (Fig.\ref{add_fig:NMR_Rabi_oscillations}\,c-d). 
\\

\section*{S12: Linear quadrupolar Stark effect ionized nucleus}

Here we present the data from which the LSQE on the ionised nucleus was extracted. The data in Fig.~\ref{Supp:LQSE_ionized} is obtained from an NMR spectrum for the ionised donor, taken as a function of the DC bias amplitude $V_{\mathrm{DC}}^{\mathrm{DG1}}$. The inset shows a linear fit to the data, where the slope has been scaled by a factor of $\left(m_I-1/2\right) = 3$ for this transition.

\label{Supp:LQSE_ionized}
\begin{figure}[ht]
    \centering
    \includegraphics{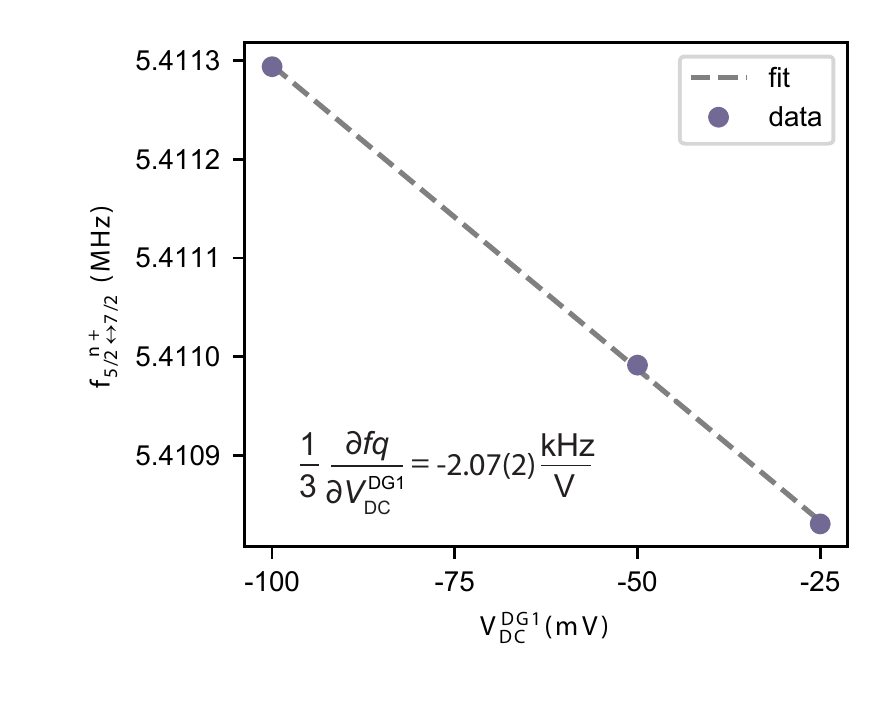}
    \caption{ \textbf{LQSE $\Delta m = 1$ for the ionised \Sb donor}. Linear quadrupolar Stark effect measured for the transition $\ket{5/2}\leftrightarrow\ket{7/2}$ in the ionised nucleus. 
 } 		
\label{add_fig:LQSE_ionised}
\end{figure}

\newpage

\bibliography{supp}